\documentclass[a4paper,11pt]{article}

\usepackage{jheppub} 

\usepackage{multirow}
\usepackage{amsmath}
\usepackage{xspace}
\usepackage{slashed}
\usepackage{bbold}
\usepackage{setspace}
\usepackage{longtable}
\usepackage{booktabs}
\usepackage{array}
\usepackage{lscape}
\usepackage[export]{adjustbox}
\usepackage{subcaption}
\usepackage{dashrule}

\usepackage[compat=1.1.0]{tikz-feynman}

\usepackage{xcolor}
\definecolor{tit}{rgb}{0.1,0.2,0.4}
\definecolor{pol}{rgb}{0,0.4,0}
\definecolor{verde}{cmyk}{0.92,0,0.59,0.25}
\definecolor{blue-plot}{rgb}{0.16,0.75,0.83}
\definecolor{red-plot}{rgb}{0.83,0,0}

\renewcommand{\arraystretch}{1.2}

\newcommand{\eq}[1]{\begin{align} #1 \end{align}}

\newcommand{\eqa}[1]{\begin{align} #1 \end{align}}

\newcommand{\av}[1]{\langle #1 \rangle}

\newcommand{\GeV}{\,{\rm GeV}}

\newcommand{\op}{\mathcal{O}}
\newcommand{\Eq}[1]{Eq.~(\ref{#1})}
\newcommand{\Eqs}[2]{Eqs.~(\ref{#1})-(\ref{#2})}
\newcommand{\Sec}[1]{Section~\ref{#1}}
\newcommand{\App}[1]{Appendix~\ref{#1}}
\newcommand{\Reff}[1]{Ref.~\cite{#1}}
\newcommand{\Refs}[1]{Refs.~\cite{#1}}
\newcommand{\Tab}[1]{Table~\ref{#1}}

\newcommand{\hc}{\mathrm{h.c.}}

\newcommand{\rut}[1]{{\tt \color{verde} #1}}
\newcommand{\pkg}[1]{{\tt #1}\xspace}
\newcommand{\dsix}{\pkg{DsixTools}}

\usepackage{float}

\usepackage{braket,slashed,bm}
\usepackage{array,multirow}
\usepackage[normalem]{ulem}

\usepackage[T1]{fontenc} 

\usepackage{mathrsfs}

\usepackage{booktabs}
\usepackage{adjustbox}
\usepackage{mathtools}

\usepackage{soul}

\usepackage{tikz}
\usetikzlibrary{arrows,decorations.pathmorphing,backgrounds,positioning,fit,petri,automata,shadows,calendar,mindmap,
decorations.markings,calc,arrows.meta,bending}

\definecolor{labelkey}{rgb}{0,0.5,0.0}

\usepackage{xparse}
\usepackage{etoolbox}

\usepackage{feynmp}
\usepackage{
}

\def\sc{{\overline s}}

\renewcommand{\O}{\mathcal{O}}
\newcommand{\cS}{\mathcal{S}}

\NewDocumentCommand{\Op}{ m m O{} o }{
	\O^{\ifblank{#3}{}{#3,}#2 }_{\IfNoValueTF{#4}{#1}{\substack{#1\\#4}}}
}
\NewDocumentCommand{\tOp}{ m m O{} o }{
	\widetilde{\O}^{\ifblank{#3}{}{#3,}#2 }_{\IfNoValueTF{#4}{#1}{\substack{#1\\#4}}}
}
\NewDocumentCommand{\Sym}{ m m O{} o }{
	\cS^{\ifblank{#3}{}{#3,}#2 }_{\IfNoValueTF{#4}{#1}{\substack{#1\\#4}}}
}

\NewDocumentCommand{\lwc}{ m m O{} o }{
	L^{\ifblank{#3}{}{#3,}#2 }_{\IfNoValueTF{#4}{#1}{\substack{#1\\#4}}}
}
\NewDocumentCommand{\dlwc}{ m m O{} o }{
	{\dot L}^{\ifblank{#3}{}{#3,}#2 }_{\IfNoValueTF{#4}{#1}{\substack{#1\\#4}}}
}

\allowdisplaybreaks

\preprint{CERN-TH-2025-007}

\title{Two-Loop Anomalous Dimensions in the LEFT: Dimension-Six Four-Fermion Operators in NDR}

\author{Jason Aebischer{$^{\, a}$},
Pol Morell{$^{\, b,c}$},
Marko Pesut{$^{\, d}$}
and
Javier Virto{$^{\, b,c}$}}

\affiliation{{$^{\, a}$} 
Theoretical Physics Department, CERN, 1211 Geneva 23, Switzerland}

\affiliation{{$^{\, b}$}
Departament de Física Quàntica i Astrofísica, Universitat de Barcelona,\\
Martí Franquès 1, 08028 Barcelona, Spain}

\affiliation{
{$^{\, c}$}
Institut de Ciències del Cosmos (ICCUB), Universitat de Barcelona,\\
Martí Franquès, 1, 08028 Barcelona, Spain
}

\affiliation{{$^{\, d}$} 
Physik-Institut, Universit\"at Z\"urich, CH-8057 Z\"urich, Switzerland\\[-1mm]}

\emailAdd{jason.aebischer@cern.ch}
\emailAdd{pmorell@icc.ub.edu}
\emailAdd{marko.pesut@physik.uzh.ch}
\emailAdd{jvirto@icc.ub.edu}

\abstract{
We derive the complete set of two-loop anomalous dimensions describing the mixing of four-fermion operators in the Low Energy Effective Field Theory (LEFT). The calculation is performed in Naive Dimensional Regularization with anticommuting~$\gamma_5$ (the NDR scheme), and the results are given in the ``JMS basis'' of dimension-six operators. The derivation relies on known results for UV poles in two-loop diagrams in QCD, which are then used to derive the two-loop Anomalous Dimension Matrix (ADM) for the full set of four-fermion operators including $O(\alpha_s^2)$, $O(\alpha_s\alpha)$ and $O(\alpha^2)$ corrections. The method employed is an extension of a common approach to deal with traces containing $\gamma_5$ in NDR. Our results have been implemented in the public code \texttt{DsixTools}. We also discuss and provide the results in the LEFT with 5, 4 and 3 active quark flavors.
}

\begin{document}
\maketitle
\flushbottom

\section{Introduction}
\label{sec:intro}

With the exception of the top quark, with a Yukawa coupling of order one, all the other fermions in the Standard Model (SM) have small Yukawa couplings. This leads to a hierarchy between the masses of Electroweak-scale particles (the $W$ and $Z$ bosons, the Higgs boson and the top quark) and the masses of all the other SM fermions, $\Lambda_\text{EW}\sim v \gg m_b$.
Any observables where all kinematic invariants $s_{ij}\equiv p_i\cdot p_j$ are small with respect to $\Lambda_\text{EW}$ can then be expanded in powers of $s_{ij}/\Lambda_\text{EW}^2$ and $m/\Lambda_\text{EW}$, with $m$ the mass of a light fermion, and with expansion coefficients containing large logarithms of these ratios, e.g.~$\log{(m_b/m_W)}$. This expansion can be framed in the context of an Effective Field Theory (EFT), which at the same time implements scale factorization and allows to resum the large logarithms.

This EFT is a generalization of Fermi's theory of weak interactions~\cite{Fermi:1933jpa}, 
and it is described by an effective Lagrangian built out of ``light'' SM fields, with local operators invariant under $SU(3)_c\times U(1)_\text{em}$ gauge transformations, and organized in increasing canonical mass dimension. Leading terms of dimension four or less give rise to the renormalizable theory of light quarks and leptons interacting with photons and gluons; dimension-five terms give rise to electromagnetic and chromomagnetic dipole interactions, and dimension-six terms contain (for example) local four-fermion interactions of the Fermi type.

Since at the renormalizable level this EFT contains no weak interactions, it is in a sense a low-energy effective theory for weak interactions, and has therefore been called the ``Weak effective Hamiltonian''~\cite{Buchalla:1995vs,Buras:2011we}, or the ``Weak Effective Theory'' (WET)~\cite{Aebischer:2017gaw}. However, it is also the low-energy EFT of any quantum field theory which shares the field content and gauge symmetry with the SM below the Electroweak (EW) scale. This includes Beyond-the-SM~(BSM) theories where all the new particles are heavier than the EW scale. Thus, it has also recently been called the ``Low-Energy Effective Field Theory'' (LEFT)~\cite{Jenkins:2017jig}.

The LEFT has been studied in detail as the low-energy EFT of the SM~\cite{Grinstein:1987vj,Grigjanis:1988iq,Grigjanis:1989py,Grinstein:1990tj,Buras:1990fn,Cella:1990sh,Cella:1990bn,Buras:1989xd,Dugan:1990df,Buras:1991jm,Buras:1992tc,Buras:1992zv,Ciuchini:1993vr,Ciuchini:1993ks,Misiak:1994zw,Herrlich:1994kh,Chetyrkin:1996vx,Chetyrkin:1997gb,Bobeth:1999mk,Buras:2000if,Gambino:2003zm,Bobeth:2003at,Gorbahn:2004my,Huber:2005ig,Bobeth:2013tba}. In this case, only a subset of the most general set of effective operators is present. However, motivated by the contemporary automation philosophy~\cite{Proceedings:2019rnh,Aebischer:2023nnv}, efforts are now being made to set up the LEFT in its entirety both to higher orders in the EFT expansion~\cite{Murphy:2020cly,Hamoudou:2022tdn}, and to higher orders in the loop expansion. In the latter case, a full one-loop calculation of anomalous dimensions was presented in~Refs.~\cite{Aebischer:2017gaw,Jenkins:2017dyc}, and the complete one-loop matching to the Standard Model EFT (SMEFT) was presented in~\cite{Dekens:2019ept}.
Two-loop anomalous dimensions for a set of flavor-changing four-quark operators can be found in~\Refs{Buras:2000if,Morell:2024aml,Aebischer:2021raf}, and very recently, the full set of two-loop dimension-five contributions have been calculated~\cite{Naterop:2023dek,Naterop:2024ydo}.

\bigskip

The purpose of this article is to 
complete the two-loop anomalous dimensions for the full set of four-fermion operators in the LEFT.
We achieve this by collecting and adapting known results scattered throughout the literature and expressing them consistently in a suitable operator basis.
This task is particularly challenging, partly because beyond the one-loop order anomalous dimensions are scheme dependent.
This means that one needs to properly perform a change of scheme whenever necessary, for example in the cases where our operator basis differs from the one used in the calculation by a transformation involving evanescent terms~\cite{Herrlich:1994kh,Aebischer:2022aze,Aebischer:2022rxf,Aebischer:2023djt,Aebischer:2024xnf}. Furthermore, following the lead of the first complete account of the dimension-six LEFT~\cite{Jenkins:2017jig} we will frame all our results in the so-called ``JMS'' operator basis.
This is somewhat arbitrary, and other choices may appear advantageous in various contexts. For example, the ``Bern'' basis of~\cite{Aebischer:2017gaw,Aebischer:2017ugx} (based on the earlier ``CMM'' basis for SM operators~\cite{Chetyrkin:1996vx,Chetyrkin:1997gb}) has the advantage of avoiding closed fermion loops with $\gamma_5$ factors in flavor-changing transitions.
However, this feature is no longer satisfied in flavor-conserving ones.
The ``BMU'' basis~\cite{Buras:2000if} has the advantage of preserving the SM set of operators, but this also complicates significantly the structure of BSM operators, and hides certain flavor symmetries of matching conditions and anomalous dimensions~\cite{Morell:2024aml}.
Taking into account that the tools for EFT calculations are mostly automated already (e.g.~\cite{Celis:2017hod,Fuentes-Martin:2020zaz,Aebischer:2018bkb,Proceedings:2019rnh,Aebischer:2023nnv,Aebischer:2024csk,DiNoi:2022ejg}), we feel that the most convenient choice is to stick to JMS.

The structure of this paper is the following. We begin in~\Sec{sec:LEFT} reviewing the LEFT, its renormalization at the two-loop level, and changes of basis at next-to-leading order.
In~\Sec{sec:RG-Sectors} we provide the classification of all four-fermion dimension-six operators in the LEFT in terms of RG-invariant sectors. 
In~\Sec{sec:details} we discuss the various approaches followed to reconstruct the full set of two-loop anomalous dimensions for all four-fermion dimension-six operators in the LEFT.
The results and their implementation are briefly explored numerically in ~\Sec{sec:numerics}, where we show a number of examples. 
We conclude in~\Sec{sec:conclusions}.
A few appendices at the end contain details concerning evanescent operators, the tables of pole coefficients, and a collection of the building sub-blocks, or ``ADM seeds'', for the penguin contributions.

\section{The LEFT at two-loop order}
\label{sec:LEFT}

\subsection{Definition of the Effective Theory}
\label{sec:defLEFT}

The LEFT is the EFT invariant under the $SU(3)_c \times U(1)_\text{em}$ gauge group and containing the known degrees of freedom lighter than the EW scale: quarks, leptons, photons, and gluons.
Depending on the cut-off scale, some of the fermions may be decoupled. One can thus distinguish between different versions of the theory, LEFT$(n_q)$, containing $n_q=n_u+n_d$ quarks ($n_u$ up-type and $n_d$ down-type quarks), $n_e$ charged leptons, and $n_\nu$ left-handed neutrinos. We will consider three cases of interest:
\begin{itemize}
\item LEFT(5): $n_u=2$, $n_d=n_e=n_\nu=3$,\quad with matter fields $\{u,c,d,s,b,e,\mu,\tau,\nu_{e,\mu,\tau}\}$\ ;
\item LEFT(4): $n_u=n_d=2$, $n_e=n_\nu=3$,\quad with matter fields $\{u,c,d,s,e,\mu,\tau,\nu_{e,\mu,\tau}\}$\ ;
\item LEFT(3): $n_u=1$, $n_d=n_e=2$, $n_\nu=3$,\quad with matter fields $\{u,d,s,e,\mu,\nu_{e,\mu,\tau}\}$\ .
\end{itemize}
The LEFT(5) contains the EFT of the SM below the EW scale after integrating out the Higgs boson, the massive $W^\pm$ and $Z$ gauge bosons and the top quark, but it is also the EFT of more general theories with new degrees of freedom at or above the EW scale.  The LEFT(4) and LEFT(3) are obtained when crossing the $b$- and $c$-quark thresholds at around $\mu_b\sim 5\GeV$ and $\mu_c\sim 1.5\GeV$. Given the closeness of charm-quark and tau-lepton masses, we will assume they are integrated out at the same threshold $\mu_c$. By `LEFT' we will, from here on, denote any of the LEFT$(n_q)$ theories.

The Lagrangian of the LEFT is given by
\begin{equation} \label{eq:LEFT}
\mathcal{L}_{\rm LEFT} = \mathcal{L}_{\rm QCD+QED}^{(n_q)} + \sum_k L_k^{(3)} \op_k^{(3)} + \sum_k L_k^{(5)} \op_k^{(5)}  + \sum_k L_k^{(6)} \op_k^{(6)} + O(1/\Lambda_{\rm EW}^3)
\ .
\end{equation}
The (dimensionful) Wilson coefficients $L_k^{(5)}$ and $L_k^{(6)}$ are implicitly suppressed by $1/\Lambda_{\rm EW}$ and $1/\Lambda_{\rm EW}^2$, respectively,
where $\Lambda_{\rm EW}$ is the electroweak scale.  A complete basis of operators up to dimension six was introduced in~\Reff{Jenkins:2017jig}. The explicit form of the operators $\op_k^{(d)}$ in this basis (the ``JMS basis'') and their symmetries can be checked easily from \dsix~\cite{Celis:2017hod,Fuentes-Martin:2020zaz} (e.g. with the command \rut{LEFTOperatorsGrid[]}). 

We use the following conventions:\footnote{These conventions differ from those in~\cite{Jenkins:2017jig,Jenkins:2017dyc}, in the definition of the fermion mass matrices and the name of the strong gauge coupling.} 
The QCD and QED Lagrangian is given by
\begin{align} 
\mathcal{L}_{\rm QCD+QED}^{(n_q)} &= -\frac{1}{4} G_{\mu \nu}^A G^{A, \mu \nu} - \frac{1}{4} F_{\mu \nu} F^{\mu \nu} + \theta_{\rm QCD} \, \frac{g_s^2}{32 \, \pi^2} \widetilde G_{\mu \nu}^{A}  G^{A,\mu \nu} + \theta_{\rm QED} \, \frac{e^2}{32 \, \pi^2} \widetilde F_{\mu \nu}  F^{\mu \nu} \nonumber \\
& + \sum_{\psi = u,d,e,\nu_L} \overline \psi \, i \slashed{D} \psi -  
\sum_{\psi = u,d,e} \big(\,\overline \psi_L M_\psi \psi_R + \hc \big) \ ,
\label{eq:QCDQED}
\end{align}
where $\psi_{L,R} = P_{L,R}\,\psi = \frac12(1\mp \gamma_5) \psi$.
Implicit in the second line is a sum over flavors: the Dirac mass matrices $M_{u,d,e}$ are, respectively, $n_u \times n_u$, $n_d \times n_d$ and $n_e \times n_e$ matrices in flavor space.
The covariant derivative is defined as
\begin{equation}
D_\mu = \partial_\mu + i g_s T^A G_\mu^A + i e\, Q A_\mu \; ,
\end{equation}
where $g_s$ and $e$ are the $SU(3)_c$ and $U(1)_\text{em}$ gauge couplings,
respectively. The gauge field tensors are defined as usual,
\begin{align}
G_{\mu \nu}^A & = \partial_\mu G_\nu^A - \partial_\nu G_\mu^A - g_s f^{ABC} G_\mu^B G_\nu^C \; , \\
F_{\mu \nu} & = \partial_\mu F_\nu - \partial_\nu F_\mu \; ,
\end{align}
with covariant derivatives
\begin{align}
\left( D_\rho G_{\mu \nu} \right)^A & = \partial_\rho G_{\mu \nu}^A - g_s f^{ABC} G_\rho^B G_{\mu \nu}^C \; , \\
\left( D_\rho F_{\mu \nu} \right) & = \partial_\rho F_{\mu \nu} \, .
\end{align}
Finally, the dual tensors are defined as $\widetilde X_{\mu\nu} = \frac{1}{2}
\epsilon_{\mu \nu \rho \sigma} X^{\rho \sigma}$ (with $\epsilon_{0123}
= +1$).

At mass dimension three, aside from the Dirac mass terms inside the QCD+QED Lagrangian, there is one single operator (up to flavor): $(\op_{\nu})_{ij}=\nu_{Li}^T C \nu_{Lj}$. The dimension-five operators $\op_k^{(5)}$ are QCD and QED dipole operators,
\eq{
\op_{\substack{\vphantom{}qG\\ij}} 
= (\bar{q}_{Li}\sigma^{\mu\nu}T^Aq_{Rj})\,G^A_{\mu\nu} \;, 
\ \  
\op_{\substack{\vphantom{}\psi\gamma\\ij}} = (\bar{\psi}_{Li}\sigma^{\mu\nu}\psi_{Rj})\,F_{\mu\nu} \;, 
\ \ 
\op_{\substack{\vphantom{}\nu\gamma\\ij}} = (\nu^T_{Li}C\sigma^{\mu\nu}\nu_{Lj})\,F_{\mu\nu}\ ,
}
with $q=u,d$ and $\psi=u,d,e$. 
At dimension six there are two triple-gluon operators,
\eq{
\op_{G} = f^{ABC}G_\mu^{A\hspace{0.2mm}\nu}G_\nu^{B\rho}G_\rho^{C\mu} \;, \quad \op_{\widetilde G} = f^{ABC}\widetilde{G}_\mu^{A\hspace{0.2mm}\nu}G_\nu^{B\rho}G_\rho^{C\mu} \;,
}
and a large number of four-fermion operators of the form
\eq{
\op_{\psi^4} \sim \bar \psi_1 \Gamma \psi_2 \,\bar\psi_3 \Gamma \psi_4 \ .
}
In this work we focus on these dimension-six four-fermion operators.
There are 87 different types of such operators, in the classification of~\Reff{Jenkins:2017jig} (see~Tables~\ref{tab:LEFTops1} and~\ref{tab:LEFTops2}). In the LEFT(5) -- including flavor indices-- there are
$235$ hermitian and $2\times 2819$ non-hermitian operators, with a total of $5873$ real paramaters.
These numbers, as well as the corresponding ones for LEFT(4) and LEFT(3), can be extracted from~Tables~\ref{tab:RG-sectors I} and~\ref{tab:RG-sectors II}. In~\Sec{sec:RG-Sectors} we provide a complete classification of the four-fermion operators in sectors invariant under renormalization.

\begin{table}
\renewcommand{\arraystretch}{1.37}
\begin{center}
\begin{adjustbox}{width=1\textwidth,center=\textwidth}
\begin{tabular}{|c|c||c|c||c|c|}
\hline
\multicolumn{2}{|c||}{$\left( \bar L L \right) \left( \bar L L \right)$} & \multicolumn{2}{|c||}{$\left( \bar L L \right) \left( \bar R R \right)$} & \multicolumn{2}{|c|}{$\left( \bar L R \right) \left( \bar L R \right)$} \\
\hline
$\op_{\nu \nu}^{V,LL}$ & $\left( \bar \nu_L \gamma_\mu \nu_L \right) \left( \bar \nu_L \gamma^\mu \nu_L \right)$ & $\op_{\nu e}^{V,LR}$ & $\left( \bar \nu_L \gamma_\mu \nu_L \right) \left( \bar e_R \gamma^\mu e_R \right)$ & $\op_{ee}^{S,RR}$ & $\left( \bar e_L e_R \right) \left( \bar e_L e_R \right)$ \\
$\op_{ee}^{V,LL}$ & $\left( \bar e_L \gamma_\mu e_L \right) \left( \bar e_L \gamma^\mu e_L \right)$ & $\op_{e e}^{V,LR}$ & $\left( \bar e_L \gamma_\mu e_L \right) \left( \bar e_R \gamma^\mu e_R \right)$ & $\op_{eu}^{S,RR}$ & $\left( \bar e_L e_R \right) \left( \bar u_L u_R \right)$ \\
$\op_{\nu e}^{V,LL}$ & $\left( \bar \nu_L \gamma_\mu \nu_L \right) \left( \bar e_L \gamma^\mu e_L \right)$ & $\op_{\nu u}^{V,LR}$ & $\left( \bar \nu_L \gamma_\mu \nu_L \right) \left( \bar u_R \gamma^\mu u_R \right)$ & $\op_{eu}^{T,RR}$ & $\left( \bar e_L \sigma_{\mu \nu} e_R \right) \left( \bar u_L \sigma^{\mu \nu} u_R \right)$ \\
$\op_{\nu u}^{V,LL}$ & $\left( \bar \nu_L \gamma_\mu \nu_L \right) \left( \bar u_L \gamma^\mu u_L \right)$ & $\op_{\nu d}^{V,LR}$ & $\left( \bar \nu_L \gamma_\mu \nu_L \right) \left( \bar d_R \gamma^\mu d_R \right)$ & $\op_{ed}^{S,RR}$ & $\left( \bar e_L e_R \right) \left( \bar d_L d_R \right)$ \\
$\op_{\nu d}^{V,LL}$ & $\left( \bar \nu_L \gamma_\mu \nu_L \right) \left( \bar d_L \gamma^\mu d_L \right)$ & $\op_{e u}^{V,LR}$ & $\left( \bar e_L \gamma_\mu e_L \right) \left( \bar u_R \gamma^\mu u_R \right)$ & $\op_{ed}^{T,RR}$ & $\left( \bar e_L \sigma_{\mu \nu} e_R \right) \left( \bar d_L \sigma^{\mu \nu} d_R \right)$ \\
$\op_{e u}^{V,LL}$ & $\left( \bar e_L \gamma_\mu e_L \right) \left( \bar u_L \gamma^\mu u_L \right)$ & $\op_{e d}^{V,LR}$ & $\left( \bar e_L \gamma_\mu e_L \right) \left( \bar d_R \gamma^\mu d_R \right)$ & $\op_{\nu e d u}^{S,RR}$ & $\left( \bar \nu_L e_R \right) \left( \bar d_L u_R \right)$ \\
$\op_{e d}^{V,LL}$ & $\left( \bar e_L \gamma_\mu e_L \right) \left( \bar d_L \gamma^\mu d_L \right)$ & $\op_{u e}^{V,LR}$ & $\left( \bar u_L \gamma_\mu u_L \right) \left( \bar e_R \gamma^\mu e_R \right)$ & $\op_{\nu e d u}^{T,RR}$ & $\left( \bar \nu_L \sigma_{\mu \nu} e_R \right) \left( \bar d_L \sigma^{\mu \nu} u_R \right)$ \\
$\op_{\nu e d u}^{V,LL}$ & $\left( \bar \nu_L \gamma_\mu e_L \right) \left( \bar d_L \gamma^\mu u_L \right)$ & $\op_{d e}^{V,LR}$ & $\left( \bar d_L \gamma_\mu d_L \right) \left( \bar e_R \gamma^\mu e_R \right)$ & $\op_{u u}^{S1,RR}$ & $\left( \bar u_L u_R \right) \left( \bar u_L u_R \right)$ \\
$\op_{u u}^{V,LL}$ & $\left( \bar u_L \gamma_\mu u_L \right) \left( \bar u_L \gamma^\mu u_L \right)$ & $\op_{\nu e d u}^{V,LR}$ & $\left( \bar \nu_L \gamma_\mu e_L \right) \left( \bar d_R \gamma^\mu u_R \right)$ & $\op_{u u}^{S8,RR}$ & $\left( \bar u_L T^A u_R \right) \left( \bar u_L T^A u_R \right)$ \\
$\op_{d d}^{V,LL}$ & $\left( \bar d_L \gamma_\mu d_L \right) \left( \bar d_L \gamma^\mu d_L \right)$ & $\op_{u u}^{V1,LR}$ & $\left( \bar u_L \gamma_\mu u_L \right) \left( \bar u_R \gamma^\mu u_R \right)$ & $\op_{u d}^{S1,RR}$ & $\left( \bar u_L u_R \right) \left( \bar d_L d_R \right)$ \\
$\op_{u d}^{V1,LL}$ & $\left( \bar u_L \gamma_\mu u_L \right) \left( \bar d_L \gamma^\mu d_L \right)$ & $\op_{u u}^{V8,LR}$ & $\left( \bar u_L \gamma_\mu T^A u_L \right) \left( \bar u_R \gamma^\mu T^A u_R \right)$ & $\op_{u d}^{S8,RR}$ & $\left( \bar u_L T^A u_R \right) \left( \bar d_L T^A d_R \right)$ \\
$\op_{u d}^{V8,LL}$ & $\left( \bar u_L \gamma_\mu T^A u_L \right) \left( \bar d_L \gamma^\mu T^A d_L \right)$ & $\op_{u d}^{V1,LR}$ & $\left( \bar u_L \gamma_\mu u_L \right) \left( \bar d_R \gamma^\mu d_R \right)$ & $\op_{d d}^{S1,RR}$ & $\left( \bar d_L d_R \right) \left( \bar d_L d_R \right)$ \\
\cline{1-2}
\multicolumn{2}{|c||}{$\left( \bar R R \right) \left( \bar R R \right)$} & $\op_{u d}^{V8,LR}$ & $\left( \bar u_L \gamma_\mu T^A u_L \right) \left( \bar d_R \gamma^\mu T^A d_R \right)$ & $\op_{d d}^{S8,RR}$ & $\left( \bar d_L T^A d_R \right) \left( \bar d_L T^A d_R \right)$ \\
\cline{1-2}
$\op_{e e}^{V,RR}$ & $\left( \bar e_R \gamma_\mu e_R \right) \left( \bar e_R \gamma^\mu e_R \right)$ & $\op_{d u}^{V1,LR}$ & $\left( \bar d_L \gamma_\mu d_L \right) \left( \bar u_R \gamma^\mu u_R \right)$ & $\op_{uddu}^{S1,RR}$ & $\left( \bar u_L d_R \right) \left( \bar d_L u_R \right)$ \\
$\op_{e u}^{V,RR}$ & $\left( \bar e_R \gamma_\mu e_R \right) \left( \bar u_R \gamma^\mu u_R \right)$ & $\op_{d u}^{V8,LR}$ & $\left( \bar d_L \gamma_\mu T^A d_L \right) \left( \bar u_R \gamma^\mu T^A u_R \right)$ & $\op_{uddu}^{S8,RR}$ & $\left( \bar u_L T^A d_R \right) \left( \bar d_L T^A u_R \right)$ \\
\cline{5-6}
$\op_{e d}^{V,RR}$ & $\left( \bar e_R \gamma_\mu e_R \right) \left( \bar d_R \gamma^\mu d_R \right)$ & $\op_{d d}^{V1,LR}$ & $\left( \bar d_L \gamma_\mu d_L \right) \left( \bar d_R \gamma^\mu d_R \right)$ & \multicolumn{2}{|c|}{$\left( \bar L R \right) \left( \bar R L \right)$} \\
\cline{5-6}
$\op_{u u}^{V,RR}$ & $\left( \bar u_R \gamma_\mu u_R \right) \left( \bar u_R \gamma^\mu u_R \right)$ & $\op_{d d}^{V8,LR}$ & $\left( \bar d_L \gamma_\mu T^A d_L \right) \left( \bar d_R \gamma^\mu T^A d_R \right)$ & $\op_{eu}^{S,RL}$ & $\left( \bar e_L e_R \right) \left( \bar u_R u_L \right)$ \\
$\op_{d d}^{V,RR}$ & $\left( \bar d_R \gamma_\mu d_R \right) \left( \bar d_R \gamma^\mu d_R \right)$ & $\op_{uddu}^{V1,LR}$ & $\left( \bar u_L \gamma_\mu d_L \right) \left( \bar d_R \gamma^\mu u_R \right)$ & $\op_{ed}^{S,RL}$ & $\left( \bar e_L e_R \right) \left( \bar d_R d_L \right)$ \\
$\op_{u d}^{V1,RR}$ & $\left( \bar u_R \gamma_\mu u_R \right) \left( \bar d_R \gamma^\mu d_R \right)$ & $\op_{uddu}^{V8,LR}$ & $\left( \bar u_L \gamma_\mu T^A d_L \right) \left( \bar d_R \gamma^\mu T^A u_R \right)$ & $\op_{\nu e d u}^{S,RL}$ & $\left( \bar \nu_L e_R \right) \left( \bar d_R u_L \right)$ \\
$\op_{u d}^{V8,RR}$ & $\left( \bar u_R \gamma_\mu T^A u_R \right) \left( \bar d_R \gamma^\mu T^A d_R \right)$ & & & & \\
\hline
\end{tabular}
\end{adjustbox}
\end{center}
\vspace{-2mm}
\caption{\it LEFT Baryon and Lepton number conserving dim-6, four-fermion operators.\\[4mm]
\label{tab:LEFTops1}}
\end{table}

\begin{table}
\renewcommand{\arraystretch}{1.37}
\begin{center}
\small
\begin{tabular}{|c|c||c|c||c|c|}
\hline
\multicolumn{2}{|c||}{$\Delta L = 2$} & \multicolumn{2}{|c||}{$\Delta B = \Delta L = 1$} & \multicolumn{2}{|c|}{$\Delta B = - \Delta L = 1$} \\
\hline
$\op_{\nu e}^{S,LL}$ & $\left( \nu_L^T C \nu_L \right) \left( \bar e_R e_L \right)$ & $\op_{udd}^{S,LL}$ & $\left( u_L^T C d_L \right) \left( d_L^T C \nu_L \right)$ & $\op_{ddd}^{S,LL}$ & $\left( d_L^T C d_L \right) \left( \bar e_R d_L \right)$ \\
$\op_{\nu e}^{T,LL}$ & $\left( \nu_L^T C \sigma_{\mu \nu} \nu_L \right) \left( \bar e_R \sigma^{\mu \nu} e_L \right)$ & $\op_{duu}^{S,LL}$ & $\left( d_L^T C u_L \right) \left( u_L^T C e_L \right)$ & $\op_{udd}^{S,LR}$ & $\left( u_L^T C d_L \right) \left( \bar \nu_L d_R \right)$ \\
$\op_{\nu e}^{S,LR}$ & $\left( \nu_L^T C \nu_L \right) \left( \bar e_L e_R \right)$ & $\op_{uud}^{S,LR}$ & $\left( u_L^T C u_L \right) \left( d_R^T C e_R \right)$ & $\op_{ddu}^{S,LR}$ & $\left( d_L^T C d_L \right) \left( \bar \nu_L u_R \right)$ \\
$\op_{\nu u}^{S,LL}$ & $\left( \nu_L^T C \nu_L \right) \left( \bar u_R u_L \right)$ & $\op_{duu}^{S,LR}$ & $\left( d_L^T C u_L \right) \left( u_R^T C e_R \right)$ & $\op_{ddd}^{S,LR}$ & $\left( d_L^T C d_L \right) \left( \bar e_L d_R \right)$ \\
$\op_{\nu u}^{T,LL}$ & $\left( \nu_L^T C \sigma_{\mu \nu} \nu_L \right) \left( \bar u_R \sigma^{\mu \nu} u_L \right)$ & $\op_{uud}^{S,RL}$ & $\left( u_R^T C u_R \right) \left( d_L^T C e_L \right)$ & $\op_{ddd}^{S,RL}$ & $\left( d_R^T C d_R \right) \left( \bar e_R d_L \right)$ \\
$\op_{\nu u}^{S,LR}$ & $\left( \nu_L^T C \nu_L \right) \left( \bar u_L u_R \right)$ & $\op_{duu}^{S,RL}$ & $\left( d_R^T C u_R \right) \left( u_L^T C e_L \right)$ & $\op_{udd}^{S,RR}$ & $\left( u_R^T C d_R \right) \left( \bar \nu_L d_R \right)$ \\
$\op_{\nu d}^{S,LL}$ & $\left( \nu_L^T C \nu_L \right) \left( \bar d_R d_L \right)$ & $\op_{dud}^{S,RL}$ & $\left( d_R^T C u_R \right) \left( d_L^T C \nu_L \right)$ & $\op_{ddd}^{S,RR}$ & $\left( d_R^T C d_R \right) \left( \bar e_L d_R \right)$ \\
\cline{5-6}
$\op_{\nu d}^{T,LL}$ & $\left( \nu_L^T C \sigma_{\mu \nu} \nu_L \right) \left( \bar d_R \sigma^{\mu \nu} d_L \right)$ & $\op_{ddu}^{S,RL}$ & $\left( d_R^T C d_R \right) \left( u_L^T C \nu_L \right)$ & \multicolumn{2}{|c|}{$\Delta L = 4$} \\
\cline{5-6}
$\op_{\nu d}^{S,LR}$ & $\left( \nu_L^T C \nu_L \right) \left( \bar d_L d_R \right)$ & $\op_{duu}^{S,RR}$ & $\left( d_R^T C u_R \right) \left( u_R^T C e_R \right)$ & $\op_{\nu \nu}^{S,LL}$ & $\left( \nu_L^T C \nu_L \right) \left( \nu_L^T C \nu_L \right)$ \\
$\op_{\nu e d u}^{S,LL}$ & $\left( \nu_L^T C e_L \right) \left( \bar d_R u_L \right)$ & & & & \\
$\op_{\nu e d u}^{T,LL}$ & $\left( \nu_L^T C \sigma_{\mu \nu} e_L \right) \left( \bar d_R \sigma^{\mu \nu} u_L \right)$ & & & & \\
$\op_{\nu e d u}^{S,LR}$ & $\left( \nu_L^T C e_L \right) \left( \bar d_L u_R \right)$ & & & & \\
$\op_{\nu e d u}^{V,RL}$ & $\left( \nu_L^T C \gamma_\mu e_R \right) \left( \bar d_L \gamma^\mu u_L \right)$ & & & & \\
$\op_{\nu e d u}^{V,RR}$ & $\left( \nu_L^T C \gamma_\mu e_R \right) \left( \bar d_R \gamma^\mu u_R \right)$ & & & & \\
\hline
\end{tabular}
\end{center}
\vspace{-2mm}
\caption{\it LEFT Baryon and Lepton number violating  dim-6, four-fermion operators.\\[-3mm]
\label{tab:LEFTops2}}
\end{table}

An important issue arises in the conventions used for the Wilson coefficients of operators of the form
\eq{
\op_{\substack{\psi\psi\\{ijkl}}}^{\,\Gamma_1\Gamma_2} \;\sim\; (\bar{\psi}_i\Gamma_1\psi_j)(\bar{\psi}_k\Gamma_2\psi_l)\ ,
}
as discussed e.g. in~Refs.~\cite{Aebischer:2018iyb,Fuentes-Martin:2020zaz}.
For example, the set of operators
\eq{
\Op{uu}{LL}[V][ijkl] = 
(\bar{u}_i \gamma^\mu P_L u_j)(\bar{u}_k \gamma_\mu P_L u_l)
}
satisfies
\eq{
\Op{uu}{LL}[V][ijkl] = \Op{uu}{LL}[V][klij]\ ,
}
and in this sense the JMS basis is redundant.
Thus the theory only knows about the combination
of Wilson coefficients
\eq{
\lwc{uu}{LL}[V][ijkl] + \lwc{uu}{LL}[V][klij]\ .
}
JMS~\cite{Jenkins:2017jig} choose to impose that these two Wilson coefficients be equal, and thus
\eq{
\label{eq:Ouu}
\mathcal{L}_{\text{LEFT}} \supset \sum_{i,j,k,l} \lwc{uu}{LL}[V][ijkl]\,\Op{uu}{LL}[V][ijkl] = 
\sum_{S}
(2 - \delta_{ik}\delta_{jl})\lwc{uu}{LL}[V][ijkl]\,\Op{uu}{LL}[V][ijkl] \ .
}
where $S=\{(i,j,k,l)\,|\,i<k\}\cup \{(i,j,k,l)\,|\, i=k\text{ and } j\le l\}$.
We prefer to restrict the operator basis to the operators appearing in the last term in the above equation, and to include the factors similar to $(2 - \delta_{ik}\delta_{jl})$ in the Wilson coefficients.
This only affects the beta functions of the following sets of operators,
\begin{align}
\Op{uu}{LL}[V][ijkl] = \Op{uu}{LL}[V][klij] 
\ , \quad
\Op{dd}{LL}[V][ijkl] = \Op{dd}{LL}[V][klij] 
\ , \quad
\Op{ee}{LL}[V][ijkl] = \Op{ee}{LL}[V][klij] 
\ , \quad
\Op{\nu\nu}{LL}[V][ijkl] = \Op{\nu\nu}{LL}[V][klij] 
\ ,
\end{align}
as well as the corresponding $\op^{V,RR}$ operators.
Furthermore, in the last two cases one also has the four-dimensional Fierz relations,
\eq{
\Op{ee}{LL}[V][ijkl] = \Op{ee}{LL}[V][ilkj] \;, \qquad \Op{\nu\nu}{LL}[V][ijkl] = \Op{\nu\nu}{LL}[V][ilkj] \;,
}
which are also imposed onto the (running) Wilson coefficients in order to define a renormalization scheme consistent with the standard treatment of evanescent operators~\cite{Buras:1989xd,Dugan:1990df,Herrlich:1994kh}.
Thus, we will work with the conventions
\begin{align}
\label{eq:JMS Red-Factors I}
\lwc{uu}{LL}[V][ijkl]\Big|_{\text{us}} &= (2 - \delta_{ik}\delta_{jl})\lwc{uu}{LL}[V][ijkl]\Big|_{\text{JMS}} \ , &
\lwc{dd}{LL}[V][ijkl]\Big|_{\text{us}} &= (2 - \delta_{ik}\delta_{jl})\lwc{dd}{LL}[V][ijkl]\Big|_{\text{JMS}} \ , \\
\label{eq:JMS Red-Factors II}
\lwc{ee}{LL}[V][ijkl]\Big|_{\text{us}} &= (4 - 2\delta_{ik} - 2\delta_{jl})\lwc{ee}{LL}[V][ijkl]\Big|_{\text{JMS}} \ , &
\lwc{\nu\nu}{LL}[V][ijkl]\Big|_{\text{us}} &= (4 - 2\delta_{ik} - 2\delta_{jl})\lwc{\nu\nu}{LL}[V][ijkl]\Big|_{\text{JMS}} \ ,
\end{align}
(and the same for $L^{V,RR}$)
such that the corresponding terms in the Lagrangian are equal,
\begin{align}
\sum_{i,j,k,l} \lwc{uu}{LL}[V][ijkl]\Big|_{\text{JMS}}\,\Op{uu}{LL}[V][ijkl] 
& =
\sum_{S}\lwc{uu}{LL}[V][ijkl]\Big|_{\text{us}}\,\Op{uu}{LL}[V][ijkl] \ , \\
\sum_{i,j,k,l} \lwc{dd}{LL}[V][ijkl]\Big|_{\text{JMS}}\,\Op{dd}{LL}[V][ijkl] 
& =
\sum_{S}\lwc{dd}{LL}[V][ijkl]\Big|_{\text{us}}\,\Op{dd}{LL}[V][ijkl] \ , \\
\sum_{i,j,k,l} \lwc{ee}{LL}[V][ijkl]\Big|_{\text{JMS}}\,\Op{ee}{LL}[V][ijkl] 
& = \label{eq:SymRules-ee}
\sum_{S'}\lwc{ee}{LL}[V][ijkl]\Big|_{\text{us}}\left(\frac12\Op{ee}{LL}[V][ijkl] + \frac12\Op{ee}{LL}[V][ilkj]\right) \ , \\
\sum_{i,j,k,l} \lwc{\nu\nu}{LL}[V][ijkl]\Big|_{\text{JMS}}\,\Op{\nu\nu}{LL}[V][ijkl] 
& = \label{eq:SymRules-nunu}
\sum_{S'}\lwc{\nu\nu}{LL}[V][ijkl]\Big|_{\text{us}}\left(\frac12\Op{\nu\nu}{LL}[V][ijkl] + \frac12\Op{\nu\nu}{LL}[V][ilkj]\right) \ , 
\end{align}
where the set $S$ is given after~\Eq{eq:Ouu} and $S'=\{(i,j,k,l)\,|\,i \leq k \text{ and } j\leq l\}$.
Note that $S$ has $n_i^2(n_i^2+1)/2$ elements for $n_i=n_u$ or~$n_d$, and $S'$ has $n_i^2(n_i+1)^2/4$ for $n_i=n_e$ or~$n_\nu$.
The corresponding full set of independent Wilson coefficients is presented in~\Sec{sec:RG-Sectors}. We will reintroduce the factors in~\Eqs{eq:JMS Red-Factors I}{eq:JMS Red-Factors II} in the beta functions given in the ancillary files, in order to be consistent with the JMS conventions.

\subsection{Renormalization and scale dependence}

The renormalized Lagrangian of the LEFT($n_q$) is given by
\eq{ 
\label{eq:LEFT_ren}
\mathcal{L}_{\rm LEFT} = \mathcal{L}_{\rm QCD+QED} + \sum_{k,l} L_l \,Z_{lk} \,Z_{\op_k} \,\op_k \, + \sum_{k,l,s} L_s \, L_l \,Z_{slk} \,Z_{\op_k} \,\op_k + O(1/\Lambda^3_\text{EW}) 
\ ,
}
where the last term quadratic in Wilson coefficients is due to double insertions of dimension-five operators.
The (renormalized) operators $\op_k = \{ Q_k , E_k \}$ include physical ($Q_k$) as well as evanescent ($E_k$) operators, the latter needed for renormalization in $d=4-2\epsilon$ dimensions~\cite{Buras:1989xd,Dugan:1990df,Herrlich:1994kh}.
The $L_k$ are now renormalized Wilson coefficients, and $Z_{lk}$ (also $Z_{slk}$) denote the renormalization constants of the Wilson coefficients, which generate operator mixing. The renormalization factors $Z_{\op_k}$ take care of the wave-function renormalization of the fields contained in $\op_k$.

We regularize all UV divergencies in naive dimensional regularization (NDR) with anticommuting $\gamma_5$, in $d=4-2\epsilon$ dimensions. These divergencies are then subtracted 
using the~\mbox{BW-$\overline{\text{MS}}$} scheme, that is, $\overline{\text{MS}}$ for physical operators, and the ``Buras-Weisz prescription'' for evanescent operators~\cite{Buras:1989xd,Dugan:1990df,Herrlich:1994kh}. 
Our choice of evanescent operators is given in~Appendix~\ref{app:EV Basis}.

Multiple insertions do not contribute to the mixing among four-fermion operators at dimension six, and thus we drop these contributions from now on.
The dependence of the renormalized Wilson coefficients on the renormalization scale $\mu$ is governed by the RGEs,
\eq{
\label{eq:RGE}
\frac{dL_i}{d\log\mu} = \gamma_{ji}\,L_j = \frac{1}{16 \pi^2} \, \beta_i \equiv \frac{1}{16 \pi^2} \, \dot{L}_i \;,
}
where $\gamma_{ij}$ are the components of the Anomalous Dimension Matrix (ADM), denoted by $\hat\gamma$, and $\beta_i$ are the beta functions.\footnote{
We will only consider beta functions and anomalous dimensions of physical operators, and $\hat \gamma$ will always refer to the matrix of coefficients $\gamma_{ij}$ with $(i,j)$ corresponding to physical operators~\cite{Buras:1989xd}.
}
Following~\Reff{Jenkins:2017dyc} we also use the notation $\beta_i\equiv \dot{L}_i$. 
The ADM can be expanded perturbatively,
\eq{
\label{eq:ADM Expansion}
\hat \gamma = \tilde \alpha_s \,\hat \gamma^{(1,0)} + \tilde \alpha \,\hat \gamma^{(0,1)} + \tilde \alpha_s \tilde \alpha \,\hat \gamma^{(1,1)} + \tilde \alpha_s^2 \,\hat \gamma^{(2,0)} + \tilde \alpha^2 \,\hat \gamma^{(0,2)} + \cdots
\; ,
}
where $\hat\gamma^{(n,m)}$ are the expansion coefficients of $\hat\gamma$ in powers of $\tilde \alpha_s \equiv \alpha_s/4\pi = g_s^2/16\pi^2$ and $\tilde \alpha \equiv \alpha/4\pi = e^2/16\pi^2$. 
Denoting by $\hat Z$ the renormalization matrix with entries $Z_{ij}$, we have (see e.g.~\Reff{Gambino:2003zm,Gorbahn:2004my})
\eq{
\hat\gamma = \hat Z \frac{d \hat Z^{-1}}{d\log\mu}
\; ,
}
with $\hat Z$ depending on the renormalization scale through its expansion in $\tilde \alpha_{s}(\mu)$ and $\tilde \alpha(\mu)$,
\eq{
\label{eq:Z Expansion}
\hat Z = \sum_{n,m = 0}^\infty \sum_{\ell=0}^{n+m} \frac{\tilde \alpha_s^{n} \,\tilde \alpha^{m}}{\epsilon^\ell} \,\hat Z^{(n,m;\ell)}\ ,
}
with $Z^{(0,0;0)}_{ij}=\delta_{ij}$, trivially.
In the BW-$\overline{\text{MS}}$ scheme, $Z_{ij}^{(n,m;0)}=0$ whenever $i$ refers to a physical operator or $j$ refers to an evanescent one. 
With this notation at hand, one finds (see e.g.~\cite{Gambino:2003zm,Gorbahn:2004my})
\eqa{
\hat\gamma^{(1,0)} &= 2\hat Z^{(1,0;1)}\ , 
\label{gamma10Z}\\
\hat\gamma^{(0,1)} &= 2\hat Z^{(0,1;1)}\ , 
\label{gamma01Z}\\
\hat\gamma^{(2,0)} &= 4\hat Z^{(2,0;1)} - 2\hat Z^{(1,0;1)}\hat Z^{(1,0;0)}\ ,
\label{gamma20Z}\\
\hat\gamma^{(0,2)} &= 4\hat Z^{(0,2;1)} - 2\hat Z^{(0,1;1)}\hat Z^{(0,1;0)}\ ,
\label{gamma02Z}\\
\hat\gamma^{(1,1)} &= 4\hat Z^{(1,1;1)} - 2\hat Z^{(1,0;1)}\hat Z^{(0,1;0)} - 2\hat Z^{(0,1;1)}\hat Z^{(1,0;0)}\ .
\label{gamma11Z}
}
The renormalization constants can be calculated in the $\overline{\text{MS}}$ scheme in terms of the (amputated) renormalized\footnote{By renormalized matrix elements or diagrams we mean with respect to $\mathcal{L}_{\rm QCD + QED}$, i.e. including only the bare diagrams and the counterterms from the $\text{dim} \leq 4$ Lagrangian, and not including countertems of higher dimension.} matrix elements of the operators $Q_i$. At any loop order, we write
\eqa{
\av{Q_i} &=
\sum_{n,m=0}^\infty
\tilde\mu^{2\epsilon(n+m)}\, \tilde \alpha_s^n \,\tilde \alpha^m \av{Q_i}^{(n,m)}\ ,
\\
\label{eq:ME(Qi)}
\av{Q_i}^{(n,m)} &= 
\sum^{n+m}_{k = 0} \frac{1}{\epsilon^{k}}
\left[ a^{(n,m;k)}_{Q_iQ_j} \langle Q_j\rangle^{(0)}
+ a^{(n,m;k)}_{Q_iE_j} \langle E_j \rangle^{(0)} \right] 
\ ,
}
where $\tilde\mu^2 \equiv \mu^2 e^{\gamma_E}/4\pi$, with $\mu$ the $\overline{\text{MS}}$ scale, and $a_{Q_i \op_j}^{(m,n;k)}$ being finite coefficient matrices. 
The $a_{Q_i \op_j}^{(m,n;k)}$ arise from the $1/\epsilon^k$ poles of the renormalized $(n+m)$-loop diagrams with an insertion of operator $Q_i$.
Renormalization then leads to~\cite{Buras:2000if,Gambino:2003zm,Gorbahn:2004my}: 
\eqa{
\label{eq:MS-bar Z10}
\hat Z^{(1,0;1)} =& -\hat a^{(1,0;1)} - \hat Z_\op^{(1,0;1)}\; , 
\\
\label{eq:MS-bar Z01}
\hat Z^{(0,1;1)} =& -\hat a^{(0,1;1)} - \hat Z_\op^{(0,1;1)}\; , 
\\
\label{eq:MS-bar Z20}
\hat Z^{(2,0;1)} =& -\hat a^{(2,0;1)} + \hat a^{(1,0;1)} \hat a^{(1,0;0)} - \hat Z^{(1,0;0)} \hat a^{(1,0;1)} - \hat Z_\op^{(2,0;1)}\; , 
\\
\label{eq:MS-bar Z02}
\hat Z^{(0,2;1)} =& -\hat a^{(0,2;1)} + \hat a^{(0,1;1)} \hat a^{(0,1;0)} - \hat Z^{(0,1;0)} \hat a^{(0,1;1)}  - \hat Z_\op^{(0,2;1)}\; , 
\\
\label{eq:MS-bar Z11}
\hat Z^{(1,1;1)} =& -\hat a^{(1,1;1)} + \hat a^{(1,0;1)} \hat a^{(0,1;0)} + \hat a^{(0,1;1)} \hat a^{(1,0;0)} - \hat Z^{(1,0;0)} \hat a^{(0,1;1)} \nonumber \\ 
&  -\, \hat Z^{(0,1;0)} \hat a^{(1,0;1)}  - \hat Z_\op^{(1,1;1)} \; ,  
}
up to two loops in $\tilde \alpha_s$ and $\tilde \alpha$. The wave-function renormalization constants are given by $(\hat Z_\op)_{ij} = Z_{Q_i} \delta_{Q_iQ_j}$. In the case of four-fermion operators,  
$Q_i=\bar{\psi}_1\Gamma \psi_2\,\bar{\psi}_3 \Gamma' \psi_4$, one has $Z_{Q_i} = Z_{\psi_1}^{1/2}Z_{\psi_2}^{1/2}Z_{\psi_3}^{1/2}Z_{\psi_4}^{1/2}$, and therefore
\eq{
Z_{Q_i}^{(n,m;1)} = \frac{1}{2} Z_{\psi_1}^{(n,m;1)} + \frac{1}{2} Z_{\psi_2}^{(n,m;1)} + \frac{1}{2} Z_{\psi_3}^{(n,m;1)} + \frac{1}{2} Z_{\psi_4}^{(n,m;1)} \;,
}
with the notation of~\Eq{eq:Z Expansion}.
The individual wave-function renormalization constants are flavor-universal (represented by a generic fermion field $\psi$), although only quarks are sensitive to $\tilde \alpha_s$ corrections, and $\alpha$ corrections depend on the electric charges $Q_{\psi}$,
\begin{gather}
    Z_q^{(1,0;1)} = -C_F \;, \quad  Z_q^{(1,1;1)} = \frac{3}{2} C_F Q_q^2 \;, \quad Z_q^{(2,0;1)} = C_F\bigg[ \,\frac{3}{4}C_F - \frac{17}{4}N_c + \frac{1}{2}n_q\, \bigg] \;, \\
    \qquad Z_{\psi}^{(0,1;1)} = -Q_{\psi}^2 \;, \quad Z_{\psi}^{(0,2;1)} =Q_{\psi}^2 \left[\frac{3}{4} Q_{\psi}^2 + \frac{1}{9} N_c (4 n_u + n_d) + n_e \right] \ .
\end{gather}
Going back to Eqs.~(\ref{eq:MS-bar Z20}), (\ref{eq:MS-bar Z02}) and (\ref{eq:MS-bar Z11}), in the Buras-Weisz scheme for evanescent operators one has $Z^{(1,0)}_{ij}=-a^{(1,0)}_{E_iQ_j}$ for $(i,j)$ being (evanescent, physical), and zero otherwise. Inserting the corresponding expressions for the renormalization matrices into~Eqs.\,(\ref{gamma10Z})-(\ref{gamma11Z}) one finds, again up to two-loop order,
\eqa{
\gamma^{(1,0)}_{ij} =&
-2 a^{(1,0;1)}_{Q_iQ_j} - 2Z_{Q_i}^{(1,0;1)} \delta_{Q_iQ_j} \; ,
\label{eq:gamma10a} \\
\gamma^{(0,1)}_{ij} =&
-2 a^{(0,1;1)}_{Q_iQ_j} - 2Z_{Q_i}^{(0,1;1)} \delta_{Q_iQ_j} \; ,
\label{eq:gamma01a} \\
\gamma^{(2,0)}_{ij} =&
-4 a^{(2,0;1)}_{Q_iQ_j} + 4 a^{(1,0;1)}_{Q_iQ_k} a^{(1,0;0)}_{Q_kQ_j} + 2 a^{(1,0;1)}_{Q_iE_k} a^{(1,0;0)}_{E_kQ_j} - 4Z_{Q_i}^{(2,0;1)} \delta_{Q_iQ_j}\; , \qquad
\label{eq:gamma20a} \\
\gamma^{(0,2)}_{ij} =&
-4 a^{(0,2;1)}_{Q_iQ_j} + 4 a^{(0,1;1)}_{Q_iQ_k} a^{(0,1;0)}_{Q_kQ_j} + 2 a^{(0,1;1)}_{Q_iE_k} a^{(0,1;0)}_{E_kQ_j} - 4Z_{Q_i}^{(0,2;1)} \delta_{Q_iQ_j}\; , \qquad
\label{eq:gamma02a} \\
\gamma^{(1,1)}_{ij} =&
-4 a^{(1,1;1)}_{Q_iQ_j} + 4 a^{(1,0;1)}_{Q_iQ_k} a^{(0,1;0)}_{Q_kQ_j} + 4 a^{(0,1;1)}_{Q_iQ_k} a^{(1,0;0)}_{Q_kQ_j} + 2 a^{(1,0;1)}_{Q_iE_k} a^{(0,1;0)}_{E_kQ_j} 
\nonumber \\ 
& + 2 a^{(0,1;1)}_{Q_iE_k} a^{(1,0;0)}_{E_kQ_j} - 4Z_{Q_i}^{(1,1;1)} \delta_{Q_iQ_j}\; .
\label{eq:gamma11a} 
}
The complete set of one-loop beta functions for the LEFT was computed in~\Reff{Jenkins:2017dyc}. They can be obtained obtained from \dsix~\cite{Celis:2017hod,Fuentes-Martin:2020zaz} via the command \rut{$\beta$[{\it parameter}]}.

\subsection{Change of Operator Basis}

It will be very relevant in the present paper to consider different bases of operators in the LEFT, and derive the two-loop ADM in one basis given the ADM in another. 
Given two different operator bases $\{Q_k,E_k\}$ and $\{Q'_k,E_k\}$, the physical operators $Q_k'$ in the second basis can always be written as a linear combination of physical ($Q_k$) and evanescent ($E_k$) operators in the first one, 
\eq{
\label{eq:Operator Transformation}
\vec{Q}' = \hat{R} \left( \vec{Q} + \hat{W} \vec{E} \right) \ ,
}
where $\hat{R}$ and $\hat{W}$ are numerical rotation matrices.
As is well known~\cite{Buras:1991jm, Herrlich:1994kh, Chetyrkin:1997gb,Gorbahn:2004my}, a change of basis that mixes physical and evanescent operators introduces a shift in the renormalization scheme. The transformation rules for anomalous dimensions  must then include a series of finite corrections that compensate for this effect in order to obtain the results in the original renormalization scheme. At one loop, the transformation of the ADMs turns out to be trivial,
\eqa{
\label{eq:LO ADM Transformation}
\hat{\gamma}'^{(1,0)} =& \hat{R} \hat{\gamma}^{(1,0)} \hat{R}^{-1} \;, \\
\hat{\gamma}'^{(0,1)} =&\hat{R} \hat{\gamma}^{(0,1)} \hat{R}^{-1} \;,
}
reflecting the scheme-independence of the one-loop ADMs. At two loops, however, the ADMs depend on the renormalization scheme through the (scheme-dependent) one-loop finite amplitudes. The transformation rules read~\cite{Buras:1992tc,Buras:1992zv}
\eqa{
\label{eq:NLO ADM Transformation 20}
\hat{\gamma}'^{(2,0)} =& \hat{R} \hat{\gamma}^{(2,0)} \hat{R}^{-1} + 4 Z_{g_s}^{(1,0;1)} \Delta\hat{r}^{(1,0)} - \left[\Delta\hat{r}^{(1,0)} , \hat{\gamma}'^{(1,0)}\right] \;, \\
\label{eq:NLO ADM Transformation 02}
\hat{\gamma}'^{(0,2)} =& \hat{R} \hat{\gamma}^{(0,2)} \hat{R}^{-1} + 4 Z_e^{(0,1;1)} \Delta\hat{r}^{(0,1)} - \left[\Delta\hat{r}^{(0,1)} , \hat{\gamma}'^{(0,1)}\right] \;, \\
\label{eq:NLO ADM Transformation 11}
\hat{\gamma}'^{(1,1)} =& \hat{R} \hat{\gamma}^{(1,1)} \hat{R}^{-1} - \left[\Delta\hat{r}^{(1,0)} , \hat{\gamma}'^{(0,1)}\right] - \left[\Delta\hat{r}^{(0,1)} , \hat{\gamma}'^{(1,0)}\right] \;,
}
where
\eqa{
Z_{g_s}^{(1,0;1)} =&\, \frac13n_q - \frac{11}{6}N_c\ ,
\\
Z_e^{(0,1;1)} =&\, \frac{2}{3} \Big( Q_u^2 n_u N_c + Q_d^2 n_d N_c + Q_e^2 n_e \Big)\ ,
}
are the one-loop renormalization constants of $g_s$ and $e$, respectively, defined by
\eqa{
\frac{d\tilde \alpha_s}{d\log\mu} =&\, 4\tilde \alpha_s^2 Z_{g_s}^{(1,0;1)} + \cdots\ ,
\\
\frac{d\tilde \alpha}{d\log\mu} =&\, 4\tilde \alpha^2 Z_{e}^{(0,1;1)} + \cdots\ .
}
The one-loop matrices $\Delta\hat{r}$ contain the shift in the renormalization scheme
\eqa{
\Delta r^{(1,0)}_{ij} \equiv \,\,& 
-R_{Q'_iQ_k}W_{Q_kE_l}\,a^{(1,0;0)}_{E_lQ_r}R^{-1}_{Q_rQ'_j}
= 
R_{Q'_iQ_k} \,a^{(1,0;0)}_{Q_kQ_l} R^{-1}_{Q_lQ'_j} - a^{(1,0;0)}_{Q'_iQ'_j}\;, \\
\Delta r^{(0,1)}_{ij} \equiv \,\,& 
-R_{Q'_iQ_k}W_{Q_kE_l}\,a^{(0,1;0)}_{E_lQ_r}R^{-1}_{Q_rQ'_j}
= 
R_{Q'_iQ_k} \,a^{(0,1;0)}_{Q_kQ_l} R^{-1}_{Q_lQ'_j} - a^{(0,1;0)}_{Q'_iQ'_j}\;,
}
where the equalities follow from~\Eq{eq:Operator Transformation}. Working with the matrices $\Delta\hat{r}$ in the form of the right-hand side of the above equations allows for a change of basis at the two-loop order without having to work directly with the evanescent operators, although the specification of an evanescent basis is still necessary to determine the finite amplitudes $\hat{a}^{(m,n;0)}$.


\section{RG-invariant sectors}
\label{sec:RG-Sectors}

Given that all dimension-four couplings in the LEFT Lagrangian are flavor-conserving (and ignoring dimension-five terms), the ADM of four-fermion operators has a block-diagonal structure, each block corresponding to an RG-invariant sector defined by the flavor quantum numbers of the operators~(see e.g.~\Reff{Aebischer:2017gaw}). 
Each of these  will be treated as a ``separate'' ADM: the ``ADM of an RG-invariant sector''. We thus have one ADM for each sector, each one possibly broken into several smaller sub-blocks, depending on the details of the sector.

These sectors are classified in terms of the flavor transition they mediate, for which we use the following terminology:
\begin{itemize}
\item {\boldmath$\Delta F = 2$} : e.g. $(\bar d s)(\bar d s)$,
\item {\boldmath$\Delta F = 1.5$} : e.g. $(\bar d b)(\bar d s)$ ,
\item {\boldmath$\Delta F = 1$} : e.g. $(\bar d c)(\bar u s)$ ,
\item {\boldmath$\Delta F = 1^{\bar{f}f}$} : e.g. $(\bar d s)(\bar u u)$ ,
\item {\boldmath$\Delta F = 0$} : e.g. $(\bar d d)(\bar e e)$ .
\end{itemize}
In addition, there are lepton-number violating sectors, $\Delta L=2,4$, and Baryon-number violating sectors, $|\Delta B|=1$.
The full list of sectors is given in Tables~\ref{tab:RG-sectors I}~and~\ref{tab:RG-sectors II}, which provide a summary of the sectors, their multiplicity, and the number of sub-blocks in each of them. 
These sectors match those of~\Reff{Aebischer:2017ugx}. 
We also remark that these sectors are defined for our non-redundant LEFT basis, which differs from the JMS conventions in~\Refs{Jenkins:2017jig,Jenkins:2017dyc} as detailed at the end of~\Sec{sec:defLEFT}. 
All sectors are composed of non-hermitian operators, except one: the operators in the last $\Delta F=0$ sub-block in~\Tab{tab:RG-sectors II} are hermitian (and thus their Wilson coefficients are real). In the LEFT(5), this sub-block is 235-dimensional. The full Lagrangian contains all the operators listed below, plus their hermitian conjugates. We do not display the hermitian conjugates because their structure is identical and their ADMs are the same.

In the following, we list explicitly the full set of RG-invariant sub-blocks in each sector.

\begin{table}
\centering
\setlength{\tabcolsep}{20pt}
\begin{tabular}{@{}c c c c@{}}
\toprule[0.7mm]
Transition & Flavor & Multiplicity & Sector Structure 
\\
\midrule[0.7mm] 
$\Delta F=2$ &
$\bar d_i d_j\bar d_i d_j$ &
$\frac{1}{2}n_d(n_d-1)$ & 
$2\times\{1\} + 3\times\{2\}$
\\
\midrule 
$\Delta F=2$ &
$\bar u_i u_j\bar u_i u_j$ &
$\frac{1}{2}n_u(n_u-1)$ &
$2\times\{1\} + 3\times\{2\}$
\\
\midrule 
$\Delta F=2$ &
$\bar e_i e_j\bar e_i e_j$ &
$\frac{1}{2}n_e(n_e-1)$ & 
$5\times\{1\}$
\\
\midrule 
$\Delta F=2$ &
$\bar \nu_i \nu_j\bar \nu_i \nu_j$ &
$\frac{1}{2}n_\nu(n_\nu-1)$ & 
$1\times\{1\}$
\\
\midrule 
$\Delta F=1.5$ &
$\bar d_i d_j\bar d_k d_j$ &
$\frac{1}{2}n_d(n_d-1)(n_d-2)$ & 
$2\times\{1\} + 4\times\{2\}$
\\
\midrule 
$\Delta F=1.5$ &
$\bar e_i e_j\bar e_k e_j$ &
$\frac{1}{2}n_e(n_e-1)(n_e-2)$ & 
$6\times\{1\}$
\\
\midrule 
$\Delta F=1.5$ &
$\bar \nu_i \nu_j\bar \nu_k \nu_j$ &
$\frac{1}{2}n_\nu(n_\nu-1)(n_\nu-2)$ & 
$1\times\{1\}$
\\
\midrule 
$\Delta F=1$ &
$\bar u_i u_j\bar d_k d_l$ &
$\frac{1}{2}n_u(n_u-1)n_d(n_d-1)$ & 
$6\times\{2\} + 2\times\{4\}$
\\
\midrule 
$\Delta F=1$ &
$\bar e_i e_j\bar d_k d_l$ &
$\frac{1}{2}n_e(n_e-1)n_d(n_d-1)$ & 
$6\times\{1\} + 2\times\{2\}$
\\
\midrule 
$\Delta F=1$ &
$\bar e_i e_j\bar u_k u_l$ &
$\frac{1}{2}n_e(n_e-1)n_u(n_u-1)$ & 
$6\times\{1\} + 2\times\{2\}$
\\
\midrule 
$\Delta F=1$ &
$\bar \nu_i \nu_j\bar d_k d_l$ &
$\frac{1}{2}n_\nu(n_\nu-1)n_d(n_d-1)$ & 
$2\times\{1\}$
\\
\midrule 
$\Delta F=1$ &
$\bar \nu_i \nu_j\bar u_k u_l$ &
$\frac{1}{2}n_\nu(n_\nu-1)n_u(n_u-1)$ & 
$2\times\{1\}$
\\
\midrule 
$\Delta F=1$ &
$\bar \nu_i \nu_j\bar e_k e_l$ &
$\frac{1}{2}n_\nu(n_\nu-1)n_e(n_e-1)$ & 
$2\times\{1\}$
\\
\midrule 
$\Delta F=1$ &
$\bar \nu_i e_j\bar d_k u_l$ &
$ n_\nu n_e n_d n_u$ & 
$3\times\{1\} + 1\times\{2\}$
\\
\midrule  
$\Delta L=2$ &
$\bar \nu^c_i \nu_j \bar f_k f_l$ &
$ \frac{1}{2}n_\nu(n_\nu-1) (n_d^2 + n_u^2 + n_e^2)$ &
$1 \times \{1\} + 1 \times \{2\} $
\\
\midrule  
$\Delta L=2$ &
$\bar \nu^c_i \nu_i \bar f_k f_l$ &
$ n_{\nu} (n_d^2 + n_u^2 + n_e^2)$ &
$2 \times \{1\}$
\\
\midrule  
$\Delta L=2$ &
$\bar \nu^c_i e_j \bar d_k u_l$ &
$n_\nu n_e n_d n_u$ &
$3 \times \{1\} + 1 \times \{2\}$
\\
\midrule  
$\Delta L=4$ &
$\bar \nu^c_i \nu_j \bar \nu^c_k \nu_l$ &
$\frac{1}{12}n_\nu^2(n_\nu^2-1)$ &
$1 \times \{1\}$
\\
\midrule  
$|\Delta B| = 1$ &
$\bar d^c_i u_j\bar u^c_k e_l$ &
$\frac12 n_u(n_u-1) n_d n_e$ &
$6\times\{1\} + 2\times\{2\}$
\\
\midrule  
$|\Delta B| = 1$ &
$\bar d^c_i u_j\bar u^c_j e_l$ &
$n_u n_d n_e$ &
$4\times\{1\}$
\\
\midrule  
$|\Delta B| = 1$ &
$\bar u^c_i d_j\bar d^c_k \nu_l$ &
$ \frac12 n_d (n_d-1) n_u n_\nu$ &
$3\times\{1\} + 1\times\{2\}$
\\
\midrule  
$|\Delta B| = 1$ &
$\bar u^c_i d_j\bar d^c_j \nu_l$ &
$ n_d n_u n_\nu$ &
$2\times\{1\}$
\\
\midrule  
$|\Delta B| = 1$ & 
$\bar d^c_i d_j\bar e^c_l d_k$ &
$\frac16 n_d (n_d-1) (n_d-2) n_e$ &
$6\times\{1\} + 2\times\{2\}$
\\
\midrule  
$|\Delta B| = 1$ & 
$\bar d^c_i d_j\bar e^c_l d_j$ &
$\frac12 n_d (n_d-1) n_e$ &
$4\times\{1\}$
\\
\midrule  
$|\Delta B| = 1$ & 
$\bar d^c_i d_j\bar e^c_l d_i$ &
$\frac12 n_d (n_d-1) n_e$ &
$4\times\{1\}$
\\
\midrule  
$|\Delta B| = 1$ &
$\bar u^c_i d_j\bar \nu^c_l d_k$ &
$\frac12 n_d (n_d - 1) n_u n_\nu$ &
$3\times\{1\} + 1\times\{2\}$
\\
\midrule  
$|\Delta B| = 1$ &
$\bar u^c_i d_j\bar \nu^c_l d_j$ &
$n_d n_u n_\nu$ &
$2\times\{1\}$
\\
\bottomrule[0.7mm]
\end{tabular}
\caption{\it Distribution of RG-invariant sectors, where $n \times \{ k\}$ represents $n$ different invariant sub-blocks, each containing $k$ operators. For instance, the ADM of the first sector contains two $1\times1$ sub-blocks, and three $2\times2$ sub-blocks. We use the notation $\overline \psi^c \equiv \psi^T C$.}
\label{tab:RG-sectors I}
\end{table}

\begin{table}
\centering
\setlength{\tabcolsep}{17pt}
\renewcommand{\arraystretch}{1.45}
\begin{tabular}{@{}c c c c@{}}
\toprule[0.7mm]
Transition & Flavor & Multiplicity & Sector Structure 
\\
\midrule  
$\Delta F=1^{\bar{f}f}$ &  
$\bar d_i d_j\bar{f}_k f_k$ & 
$\frac{1}{2}n_d(n_d-1)$ &
$\begin{array}{l}
    2n_e\times\{1\} + 2(n_q + n_e)\times\{2\}
    \\[2mm]
    + \, 2(n_q-2)\times\{4\}
    \\[2mm]
     + \, 2\times\{4n_q + 2n_e + n_\nu - 2\} 
\end{array}$
\\
\midrule  
$\Delta F=1^{\bar{f}f}$ &  
$\bar u_i u_j\bar{f}_k f_k$ & 
$\frac{1}{2}n_u(n_u-1)$ & 
$\begin{array}{l}
    2n_e\times\{1\} + 2(n_q + n_e)\times\{2\}
    \\[2mm]
    + \, 2(n_q-2)\times\{4\}
    \\[2mm]
     + \, 2\times\{4n_q + 2n_e + n_\nu - 2\} 
\end{array}$
\\
\midrule  
$\Delta F=1^{\bar{f}f}$ &  
$\bar e_i e_j\bar{f}_k f_k$ & 
$\frac{1}{2}n_e(n_e-1)$ &
$\begin{array}{l}
    2(n_q + n_e)\times\{1\} 
    \\[2mm]
    + \, 2(n_q + n_e - 2)\times\{2\} 
    \\[2mm]
     + \, 2\times\{2n_q + 2n_e + n_\nu\} 
\end{array}$
\\
\midrule  
$\Delta F=1^{\bar{f}f}$ &  
$\bar \nu_i \nu_j\bar{f}_k f_k$ & 
$\frac{1}{2}n_\nu(n_\nu-1)$ &
$1\times\{2n_q + 2n_e + n_\nu\}$
\\
\midrule  
$\Delta F=0$ &  
$\bar f_i f_i \bar f'_j f'_j$ &
$1$ &
$\begin{array}{l}
    \big[n_q n_e + \frac12n_e(n_e+1)\big] \times \{1\} \\[2mm]
    +\,\frac12\big[(n_q+n_e)^2+n_q-n_e\big] \times \{2\} \\[2mm]
    +\,\frac12(n_q^2 - n_q) \times \{4\} \\[2mm]
+\, 1 \times \big\{  4 n_q^2 + 4 n_q n_e 
+ n_e(2n_e + 1) \\[2mm]
\hspace{12mm} + \frac12n_\nu(n_\nu + 1) 
 + 2n_\nu (n_q + n_e)  \big\}
\end{array}$
\\
\bottomrule[0.7mm]
\end{tabular}
\caption{\it Distribution of RG-invariant sectors, where $n \times \{ k\}$ represents $n$ different invariant sub-blocks, each containing $k$ operators. Here $f,f'$ represent generic fermion fields of any type and flavor.}
\label{tab:RG-sectors II}
\end{table}

\subsection{\boldmath$\Delta F = 2$ transitions \boldmath$\left(\bar{f}_i f_j \bar{f}_i f_j\right)$}

\subsubsection{\boldmath{$\Delta F = 2 \;:\; d\text{-type} \; \left(\bar{d}_i d_j \bar{d}_i d_j\right)$}}
\label{sec:DF=2 d-type}

The RG-invariant sub-blocks of the $\Delta F = 2$ $d$-type sector read
\begin{gather*}
\label{eq:DF=2 d-type}
\bigg\{ \lwc{dd}{LL}[V][ijij]\bigg\} , \bigg\{ \lwc{dd}{RR}[V][ijij]\bigg\} , \bigg\{ \lwc{dd}{LR}[V1][ijij], \lwc{dd}{LR}[V8][ijij] \bigg\} , \bigg\{ \lwc{dd}{RR}[S1][ijij], \lwc{dd}{RR}[S8][ijij] \bigg\} , \bigg\{ \lwc{dd}{RR}[S1][jiji], \lwc{dd}{RR}[S8][jiji] \bigg\} \;,
\end{gather*}
where $i < j$. 
This sector exists for all versions of the LEFT, with $n_q \geq 3$.

\subsubsection{\boldmath{$\Delta F = 2 \;:\; u\text{-type} \; \left(\bar{u}_i u_j \bar{u}_i u_j\right)$}}
The RG-invariant sub-blocks for the $\Delta F = 2$ $u$-type sector read
\begin{gather*}
\label{eq:DF=2 u-type}
\bigg\{ \lwc{uu}{LL}[V][ijij]\bigg\} , \bigg\{ \lwc{uu}{RR}[V][ijij]\bigg\} , \bigg\{ \lwc{uu}{LR}[V1][ijij], \lwc{uu}{LR}[V8][ijij] \bigg\} , \bigg\{ \lwc{uu}{RR}[S1][ijij], \lwc{uu}{RR}[S8][ijij] \bigg\} , \bigg\{ \lwc{uu}{RR}[S1][jiji], \lwc{uu}{RR}[S8][jiji] \bigg\} \;,
\end{gather*}
where $i < j$. 
This sector exists for $n_q \geq 4$.

\subsubsection{\boldmath{$\Delta F = 2 \;:\; e\text{-type} \; \left(\bar{e}_i e_j \bar{e}_i e_j\right)$}}

The RG-invariant sub-blocks of the $\Delta F = 2$ $e$-type sector read
\begin{gather*}
\label{eq:DF=2 e-type}
\bigg\{ \lwc{ee}{LL}[V][ijij]\bigg\} , \bigg\{ \lwc{ee}{RR}[V][ijij]\bigg\} , \bigg\{ \lwc{ee}{LR}[V][ijij] \bigg\} , \bigg\{ \lwc{ee}{RR}[S][ijij] \bigg\} , \bigg\{ \lwc{ee}{RR}[S][jiji] \bigg\} \;,
\end{gather*}
where $i < j$.  
This sector exists for $n_q \geq 4$.

\subsubsection{\boldmath{$\Delta F = 2 \;:\; \nu\text{-type} \; \left(\bar{\nu}_i \nu_j \bar{\nu}_i \nu_j\right)$}}

The RG-invariant sub-blocks of the $\Delta F = 2$ $\nu$-type sector read
\begin{gather*}
\label{eq:DF=2 n-type}
\bigg\{ \lwc{\nu\nu}{LL}[V][ijij]\bigg\} \;.
\end{gather*}
where $i < j$.
This sector exists for all versions of the LEFT with $n_q \geq 3$.

\subsection{\boldmath{$\Delta F = 1.5$ transitions $\left(\bar{f}_i f_j \bar{f}_k f_j\right)$}}

\subsubsection{\boldmath{$\Delta F = 1.5 \;:\; d\text{-type} \; \left(\bar{d}_i d_j \bar{d}_k d_j\right)$}}
\label{sec:DF=1.5 d-type}

The RG-invariant sub-blocks of the $\Delta F = 1.5$ $d$-type sector read
\begin{gather*}
\label{eq:DF=1.5 d-type}
\bigg\{ \lwc{dd}{LL}[V][ijkj]\bigg\} , \bigg\{ \lwc{dd}{RR}[V][ijkj]\bigg\} , \bigg\{ \lwc{dd}{LR}[V1][ijkj], \lwc{dd}{LR}[V8][ijkj] \bigg\} , \\
\bigg\{ \lwc{dd}{LR}[V1][kjij], \lwc{dd}{LR}[V8][kjij] \bigg\} , \bigg\{ \lwc{dd}{RR}[S1][ijkj], \lwc{dd}{RR}[S8][ijkj] \bigg\} , \bigg\{ \lwc{dd}{RR}[S1][jijk], \lwc{dd}{RR}[S8][jijk] \bigg\} \;,
\end{gather*}
where $i \neq j \neq k$ and $i < k$.
This sector exists only for $n_q = 5$. The up-type version of this sector does not exist, since there are only two up-type quark flavors in the LEFT.

\subsubsection{\boldmath{$\Delta F = 1.5 \;:\; e\text{-type} \; \left(\bar{e}_i e_j \bar{e}_k e_j\right)$}}

The RG-invariant sub-blocks of the $\Delta F = 1.5$ $e$-type RG-invariant sector read
\begin{gather*}
\label{eq:DF=1.5 e-type}
\bigg\{ \lwc{ee}{LL}[V][ijkj]\bigg\} , \bigg\{ \lwc{ee}{RR}[V][ijkj]\bigg\} , \bigg\{ \lwc{ee}{LR}[V][ijkj] \bigg\} , \bigg\{ \lwc{ee}{LR}[V][kjij] \bigg\} , \bigg\{ \lwc{ee}{RR}[S][ijkj] \bigg\} , \bigg\{ \lwc{ee}{RR}[S][jijk] \bigg\} \;,
\end{gather*}
where $i \neq j \neq k$ and $i < k$.
These sector exists for $n_q \geq 4$.

\subsubsection{\boldmath{$\Delta F = 1.5 \;:\; \nu\text{-type} \; \left(\bar{\nu}_i \nu_j \bar{\nu}_k \nu_j\right)$}}

The RG-invariant sub-blocks of the $\Delta F = 1.5$ $\nu$-type sector read
\begin{gather*}
\label{eq:DF=1.5 n-type}
\bigg\{ \lwc{\nu\nu}{LL}[V][ijkj]\bigg\} \;,
\end{gather*}
where $i \neq j \neq k$ and $i < k$.
This sector exists for all versions of the LEFT, with $n_q \geq 3$.

\subsection{{\boldmath$\Delta F = 1$ transitions $\left(\bar{f}_i f_j \bar{f}'_k f'_l\right)$}}

\subsubsection{\boldmath{$\Delta F = 1 \;:\; ud\text{-type} \; \left(\bar{u}_i u_j \bar{d}_k d_l\right)$}}

The RG-invariant sub-blocks of the $\Delta F = 1$ $ud$-type sector read
\begin{gather*}
\label{eq:DF=1 ud-type}
\bigg\{ \lwc{ud}{LL}[V1][ijkl], \lwc{ud}{LL}[V8][ijkl] \bigg\} , \bigg\{ \lwc{ud}{RR}[V1][ijkl], \lwc{ud}{RR}[V8][ijkl] \bigg\} , \bigg\{ \lwc{ud}{LR}[V1][ijkl], \lwc{ud}{LR}[V8][ijkl] \bigg\} , \\ 
\bigg\{ \lwc{du}{LR}[V1][klij], \lwc{du}{LR}[V8][klij] \bigg\} ,\bigg\{ \lwc{uddu}{LR}[V1][ilkj], \lwc{uddu}{LR}[V8][ilkj] \bigg\} , \bigg\{ \lwc{uddu}{LR}[V1][jkli], \lwc{uddu}{LR}[V8][jkli] \bigg\} , \\
\bigg\{ \lwc{ud}{RR}[S1][ijkl], \lwc{ud}{RR}[S8][ijkl], \lwc{uddu}{RR}[S1][ilkj], \lwc{uddu}{RR}[S8][ilkj] \bigg\} , \bigg\{ \lwc{ud}{RR}[S1][jilk], \lwc{ud}{RR}[S8][jilk], \lwc{uddu}{RR}[S1][jkli], \lwc{uddu}{RR}[S8][jkli] \bigg\} \;,
\end{gather*}
where $i < j$ and $k \neq l$.
This sector exists for $n_q \geq 4$.

\subsubsection{\boldmath{$\Delta F = 1 \;:\; ed\text{-type} \; \left(\bar{e}_i e_j \bar{d}_k d_l\right)$}}

The RG-invariant sub-blocks of the $\Delta F = 1$ $ed$-type sector read
\begin{gather*}
\label{eq:DF=1 ed-type}
\bigg\{ \lwc{ed}{LL}[V][ijkl] \bigg\} , \bigg\{ \lwc{ed}{RR}[V][ijkl] \bigg\} , 
\bigg\{ \lwc{ed}{LR}[V][ijkl] \bigg\} , \bigg\{ \lwc{de}{LR}[V][klij] \bigg\} , \bigg\{ \lwc{ed}{RL}[S][ijkl] \bigg\} , \\
\bigg\{ \lwc{ed}{RL}[S][jilk] \bigg\} , \bigg\{ \lwc{ed}{RR}[S][ijkl], \lwc{ed}{RR}[T][ijkl] \bigg\} , \bigg\{ \lwc{ed}{RR}[S][jilk], \lwc{ed}{RR}[T][jilk] \bigg\} \;,
\end{gather*}
where $i < j$ and $k \neq l$.
This sector exists for all versions of the LEFT, with $n_q \geq 3$.

\subsubsection{\boldmath{$\Delta F = 1 \;:\; eu\text{-type} \; \left(\bar{e}_i e_j \bar{u}_k u_l\right)$}}

The RG-invariant sub-blocks of the $\Delta F = 1$ $eu$-type sector read
\begin{gather*}
\label{eq:DF=1 eu-type}
\bigg\{ \lwc{eu}{LL}[V][ijkl] \bigg\} , \bigg\{ \lwc{eu}{RR}[V][ijkl] \bigg\} , \bigg\{ \lwc{eu}{LR}[V][ijkl] \bigg\} , \bigg\{ \lwc{ue}{LR}[V][klij] \bigg\} , \bigg\{ \lwc{eu}{RL}[S][ijkl] \bigg\} , \\
\bigg\{ \lwc{eu}{RL}[S][jilk] \bigg\} , \bigg\{ \lwc{eu}{RR}[S][ijkl], \lwc{eu}{RR}[T][ijkl] \bigg\} , \bigg\{ \lwc{eu}{RR}[S][jilk], \lwc{eu}{RR}[T][jilk] \bigg\} \;,
\end{gather*}
where $i < j$ and $k \neq l$.
This sector exists for $n_q \geq 4$.

\subsubsection{\boldmath{$\Delta F = 1 \;:\; \nu d\text{-type} \; \left(\bar{\nu}_i \nu_j \bar{d}_k d_l\right)$}}

The RG-invariant sub-blocks of the $\Delta F = 1$ $\nu d$-type sector read
\begin{gather*}
\label{eq:DF=1 nd-type}
\bigg\{ \lwc{\nu d}{LL}[V][ijkl] \bigg\} , \bigg\{ \lwc{\nu d}{LR}[V][ijkl] \bigg\} \;,
\end{gather*}
where $i < j$ and $k \neq l$.
This sector exists for all versions of the LEFT, with $n_q \geq 3$.

\subsubsection{\boldmath{$\Delta F = 1 \;:\; \nu u\text{-type} \; \left(\bar{\nu}_i \nu_j \bar{u}_k u_l\right)$}}

The RG-invariant sub-blocks of the $\Delta F = 1$ $\nu u$-type sector read
\begin{gather*}
\label{eq:DF=1 nu-type}
\bigg\{ \lwc{\nu u}{LL}[V][ijkl] \bigg\} , \bigg\{ \lwc{\nu u}{LR}[V][ijkl] \bigg\} \;,
\end{gather*}
where $i < j$ and $k \neq l$.
This sector exists for all versions of the LEFT, with $n_q \geq 3$.

\subsubsection{\boldmath{$\Delta F = 1 \;:\; \nu e\text{-type} \; \left(\bar{\nu}_i \nu_j \bar{e}_k e_l\right)$}}

The RG-invariant sub-blocks of the $\Delta F = 1$ $\nu e$-type RG-invariant sector read
\begin{gather*}
\label{eq:DF=1 ne-type}
\bigg\{ \lwc{\nu e}{LL}[V][ijkl] \bigg\} , \bigg\{ \lwc{\nu e}{LR}[V][ijkl] \bigg\} \;,
\end{gather*}
where $i < j$ and $k \neq l$.
This sector exists for all versions of the LEFT, with $n_q \geq 3$.

\subsubsection{\boldmath{$\Delta F = 1 \;:\; \nu e d u\text{-type} \; \left(\bar{\nu}_i e_j \bar{d}_k u_l\right)$}}

The RG-invariant sub-blocks of the $\Delta F = 1$ $\nu e d u$-type sector read 
\begin{gather*}
\label{eq:DF=1 nedu-type}
\bigg\{ \lwc{\nu e d u}{LL}[V][ijkl] \bigg\} , \bigg\{ \lwc{\nu e d u}{LR}[V][ijkl] \bigg\} 
, \bigg\{ \lwc{\nu e d u}{RL}[S][ijkl] \bigg\} , \bigg\{ \lwc{\nu e d u}{RR}[S][ijkl], \lwc{\nu e d u}{RR}[T][ijkl] \bigg\} \;,
\end{gather*}
for any indices $i,j,k,l$.
This sector exists for all versions of the LEFT, with $n_q \geq 3$.

\subsection{\boldmath{$\Delta F = 1^{\bar f f}$ transitions $\left(\bar{f}_i f_j \bar{f}'_k f'_k\right)$}}

\subsubsection{\boldmath{$\Delta F = 1^{\bar{f}f} \;:\; d\text{-type} \; \left(\bar{d}_i d_j \bar{f}_k f_k\right)$}}
\label{sec:DF=1qq d-type}

The RG-invariant sub-blocks of the $\Delta F = 1^{\bar{f}f}$ $d$-type sector can be further separated into a vector and scalar part. The vector part contains two large invariant sub-blocks, which for $i < j$ and $k \neq i,j$ read
\begin{gather*}
\label{eq:DF=1qq d-type Vector}
\bigg\{ \lwc{ed}{LL}[V][11ij], \lwc{ed}{LL}[V][22ij], \lwc{ed}{LL}[V][33ij],  \lwc{\nu d}{LL}[V][11ij], \lwc{\nu d}{LL}[V][22ij], \lwc{\nu d}{LL}[V][33ij], \lwc{ud}{LL}[V1][11ij], \lwc{ud}{LL}[V8][11ij], \lwc{ud}{LL}[V1][22ij], \lwc{ud}{LL}[V8][22ij], \\ 
\lwc{dd}{LL}[V][ijkk], \lwc{dd}{LL}[V][ikkj], \lwc{dd}{LL}[V][ijjj], \lwc{dd}{LL}[V][ijii], \lwc{de}{LR}[V][ij11], \lwc{de}{LR}[V][ij22], \lwc{de}{LR}[V][ij33], \lwc{du}{LR}[V1][ij11], \lwc{du}{LR}[V8][ij11], \\ 
\lwc{du}{LR}[V1][ij22], \lwc{du}{LR}[V8][ij22], \lwc{dd}{LR}[V1][ijkk], \lwc{dd}{LR}[V8][ijkk], \lwc{dd}{LR}[V1][ijjj], \lwc{dd}{LR}[V8][ijjj], \lwc{dd}{LR}[V1][ijii], \lwc{dd}{LR}[V8][ijii] \bigg\} , \\ 
\bigg\{ \lwc{ed}{RR}[V][11ij], \lwc{ed}{RR}[V][22ij], \lwc{ed}{RR}[V][33ij], \lwc{ud}{RR}[V1][11ij], \lwc{ud}{RR}[V8][11ij], \lwc{ud}{RR}[V1][22ij], \lwc{ud}{RR}[V8][22ij], \lwc{dd}{RR}[V][ijkk], \lwc{dd}{RR}[V][ikkj], \\ 
\lwc{dd}{RR}[V][ijjj], \lwc{dd}{RR}[V][ijii], \lwc{ed}{LR}[V][11ij], \lwc{ed}{LR}[V][22ij], \lwc{ed}{LR}[V][33ij], \lwc{\nu d}{LR}[V][11ij], \lwc{\nu d}{LR}[V][22ij], \lwc{\nu d}{LR}[V][33ij], \lwc{ud}{LR}[V1][11ij], \lwc{ud}{LR}[V8][11ij], \\ 
\lwc{ud}{LR}[V1][22ij], \lwc{ud}{LR}[V8][22ij], \lwc{dd}{LR}[V1][kkij], \lwc{dd}{LR}[V8][kkij], \lwc{dd}{LR}[V1][jjij], \lwc{dd}{LR}[V8][jjij], \lwc{dd}{LR}[V1][iiij], \lwc{dd}{LR}[V8][iiij] \bigg\} \;.
\end{gather*}
The scalar part is separated into smaller independent sub-blocks. For indices satisfying $i \neq j$ and $k \neq i,j$ they read
\begin{gather*}
\label{eq:DF=1qq d-type Scalar}
\bigg\{ \lwc{ed}{RL}[S][11ij] \bigg\} , \bigg\{ \lwc{ed}{RL}[S][22ij] \bigg\} , \bigg\{ \lwc{ed}{RL}[S][33ij] \bigg\} , 
\bigg\{ \lwc{uddu}{LR}[V1][1ji1], \lwc{uddu}{LR}[V8][1ji1] \bigg\} , \bigg\{ \lwc{uddu}{LR}[V1][2ji2], \lwc{uddu}{LR}[V8][2ji2] \bigg\} , \\ 
\bigg\{ \lwc{dd}{LR}[V1][kjik], \lwc{dd}{LR}[V8][kjik], \bigg \} , \bigg\{ \lwc{ed}{RR}[S][11ij], \lwc{ed}{RR}[T][11ij] \bigg\} , \bigg\{ \lwc{ed}{RR}[S][22ij], \lwc{ed}{RR}[T][22ij] \bigg\} , \bigg\{ \lwc{ed}{RR}[S][33ij], \lwc{ed}{RR}[T][33ij] \bigg\} , \\ 
\bigg\{ \lwc{ud}{RR}[S1][11ij], \lwc{ud}{RR}[S8][11ij], \lwc{uddu}{RR}[S1][1ji1], \lwc{uddu}{RR}[S8][1ji1] \bigg\} , \bigg\{ \lwc{ud}{RR}[S1][22ij], \lwc{ud}{RR}[S8][22ij], \lwc{uddu}{RR}[S1][2ji2], \lwc{uddu}{RR}[S8][2ji2] \bigg\} , \\
\bigg\{ \lwc{dd}{RR}[S1][ijkk], \lwc{dd}{RR}[S8][ijkk], \lwc{dd}{RR}[S1][ikkj], \lwc{dd}{RR}[S8][ikkj] \bigg\} , \bigg\{ \lwc{dd}{RR}[S1][ijjj], \lwc{dd}{RR}[S8][ijjj] \bigg\} , \bigg\{ \lwc{dd}{RR}[S1][ijii], \lwc{dd}{RR}[S8][ijii] \bigg\}
\;. 
\end{gather*}
This sector exists for all versions of the LEFT, with $n_q \geq 3$, barring the removal of  
operators containing heavy flavors in each case.

\subsubsection{\boldmath{$\Delta F = 1^{\bar{f}f} \;:\; u\text{-type} \; \left(\bar{u}_i u_j \bar{f}_k f_k\right)$}}

The RG-invariant sub-blocks of the $\Delta F = 1^{\bar{f}f}$ $u$-type RG-invariant sector can be further separated in a vector part and a scalar one. The vector part forms two large invariant sub-blocks, which read (always with $i < j$)
\begin{gather*}
\label{eq:DF=1qq u-type Vector}
\bigg\{ \lwc{eu}{LL}[V][11ij], \lwc{eu}{LL}[V][22ij], \lwc{eu}{LL}[V][33ij],  \lwc{\nu u}{LL}[V][11ij], \lwc{\nu u}{LL}[V][22ij], \lwc{\nu u}{LL}[V][33ij], \lwc{ud}{LL}[V1][ij11], \lwc{ud}{LL}[V8][ij11], \lwc{ud}{LL}[V1][ij22], \lwc{ud}{LL}[V8][ij22], \\ 
\lwc{ud}{LL}[V1][ij33], \lwc{ud}{LL}[V8][ij33], \lwc{uu}{LL}[V][ijjj], \lwc{uu}{LL}[V][ijii], \lwc{ue}{LR}[V][ij11], \lwc{ue}{LR}[V][ij22], \lwc{ue}{LR}[V][ij33], \lwc{ud}{LR}[V1][ij11], \lwc{ud}{LR}[V8][ij11], \\ 
\lwc{ud}{LR}[V1][ij22], \lwc{ud}{LR}[V8][ij22], \lwc{ud}{LR}[V1][ij33], \lwc{ud}{LR}[V8][ij33], \lwc{uu}{LR}[V1][ijjj], \lwc{uu}{LR}[V8][ijjj], \lwc{uu}{LR}[V1][ijii], \lwc{uu}{LR}[V8][ijii] \bigg\} , \\ 
\bigg\{ \lwc{eu}{RR}[V][11ij], \lwc{eu}{RR}[V][22ij], \lwc{eu}{RR}[V][33ij], \lwc{ud}{RR}[V1][ij11], \lwc{ud}{RR}[V8][ij11], \lwc{ud}{RR}[V1][ij22], \lwc{ud}{RR}[V8][ij22], \lwc{ud}{RR}[V1][ij33], \lwc{ud}{RR}[V8][ij33], \\ 
\lwc{uu}{RR}[V][ijjj], \lwc{uu}{RR}[V][ijii], \lwc{eu}{LR}[V][11ij], \lwc{eu}{LR}[V][22ij], \lwc{eu}{LR}[V][33ij], \lwc{\nu u}{LR}[V][11ij], \lwc{\nu u}{LR}[V][22ij], \lwc{\nu u}{LR}[V][33ij], \lwc{du}{LR}[V1][11ij], \lwc{du}{LR}[V8][11ij], \\ 
\lwc{du}{LR}[V1][22ij], \lwc{du}{LR}[V8][22ij], \lwc{du}{LR}[V1][33ij],  \lwc{du}{LR}[V8][33ij], \lwc{uu}{LR}[V1][jjij], \lwc{uu}{LR}[V8][jjij], \lwc{uu}{LR}[V1][iiij], \lwc{uu}{LR}[V8][iiij] \bigg\} \;. 
\end{gather*}
The scalar part is separated into smaller independent sub-blocks, with $i \neq j$. These sub-blocks read
\begin{gather*}
\label{eq:DF=1qq u-type Scalar}
\bigg\{ \lwc{eu}{RL}[S][11ij] \bigg\} , \bigg\{ \lwc{eu}{RL}[S][22ij] \bigg\} , \bigg\{ \lwc{eu}{RL}[S][33ij] \bigg\} 
\bigg\{ \lwc{uddu}{LR}[V1][i11j], \lwc{uddu}{LR}[V8][i11j] \bigg\} , \bigg\{ \lwc{uddu}{LR}[V1][i22j], \lwc{uddu}{LR}[V8][i22j] \bigg\} , \\ 
\bigg\{ \lwc{uddu}{LR}[V1][i33j], \lwc{uddu}{LR}[V8][i33j], \bigg \} , \bigg\{ \lwc{eu}{RR}[S][11ij], \lwc{eu}{RR}[T][11ij] \bigg\} , \bigg\{ \lwc{eu}{RR}[S][22ij], \lwc{eu}{RR}[T][22ij] \bigg\} , \bigg\{ \lwc{eu}{RR}[S][33ij], \lwc{eu}{RR}[T][33ij] \bigg\} , \\  
\bigg\{ \lwc{ud}{RR}[S1][ij11], \lwc{ud}{RR}[S8][ij11], \lwc{uddu}{RR}[S1][i11j], \lwc{uddu}{RR}[S8][i11j] \bigg\} , \bigg\{ \lwc{ud}{RR}[S1][ij22], \lwc{ud}{RR}[S8][ij22], \lwc{uddu}{RR}[S1][i22j], \lwc{uddu}{RR}[S8][i22j] \bigg\} , \\
\bigg\{ \lwc{ud}{RR}[S1][ij33], \lwc{ud}{RR}[S8][ij33], \lwc{uddu}{RR}[S1][i33j], \lwc{uddu}{RR}[S8][i33j] \bigg\} , \bigg\{ \lwc{uu}{RR}[S1][ijjj], \lwc{uu}{RR}[S8][ijjj] \bigg\} , \bigg\{ \lwc{uu}{RR}[S1][ijii], \lwc{uu}{RR}[S8][ijii] \bigg\}
\;. 
\end{gather*}
This sector exists for $n_q \geq 4$, barring the removal of operators containing heavy flavors in each case.

\subsubsection{\boldmath{$\Delta F = 1^{\bar{f}f} \;:\; e\text{-type} \; \left(\bar{e}_i e_j \bar{f}_k f_k\right)$}}
\label{sec:DF=1qq e-type}

The RG-invariant sub-blocks of the $\Delta F = 1^{\bar{f}f}$ $e$-type RG-invariant sector can be further separated into a vector part and a scalar one. The vector part forms two large invariant sub-blocks, which read
\begin{gather*}
\label{eq:DF=1qq e-type Vector}
\bigg\{ \lwc{ed}{LL}[V][ij11], \lwc{ed}{LL}[V][ij22], \lwc{ed}{LL}[V][ij33], \lwc{eu}{LL}[V][ij11], \lwc{eu}{LL}[V][ij22], \lwc{\nu e}{LL}[V][11ij], \lwc{\nu e}{LL}[V][22ij], \lwc{\nu e}{LL}[V][33ij], \lwc{ee}{LL}[V][ijkk], \lwc{ee}{LL}[V][ijjj], \\ 
\lwc{ee}{LL}[V][ijii], \lwc{ed}{LR}[V][ij11], \lwc{ed}{LR}[V][ij22], \lwc{ed}{LR}[V][ij33], \lwc{eu}{LR}[V][ij11], \lwc{eu}{LR}[V][ij22], \lwc{ee}{LR}[V][ijkk], \lwc{ee}{LR}[V][ijjj], \lwc{ee}{LR}[V][ijii] \bigg\} , \\ 
\bigg\{ \lwc{ed}{RR}[V][ij11], \lwc{ed}{RR}[V][ij22], \lwc{ed}{RR}[V][ij33], \lwc{eu}{RR}[V][ij11], \lwc{eu}{RR}[V][ij22], \lwc{ee}{RR}[V][ijkk], \lwc{ee}{RR}[V][ijjj], \lwc{ee}{RR}[V][ijii], \lwc{de}{LR}[V][11ij], \lwc{de}{LR}[V][22ij], \\ 
\lwc{de}{LR}[V][33ij], \lwc{ue}{LR}[V][11ij], \lwc{ue}{LR}[V][22ij], \lwc{\nu e}{LR}[V][11ij], \lwc{\nu e}{LR}[V][22ij], \lwc{\nu e}{LR}[V][33ij], \lwc{ee}{LR}[V][kkij], \lwc{ee}{LR}[V][jjij], \lwc{ee}{LR}[V][iiij] \bigg\} \;,
\end{gather*}
with $i < j$ and $k \neq i,j$.
The scalar part is separated into smaller independent sub-blocks,
\begin{gather*}
\label{eq:DF=1qq e-type Scalar}
\bigg\{ \lwc{ed}{RL}[S][ij11] \bigg\} , \bigg\{ \lwc{ed}{RL}[S][ij22] \bigg\} , \bigg\{ \lwc{ed}{RL}[S][ij33] \bigg\} , \bigg\{ \lwc{eu}{RL}[S][ij11] \bigg\} , \bigg\{ \lwc{eu}{RL}[S][ij22] \bigg\} , \bigg\{ \lwc{ee}{LR}[V][kjik] \bigg\} , \\
\bigg\{ \lwc{ed}{RR}[S][ij11], \lwc{ed}{RR}[T][ij11] \bigg\} , \bigg\{ \lwc{ed}{RR}[S][ij22], \lwc{ed}{RR}[T][ij22] \bigg\} , \bigg\{ \lwc{ed}{RR}[S][ij33], \lwc{ed}{RR}[T][ij33] \bigg\} , \\
\bigg\{ \lwc{eu}{RR}[S][ij11], \lwc{eu}{RR}[T][ij11] \bigg\} , \bigg\{ \lwc{eu}{RR}[S][ij22], \lwc{eu}{RR}[T][ij22] \bigg\} , \bigg\{ \lwc{ee}{RR}[S][ijkk], \lwc{ee}{RR}[S][ikkj] \bigg\} , \bigg\{ \lwc{ee}{RR}[S][ijjj] \bigg\} , \bigg\{ \lwc{ee}{RR}[S][ijii] \bigg\} \;, 
\end{gather*}
with $i \neq j$ and $k \neq i,j$.
This sector exists for all versions of the LEFT, with $n_q \geq 3$, barring the removal of operators containing heavy flavors in each case.

\subsubsection{\boldmath{$\Delta F = 1^{\bar{f}f} \;:\; \nu\text{-type} \; \left(\bar{\nu}_i \nu_j \bar{f}_k f_k\right)$}}

The $\Delta F = 1^{\bar{f}f}$ $\nu$-type RG-invariant sector consists of a large invariant sub-block,
\begin{gather*}
\label{eq:DF=1qq n-type Vector}
\bigg\{ \lwc{\nu e}{LL}[V][ij11], \lwc{\nu e}{LL}[V][ij22], \lwc{\nu e}{LL}[V][ij33], \lwc{\nu d}{LL}[V][ij11], \lwc{\nu d}{LL}[V][ij22], \lwc{\nu d}{LL}[V][ij33], \lwc{\nu u}{LL}[V][ij11], \lwc{\nu u}{LL}[V][ij22], \lwc{\nu \nu}{LL}[V][ijkk], \lwc{\nu \nu}{LL}[V][ijjj], \\ 
\lwc{\nu \nu}{LL}[V][ijii], \lwc{\nu e}{LR}[V][ij11], \lwc{\nu e}{LR}[V][ij22], \lwc{\nu e}{LR}[V][ij33], \lwc{\nu d}{LR}[V][ij11], \lwc{\nu d}{LR}[V][ij22], \lwc{\nu d}{LR}[V][ij33], \lwc{\nu u}{LR}[V][ij11], \lwc{\nu u}{LR}[V][ij22] \bigg\} \;,
\end{gather*}
with $i < j$ and $k \neq i,j$.
This sector exists for all versions of the LEFT, with $n_q \geq 3$, barring the removal of operators containing heavy flavors in each case.

\subsection{\boldmath{$\Delta F = 0$ transitions $\left(\bar{f}_i f_i \bar{f}'_j f'_j\right)$}}

The $\Delta F = 0$ RG-invariant sector consists first of a set of a scalar sub-blocks. In this specific case, all allowed values of the indices $i,j$ should be understood as part of a single $\Delta F = 0$ sector. For arbitrary indices $i,j$ one finds 
\begin{gather*}
\label{eq:DF=0 Scalar I}
\bigg\{ \lwc{ed}{RL}[S][iijj] \bigg\} , \bigg\{ \lwc{eu}{RL}[S][iijj] \bigg\} ,  \bigg\{ \lwc{uddu}{LR}[V1][ijji], \lwc{uddu}{LR}[V8][ijji] \bigg\} , \bigg\{ \lwc{ed}{RR}[S][iijj], \lwc{ed}{RR}[T][iijj] \bigg\} , \\
\bigg\{ \lwc{eu}{RR}[S][iijj], \lwc{eu}{RR}[T][iijj] \bigg\} , \bigg\{ \lwc{ud}{RR}[S1][iijj], \lwc{ud}{RR}[S8][iijj], \lwc{uddu}{RR}[S1][ijji], \lwc{uddu}{RR}[S8][ijji] \bigg\} , \\
\bigg\{ \lwc{dd}{RR}[S1][iiii], \lwc{dd}{RR}[S8][iiii] \bigg\} , \bigg\{ \lwc{uu}{RR}[S1][iiii], \lwc{uu}{RR}[S8][iiii] \bigg\} , \bigg\{ \lwc{ee}{RR}[S][iiii] \bigg\} \;,
\end{gather*}
while restricting to $i < j$ one also finds the following scalar sub-blocks
\begin{gather*}
\label{eq:DF=0 Scalar II}
\bigg\{ \lwc{dd}{LR}[V1][ijji], \lwc{dd}{LR}[V8][ijji] \bigg\} , \bigg\{ \lwc{uu}{LR}[V1][ijji], \lwc{uu}{LR}[V8][ijji] \bigg\} , \bigg\{ \lwc{ee}{LR}[V][ijji] \bigg\} , \bigg\{ \lwc{ee}{RR}[S][iijj], \lwc{ee}{RR}[S][ijji] \bigg\} , \\
\bigg\{ \lwc{dd}{RR}[S1][iijj], \lwc{dd}{RR}[S8][iijj], \lwc{dd}{RR}[S1][ijji], \lwc{dd}{RR}[S8][ijji] \bigg\} , \bigg\{ \lwc{uu}{RR}[S1][iijj], \lwc{uu}{RR}[S8][iijj], \lwc{uu}{RR}[S1][ijji], \lwc{uu}{RR}[S8][ijji] \bigg\} \;.
\end{gather*}
There is also a single $\Delta F = 0$ sub-block containing the vector-like $\Delta F = 0$ operators, 
\begin{gather*}
\label{eq:DF=0 Vector}
\bigg\{ \lwc{ed}{LL}[V][1111], \lwc{ed}{LL}[V][1122], \lwc{ed}{LL}[V][1133], \lwc{ed}{LL}[V][2211], \lwc{ed}{LL}[V][2222], \lwc{ed}{LL}[V][2233], \lwc{ed}{LL}[V][3311], \lwc{ed}{LL}[V][3322], \lwc{ed}{LL}[V][3333], \lwc{eu}{LL}[V][1111], \lwc{eu}{LL}[V][1122], \\
\lwc{eu}{LL}[V][2211], \lwc{eu}{LL}[V][2222], \lwc{eu}{LL}[V][3311], \lwc{eu}{LL}[V][3322], \lwc{\nu e}{LL}[V][1111], \lwc{\nu e}{LL}[V][1122], \lwc{\nu e}{LL}[V][1133], \lwc{\nu e}{LL}[V][2211], \lwc{\nu e}{LL}[V][2222], \lwc{\nu e}{LL}[V][2233], \lwc{\nu e}{LL}[V][3311], \\
\lwc{\nu e}{LL}[V][3322], \lwc{\nu e}{LL}[V][3333], \lwc{\nu d}{LL}[V][1111], \lwc{\nu d}{LL}[V][1122], \lwc{\nu d}{LL}[V][1133], \lwc{\nu d}{LL}[V][2211], \lwc{\nu d}{LL}[V][2222], \lwc{\nu d}{LL}[V][2233], \lwc{\nu d}{LL}[V][3311], \lwc{\nu d}{LL}[V][3322], \lwc{\nu d}{LL}[V][3333], \\
\lwc{\nu u}{LL}[V][1111], \lwc{\nu u}{LL}[V][1122], \lwc{\nu u}{LL}[V][2211], \lwc{\nu u}{LL}[V][2222], \lwc{\nu u}{LL}[V][3311], \lwc{\nu u}{LL}[V][3322], \lwc{ud}{LL}[V1][1111], \lwc{ud}{LL}[V8][1111], \lwc{ud}{LL}[V1][1122], \lwc{ud}{LL}[V8][1122], \\
\lwc{ud}{LL}[V1][1133], \lwc{ud}{LL}[V8][1133], \lwc{ud}{LL}[V1][2211], \lwc{ud}{LL}[V8][2211], \lwc{ud}{LL}[V1][2222], \lwc{ud}{LL}[V8][2222], \lwc{ud}{LL}[V1][2233], \lwc{ud}{LL}[V8][2233], \lwc{dd}{LL}[V][1111], \\
\lwc{dd}{LL}[V][1122], \lwc{dd}{LL}[V][1221], \lwc{dd}{LL}[V][1133], \lwc{dd}{LL}[V][1331], \lwc{dd}{LL}[V][2222], \lwc{dd}{LL}[V][2233], \lwc{dd}{LL}[V][2332], \lwc{dd}{LL}[V][3333], \lwc{uu}{LL}[V][1111], \lwc{uu}{LL}[V][1122], \lwc{uu}{LL}[V][1221], \\
\lwc{uu}{LL}[V][2222], \lwc{ee}{LL}[V][1111], \lwc{ee}{LL}[V][1122], \lwc{ee}{LL}[V][1133], \lwc{ee}{LL}[V][2222], \lwc{ee}{LL}[V][2233], \lwc{ee}{LL}[V][3333], \lwc{\nu\nu}{LL}[V][1111], \lwc{\nu\nu}{LL}[V][1122], \lwc{\nu\nu}{LL}[V][1133], \lwc{\nu\nu}{LL}[V][2222], \\
\lwc{\nu\nu}{LL}[V][2233], \lwc{\nu\nu}{LL}[V][3333], \lwc{de}{LR}[V][1111], \lwc{de}{LR}[V][1122], \lwc{de}{LR}[V][1133], \lwc{de}{LR}[V][2211], \lwc{de}{LR}[V][2222], \lwc{de}{LR}[V][2233], \lwc{de}{LR}[V][3311], \lwc{de}{LR}[V][3322], \lwc{de}{LR}[V][3333], \\
\lwc{ue}{LR}[V][1111], \lwc{ue}{LR}[V][1122], \lwc{ue}{LR}[V][1133], \lwc{ue}{LR}[V][2211], \lwc{ue}{LR}[V][2222], \lwc{ue}{LR}[V][2233], \lwc{\nu e}{LR}[V][1111], \lwc{\nu e}{LR}[V][1122], \lwc{\nu e}{LR}[V][1133], \lwc{\nu e}{LR}[V][2211], \lwc{\nu e}{LR}[V][2222], \\
\lwc{\nu e}{LR}[V][2233], \lwc{\nu e}{LR}[V][3311], \lwc{\nu e}{LR}[V][3322], \lwc{\nu e}{LR}[V][3333], \lwc{\nu d}{LR}[V][1111], \lwc{\nu d}{LR}[V][1122], \lwc{\nu d}{LR}[V][1133], \lwc{\nu d}{LR}[V][2211], \lwc{\nu d}{LR}[V][2222], \lwc{\nu d}{LR}[V][2233], \lwc{\nu d}{LR}[V][3311], \\
\lwc{\nu d}{LR}[V][3322], \lwc{\nu d}{LR}[V][3333], \lwc{\nu u}{LR}[V][1111], \lwc{\nu u}{LR}[V][1122], \lwc{\nu u}{LR}[V][2211], \lwc{\nu u}{LR}[V][2222], \lwc{\nu u}{LR}[V][3311], \lwc{\nu u}{LR}[V][3322], \lwc{du}{LR}[V1][1111], \lwc{du}{LR}[V8][1111], \\
\lwc{du}{LR}[V1][1122], \lwc{du}{LR}[V8][1122], \lwc{du}{LR}[V1][2211], \lwc{du}{LR}[V8][2211], \lwc{du}{LR}[V1][2222], \lwc{du}{LR}[V8][2222], \lwc{du}{LR}[V1][3311], \lwc{du}{LR}[V8][3311], \lwc{du}{LR}[V1][3322], \\
\lwc{du}{LR}[V8][3322], \lwc{ed}{LR}[V][1111], \lwc{ed}{LR}[V][1122], \lwc{ed}{LR}[V][1133], \lwc{ed}{LR}[V][2211], \lwc{ed}{LR}[V][2222], \lwc{ed}{LR}[V][2233], \lwc{ed}{LR}[V][3311], \lwc{ed}{LR}[V][3322], \lwc{ed}{LR}[V][3333], \\
\lwc{eu}{LR}[V][1111], \lwc{eu}{LR}[V][1122], \lwc{eu}{LR}[V][2211], \lwc{eu}{LR}[V][2222], \lwc{eu}{LR}[V][3311], \lwc{eu}{LR}[V][3322], \lwc{ud}{LR}[V1][1111], \lwc{ud}{LR}[V8][1111], \lwc{ud}{LR}[V1][1122], \lwc{ud}{LR}[V8][1122], \\
\lwc{ud}{LR}[V1][1133], \lwc{ud}{LR}[V8][1133], \lwc{ud}{LR}[V1][2211], \lwc{ud}{LR}[V8][2211], \lwc{ud}{LR}[V1][2222], \lwc{ud}{LR}[V8][2222], \lwc{ud}{LR}[V1][2233], \lwc{ud}{LR}[V8][2233], \lwc{dd}{LR}[V1][1111], \\
\lwc{dd}{LR}[V8][1111], \lwc{dd}{LR}[V1][1122], \lwc{dd}{LR}[V8][1122], \lwc{dd}{LR}[V1][1133], \lwc{dd}{LR}[V8][1133], \lwc{dd}{LR}[V1][2211], \lwc{dd}{LR}[V8][2211], \lwc{dd}{LR}[V1][2222], \lwc{dd}{LR}[V8][2222], \\ 
\lwc{dd}{LR}[V1][2233], \lwc{dd}{LR}[V8][2233], \lwc{dd}{LR}[V1][3311], \lwc{dd}{LR}[V8][3311], \lwc{dd}{LR}[V1][3322], \lwc{dd}{LR}[V8][3322], \lwc{dd}{LR}[V1][3333], \lwc{dd}{LR}[V8][3333], \lwc{uu}{LR}[V1][1111], \\ 
\lwc{uu}{LR}[V8][1111], \lwc{uu}{LR}[V1][1122], \lwc{uu}{LR}[V8][1122], \lwc{uu}{LR}[V1][2211], \lwc{uu}{LR}[V8][2211], \lwc{uu}{LR}[V1][2222], \lwc{uu}{LR}[V8][2222], \lwc{ee}{LR}[V][1111], \lwc{ee}{LR}[V][1122], \\
\lwc{ee}{LR}[V][1133], \lwc{ee}{LR}[V][2211], \lwc{ee}{LR}[V][2222], \lwc{ee}{LR}[V][2233], \lwc{ee}{LR}[V][3311], \lwc{ee}{LR}[V][3322], \lwc{ee}{LR}[V][3333], \lwc{ed}{RR}[V][1111], \lwc{ed}{RR}[V][1122], \lwc{ed}{RR}[V][1133], \\
\lwc{ed}{RR}[V][2211], \lwc{ed}{RR}[V][2222], \lwc{ed}{RR}[V][2233], \lwc{ed}{RR}[V][3311], \lwc{ed}{RR}[V][3322], \lwc{ed}{RR}[V][3333], \lwc{eu}{RR}[V][1111], \lwc{eu}{RR}[V][1122], \lwc{eu}{RR}[V][2211], \lwc{eu}{RR}[V][2222], \\
\lwc{eu}{RR}[V][3311], \lwc{eu}{RR}[V][3322], \lwc{ud}{RR}[V1][1111], \lwc{ud}{RR}[V8][1111], \lwc{ud}{RR}[V1][1122], \lwc{ud}{RR}[V8][1122], \lwc{ud}{RR}[V1][1133], \lwc{ud}{RR}[V8][1133], \lwc{ud}{RR}[V1][2211], \\
\lwc{ud}{RR}[V8][2211], \lwc{ud}{RR}[V1][2222], \lwc{ud}{RR}[V8][2222], \lwc{ud}{RR}[V1][2233], \lwc{ud}{RR}[V8][2233], \lwc{dd}{RR}[V][1111], \lwc{dd}{RR}[V][1122], \lwc{dd}{RR}[V][1221], \lwc{dd}{RR}[V][1133], \\
\lwc{dd}{RR}[V][1331], \lwc{dd}{RR}[V][2222], \lwc{dd}{RR}[V][2233], \lwc{dd}{RR}[V][2332], \lwc{dd}{RR}[V][3333], \lwc{uu}{RR}[V][1111], \lwc{uu}{RR}[V][1122], \lwc{uu}{RR}[V][1221], \lwc{uu}{RR}[V][2222], \lwc{ee}{RR}[V][1111], \\
\lwc{ee}{RR}[V][1122], \lwc{ee}{RR}[V][1133], \lwc{ee}{RR}[V][2222], \lwc{ee}{RR}[V][2233], \lwc{ee}{RR}[V][3333] \bigg\} \,.
\end{gather*}
The Wilson coefficients from this vector sub-block are real, given that their associated operators are hermitian.

This sector exists for all versions of the LEFT, with $n_q \geq 3$, barring the removal of operators containing heavy flavors in each case.

\subsection{\boldmath{$\Delta L = 2$ transitions $(\bar\nu^c_i \nu_j \bar{f}_k f_l \; \text{or} \; \bar\nu^c_i e_j \bar d_k u_l)$}}

The RG-invariant sub-blocks of the sector of operators violating lepton number by two units read
\begin{gather*}
\label{eq:DL=2}
\bigg\{ \lwc{\nu d}{LR}[S][ijkl] \bigg\} , \bigg\{ \lwc{\nu d}{LL}[S][ijkl], \lwc{\nu d}{LL}[T][ijkl] \bigg\} , 
\bigg\{ \lwc{\nu u}{LR}[S][ijkl] \bigg\} , \bigg\{ \lwc{\nu u}{LL}[S][ijkl], \lwc{\nu u}{LL}[T][ijkl] \bigg\} , 
\bigg\{ \lwc{\nu e}{LR}[S][ijkl] \bigg\} , \bigg\{ \lwc{\nu e}{LL}[S][ijkl], \lwc{\nu e}{LL}[T][ijkl] \bigg\} , \\
\bigg\{ \lwc{\nu e d u}{RR}[V][ijkl] \bigg\} , \bigg\{ \lwc{\nu e d u}{RL}[V][ijkl] \bigg\} , \bigg\{ \lwc{\nu e d u}{LR}[S][ijkl] \bigg\} , \bigg\{ \lwc{\nu e d u}{LL}[S][ijkl], \lwc{\nu e d u}{LL}[T][ijkl] \bigg\} , \\
\bigg\{ \lwc{\nu e d u}{RR}[V][jikl] \bigg\} , \bigg\{ \lwc{\nu e d u}{RL}[V][jikl] \bigg\} , \bigg\{ \lwc{\nu e d u}{LR}[S][jikl] \bigg\} , \bigg\{ \lwc{\nu e d u}{LL}[S][jikl], \lwc{\nu e d u}{LL}[T][jikl] \bigg\} \,,
\end{gather*}
with $i < j$ and any $k,l$. For $i = j$ the sector is reduced to
\begin{gather*}
\label{eq:DL=2bis}
\bigg\{ \lwc{\nu d}{LR}[S][iikl] \bigg\} , \bigg\{ \lwc{\nu d}{LL}[S][iikl] \bigg\} , 
\bigg\{ \lwc{\nu u}{LR}[S][iikl] \bigg\} , \bigg\{ \lwc{\nu u}{LL}[S][iikl] \bigg\} , 
\bigg\{ \lwc{\nu e}{LR}[S][iikl] \bigg\} , \bigg\{ \lwc{\nu e}{LL}[S][iikl] \bigg\} , \\
\bigg\{ \lwc{\nu e d u}{RR}[V][iikl] \bigg\} , \bigg\{ \lwc{\nu e d u}{RL}[V][iikl] \bigg\} , \bigg\{ \lwc{\nu e d u}{LR}[S][iikl] \bigg\} , \bigg\{ \lwc{\nu e d u}{LL}[S][iikl], \lwc{\nu e d u}{LL}[T][iikl] \bigg\} \ .
\end{gather*}
This sector exists for all versions of the LEFT, with $n_q \geq 3$.

\subsection{\boldmath{$\Delta L = 4$ transitions $(\bar\nu^c_i \nu_j \bar\nu^c_k \nu_l)$}}

There is only one type of operator violating lepton number by four units, 
\begin{gather*}
\label{eq:DL=4}
\bigg\{ \lwc{\nu\nu}{LL}[S][ijkl] \bigg\} \;,
\end{gather*}
with $i \leq j < l$ and $i < k \leq l$.
This sector exists for all versions of the LEFT, with $n_q \geq 3$.

\subsection{\boldmath{$|\Delta B| = |\Delta L| = 1$ transitions: (baryon number violating)}}

\subsubsection{\boldmath{$\Delta B = \Delta L = 1$ \;:\; $e$-type \; $(\bar d^c_i u_j \bar u^c_k e_l \text{ or } \bar u^c_k u_j \bar d^c_i e_l)$}}

The RG-invariant sub-blocks of this baryon-number-violating RG-invariant sector read
\begin{gather*}
\label{eq:BNV+e1}
\bigg\{ \lwc{duu}{LL}[S][ijkl], \lwc{duu}{LL}[S][ikjl] \bigg\} , \bigg\{ \lwc{duu}{RR}[S][ijkl], \lwc{duu}{RR}[S][ikjl] \bigg\} , \bigg\{ \lwc{duu}{LR}[S][ijkl] \bigg\} , \bigg\{ \lwc{duu}{RL}[S][ijkl] \bigg\}, \\
\bigg\{ \lwc{duu}{LR}[S][ikjl] \bigg\} , \bigg\{ \lwc{duu}{RL}[S][ikjl] \bigg\}, \bigg\{ \lwc{uud}{LR}[S][jkil] \bigg\} , \bigg\{ \lwc{uud}{RL}[S][jkil] \bigg\} ,
\end{gather*}
for $j < k$ and any indices $i,l$. For $j = k$ this sector is reduced to
\begin{gather*}
\label{eq:BNV+e2}
\bigg\{ \lwc{duu}{LL}[S][ijjl] \bigg\} , \bigg\{ \lwc{duu}{RR}[S][ijjl] \bigg\} , \bigg\{ \lwc{duu}{LR}[S][ijjl] \bigg\} , \bigg\{ \lwc{duu}{RL}[S][ijjl] \bigg\} \ .
\end{gather*}
This sector exists for all versions of the LEFT, with $n_q \geq 3$.

\subsubsection{\boldmath{$\Delta B = \Delta L = 1$ \;:\; $\nu$-type \; $(\bar u^c_i d_j \bar d^c_k \nu_l \text{ or } \bar d^c_k u_i \bar d^c_j \nu_l \text{ or } \bar d^c_j d_k \bar u^c_i \nu_l)$}}

The RG-invariant sub-blocks of this baryon-number-violating RG-invariant sector read
\begin{gather*}
\label{eq:BNV+nu1}
\bigg\{ \lwc{udd}{LL}[S][ijkl], \lwc{udd}{LL}[S][ikjl] \bigg\} , \bigg\{ \lwc{dud}{RL}[S][jikl] \bigg\} , \bigg\{ \lwc{dud}{RL}[S][kijl] \bigg\} , \bigg\{ \lwc{ddu}{RL}[S][jkil] \bigg\}
\end{gather*}
for $j < k$ and any indices $i,l$. For $j = k$ this sector is reduced to
\begin{gather*}
\label{eq:BNV+nu1bis}
\bigg\{ \lwc{udd}{LL}[S][ijjl] \bigg\} , \bigg\{ \lwc{dud}{RL}[S][jijl] \bigg\} \ .
\end{gather*}
This sector exists for all versions of the LEFT, with $n_q \geq 3$.

\subsubsection{\boldmath{$\Delta B = -\Delta L = 1$ \;:\; $e$-type \; $(\bar d^c_i d_j \bar e^c_l d_k)$}}

The RG-invariant sub-blocks of this baryon-number-violating RG-invariant sector read
\begin{gather*}
\label{eq:BNV-e1}
\bigg\{ \lwc{ddd}{LL}[S][ijlk], \lwc{ddd}{LL}[S][iklj] \bigg\} , \bigg\{ \lwc{ddd}{RR}[S][ijlk], \lwc{ddd}{RR}[S][iklj] \bigg\} , \bigg\{ \lwc{ddd}{LR}[S][ijlk] \bigg\} , \bigg\{ \lwc{ddd}{RL}[S][ijlk] \bigg\} , \\
\bigg\{ \lwc{ddd}{LR}[S][iklj] \bigg\} , \bigg\{ \lwc{ddd}{RL}[S][iklj] \bigg\}, \bigg\{ \lwc{ddd}{LR}[S][jkli] \bigg\} , \bigg\{ \lwc{ddd}{RL}[S][jkli] \bigg\},  
\end{gather*}
for $i < j < k$ and any index $l$. This sector is reduced to
\begin{gather*}
\label{eq:BNV-e2}
\bigg\{ \lwc{ddd}{LL}[S][ijlj] \bigg\} , \bigg\{ \lwc{ddd}{RR}[S][ijlj] \bigg\} , \bigg\{ \lwc{ddd}{LR}[S][ijlj] \bigg\} , \bigg\{ \lwc{ddd}{RL}[S][ijlj] \bigg\}, 
\end{gather*}
for $i < j = k$ (for arbitrary $l$). It is instead reduced to
\begin{gather*}
\label{eq:BNV-e3}
\bigg\{ \lwc{ddd}{LL}[S][ijli] \bigg\} , \bigg\{ \lwc{ddd}{RR}[S][ijli] \bigg\} , \bigg\{ \lwc{ddd}{LR}[S][ijli] \bigg\} , \bigg\{ \lwc{ddd}{RL}[S][ijli] \bigg\},  
\end{gather*}
for $i = k < j$ (for arbitrary $l$). 
This sector exists for all versions of the LEFT, with $n_q \geq 3$.

\subsubsection{\boldmath{$\Delta B = -\Delta L = 1$ \;:\; $\nu$-type \; $(\bar u^c_i d_j \bar \nu^c_l d_k \text{ or } \bar d^c_k d_j \bar \nu^c_l u_i)$}}

The RG-invariant sub-blocks of this baryon-number-violating RG-invariant sector read
\begin{gather*}
\label{eq:BNV-nu1}
\bigg\{ \lwc{udd}{RR}[S][ijlk], \lwc{udd}{RR}[S][iklj] \bigg\} , \bigg\{ \lwc{ddu}{LR}[S][kjli] \bigg\} , \bigg\{ \lwc{udd}{LR}[S][ijlk] \bigg\} , \bigg\{ \lwc{udd}{LR}[S][iklj] \bigg\} ,
\end{gather*}
for $j<k$ and any indices $i,l$. For $j = k$ this sector is reduced to
\begin{gather*}
\label{eq:BNV-nu2}
\bigg\{ \lwc{udd}{RR}[S][ijlj]\bigg\} , \bigg\{ \lwc{udd}{LR}[S][ijlj] \bigg\} \ .
\end{gather*}
This sector exists for all versions of the LEFT, with $n_q \geq 3$.

\section{Details of the calculation and results}
\label{sec:details}

In order to obtain the two-loop ADMs for the four-fermion operators in the LEFT we have followed three different approaches. 
First, 
we have collected all the results available in the literature. These are typically given in an operator basis (and/or evanescent renormalization scheme) that differs from the one in this work. Hence, we have translated these results to the JMS basis through NLO basis changes. 
Second, 
we have performed a series of operations in flavor space, consisting in flavor-symmetrizations of the different bases, in order to extend the results in the literature to other sectors.
Third, 
we have used certain coefficients of UV~divergencies of two-loop diagrams (``pole coefficients'') that can be found in the literature.
Combined with the corresponding charge and color factors, and with the use of NLO changes of basis, we have used these pole coefficients to obtain the complete set of two-loop ADMs for all four-fermion operators, at order $O(\alpha_s^2), O(\alpha_s \alpha)$, and $O(\alpha^2)$. In doing so, we have verified the results obtained with the first and second approaches, and derived the two-loop ADMs of the sectors that remained to be computed.

\subsection{Change of basis of known results into JMS}

The list of RG-invariant sectors of the LEFT for which the two-loop ADMs have already been given in the literature includes: 
\begin{itemize}
    \item {\bf $\boldsymbol{\Delta F = 2: \;d}$-type} 
    \subitem At order $O(\alpha_s^2)$~\cite{Buras:2000if} in the BMU basis.
    \item {\bf $\boldsymbol{\Delta F = 1: \;ud}$-type} 
    \subitem At order $O(\alpha_s^2)$~\cite{Buras:2000if} in the BMU basis.
    \item {\bf $\boldsymbol{\Delta F = 1^{\bar f f}: \;d}$-type} 
    \subitem At order $O(\alpha_s^2)$~\cite{Buras:2000if} in the BMU basis. 
    \item {\bf $\boldsymbol{\Delta F = 1^{\bar f f}: \;d}$-type (only SM operators)} 
    \subitem At order $O(\alpha_s^2)$~\cite{Buras:1991jm,Buras:1992tc,Ciuchini:1993vr,Ciuchini:1993ks} and $O(\alpha_s \alpha)$~\cite{Buras:1992zv,Ciuchini:1993vr,Ciuchini:1993ks} in the BMU basis. 
    \subitem At order $O(\alpha_s^2)$~\cite{Chetyrkin:1996vx,Chetyrkin:1997gb,Gambino:2003zm,Bobeth:2003at,Gorbahn:2004my,Huber:2005ig} and $O(\alpha_s \alpha)$~\cite{Bobeth:2003at,Huber:2005ig} in the CMM basis.\footnote{The CMM basis was later completed into the ``Bern'' basis~\cite{Aebischer:2017gaw}, for all 4-fermion operators in the LEFT.} 
\end{itemize}
In all cases, the BMU basis is used together with the evanescent scheme corresponding to the ``Greek projections''~\cite{Tracas:1982gp,Buras:2000if}, as indicated in Appendix~\ref{app:EV Basis}, while the CMM basis is used in combination with a scheme that makes all explicit $O(\epsilon)$ terms vanish in the definition of evanescent operators, corresponding to $a_{\text{ev}},b_{\text{ev}},c_{\text{ev}}, ... = 0$ in~\Reff{Dekens:2019ept}. In the $d$-type $\Delta F = 1^{\bar f f}$ sector, the term ``SM operators'' refers to the 10 operators with non-vanishing SM matching conditions in either the BMU or CMM bases. They include two current-current, four QCD penguin and four QED penguin operators.

The known ADMs for the sectors $\Delta F = 2:$ $d$-type and $\Delta F = 1:$ $ud$-type, can be transformed into the JMS basis by direct application of~\Eqs{eq:NLO ADM Transformation 20}{eq:NLO ADM Transformation 11}.
For the following JMS operators in the $\Delta F = 2:$ $d$-type sector,
\eq{
\Big\{ & \Op{dd}{LL}[V][ijij], \Op{dd}{LR}[V1][ijij], \Op{dd}{LR}[V8][ijij], \Op{dd}{RR}[S1][ijij], \Op{dd}{RR}[S8][ijij] \Big\} \ , 
\nonumber
}
and the corresponding BMU operators in Section~2 of~\Reff{Buras:2000if} we find,
\eq{
\label{eq:R DF=2}
\hat R_{\Delta F = 2}=
\left(
\arraycolsep=4pt
\def\arraystretch{1}
\begin{array}{ccccc}
 1 & 0 & 0 & 0 & 0 \\
 0 & 1 & 0 & 0 & 0 \\
 0 & -\frac{1}{6} & -1 & 0 & 0 \\
 0 & 0 & 0 & 1 & 0 \\
 0 & 0 & 0 & -\frac{5}{12} & -\frac{1}{16}
\end{array}
\right) \;,
\quad
\Delta\hat r^{(1,0)}_{\Delta F = 2}=
\left(
\arraycolsep=4pt
\def\arraystretch{1}
\begin{array}{ccccc}
 0 & 0 & 0 & 0 & 0 \\
 0 & 0 & 0 & 0 & 0 \\
 0 & 0 & 0 & 0 & 0 \\
 0 & 0 & 0 & 0 & 0 \\
 0 & 0 & 0 & \frac{32}{9} & \frac{29}{6} \\
\end{array}
\right) \;,
}
leading to the following ADM in the JMS basis,
\eq{
\label{eq:ADM DF=2}
\hat \gamma^{(2,0)}_{\Delta F = 2}=
\left(
\arraycolsep=2pt
\def\arraystretch{1.25}
\begin{array}{ccccc}
 \frac{4 n_q}{9}-7 & 0 & 0 & 0 & 0 \\
 0 & -\frac{28}{3} & \frac{44 n_q}{3}-198 & 0 & 0 \\
 0 & \frac{88 n_q}{27}-92 & \frac{46 n_q}{9}-\frac{1145}{6} & 0 & 0 \\
 0 & 0 & 0 & \frac{656 n_q}{81}-\frac{1432}{9} & -\frac{8 n_q}{27}-\frac{8}{3} \\
 0 & 0 & 0 & \frac{2240}{27}-\frac{160 n_q}{243} & \frac{1081}{9}-\frac{650 n_q}{81} \\
\end{array}
\right) \ .
}
The operator $\op^{V,RR}$ has the same anomalous dimension as $\op^{V,LL}$.
The ADM for the $\Delta F = 2:$ $u$-type sector is trivially identical.

For the following JMS operators in the $\Delta F = 1:$ $ud$-type sector,
\eq{
\Big\{ \Op{ud}{LL}[V1][ijkl], \Op{ud}{LL}[V8][ijkl], \Op{ud}{LR}[V1][ijkl], \Op{ud}{LR}[V8][ijkl], \Op{uddu}{LR}[V1][ilkj], 
\Op{uddu}{LR}[V8][ilkj], \Op{ud}{RR}[S1][ijkl], \Op{ud}{RR}[S8][ijkl], \Op{uddu}{RR}[S1][ilkj], \Op{uddu}{RR}[S8][ilkj] \Big\} \ ,
\nonumber
}
and the corresponding BMU operators in Section~3 of~\Reff{Buras:2000if} we find,
\begin{gather}
\label{eq:R DF=1}
\hat R_{\Delta F = 1: \,ud\text{-type}}=
\left(
\arraycolsep=4pt
\def\arraystretch{1}
\begin{array}{cccccccccc}
 0 & 1 & 0 & 0 & 0 & 0 & 0 & 0 & 0 & 0 \\
 \frac{1}{2} & -\frac{1}{6} & 0 & 0 & 0 & 0 & 0 & 0 & 0 & 0 \\
 0 & 0 & 0 & 1 & 0 & 0 & 0 & 0 & 0 & 0 \\
 0 & 0 & \frac{1}{2} & -\frac{1}{6} & 0 & 0 & 0 & 0 & 0 & 0 \\
 0 & 0 & 0 & 0 & -2 & 0 & 0 & 0 & 0 & 0 \\
 0 & 0 & 0 & 0 & \frac{1}{3} & -1 & 0 & 0 & 0 & 0 \\
 0 & 0 & 0 & 0 & 0 & 0 & 0 & 1 & 0 & 0 \\
 0 & 0 & 0 & 0 & 0 & 0 & \frac{1}{2} & -\frac{1}{6} & 0 & 0 \\
 0 & 0 & 0 & 0 & 0 & 0 & -\frac{1}{2} & 0 & \frac{1}{8} & 0 \\
 0 & 0 & 0 & 0 & 0 & 0 & \frac{1}{12} & -\frac{1}{4} & -\frac{1}{48} & \frac{1}{16} 
\end{array}
\right) \;, \\
\label{eq:Dr DF=1}
\Delta\hat r^{(1,0)}_{\Delta F = 1: \,ud\text{-type}}=
\left(
\arraycolsep=4pt
\def\arraystretch{1}
\begin{array}{c|c}
 0_{6\times 6} & 0_{6\times 4} \\
 \hline
 0_{4\times 6} & 
\begin{array}{cccc}
 0 & 0 & 0 & 0 \\
 0 & 0 & 0 & 0 \\
 \frac{16}{3} & 2 & \frac{4}{3} & 3 \\
 \frac{16}{9} & \frac{11}{3} & \frac{2}{3} & \frac{1}{3} \\
\end{array}
\end{array}
\right) \;.
\end{gather}
leading to the following ADM in the JMS basis,
\begin{gather}
\label{eq:ADM DF=1}
\begin{aligned}
\hat \gamma^{(2,0)}_{\Delta F = 1: \,ud\text{-type}}=
&\left(
\arraycolsep=2pt
\def\arraystretch{1.25}
\begin{array}{cccccccccc}
 -\frac{28}{3} & \frac{4 n_q}{3}+7 & 0 & 0 & 0 & 0  \\
 \frac{8 n_q}{27}+\frac{14}{9} & -\frac{4 n_q}{9}-\frac{35}{3} & 0 & 0 & 0 & 0  \\
 0 & 0 & -\frac{28}{3} & \frac{44 n_q}{3}-198 & 0 & 0 \\
 0 & 0 & \frac{88 n_q}{27}-92 & \frac{46 n_q}{9}-\frac{1145}{6} & 0 & 0 \\
 0 & 0 & 0 & 0 & -\frac{28}{3} & \frac{44 n_q}{3}-198 & \\
 0 & 0 & 0 & 0 & \frac{88 n_q}{27}-92 & \frac{46 n_q}{9}-\frac{1145}{6} \\
 0 & 0 & 0 & 0 & 0 & 0  \\
 0 & 0 & 0 & 0 & 0 & 0  \\
 0 & 0 & 0 & 0 & 0 & 0  \\
 0 & 0 & 0 & 0 & 0 & 0  \\
\end{array} \right. 
\\
&\hspace{3cm}\left.
\arraycolsep=2pt
\def\arraystretch{1.25}
\begin{array}{cccccccccc}
 0 & 0 & 0 & 0 \\
 0 & 0 & 0 & 0 \\
 0 & 0 & 0 & 0 \\
 0 & 0 & 0 & 0 \\
 0 & 0 & 0 & 0 \\
 0 & 0 & 0 & 0 \\
\frac{80 n_q}{9}-280& 32-\frac{8 n_q}{9}&\frac{1088}{9}-\frac{64 n_q}{81}&\frac{16 n_q}{27}-\frac{104}{3} \\
\frac{112}{9}-\frac{16 n_q}{81}&\frac{445}{3}-\frac{202 n_q}{27}&\frac{1904}{27}-\frac{112 n_q}{243}&-\frac{44 n_q}{81}-\frac{254}{9} \\
\frac{1088}{9}-\frac{64 n_q}{81}&\frac{16 n_q}{27}-\frac{104}{3}&\frac{80 n_q}{9}-280& 32-\frac{8 n_q}{9} \\
\frac{1904}{27}-\frac{112 n_q}{243}&-\frac{44 n_q}{81}-\frac{254}{9}&\frac{112}{9}-\frac{16 n_q}{81}&\frac{445}{3}-\frac{202 n_q}{27} \\
\end{array}
\right) \;.
\end{aligned}
\end{gather}
For the $d$-type $\Delta F = 1^{\bar f f}$ sector, the complete $O(\alpha_s^2)$ ADM is known from~\Reff{Buras:2000if} (taking also into account the correction found recently in~\Reff{Morell:2024aml}), and a change of basis to JMS can thus be performed as before.
Given their size, we refrain from writing the explicit matrices here, but we remark that our matrix $\Delta \hat r^{(1,0)}$ has a discrepancy with respect to~Eq.~(22) in~\Reff{Aebischer:2021raf}, where a similar transformation has been worked out.\footnote{
Our results, however, are consistent with the  results in~\Reff{Aebischer:2022tvz}.
}
We thus limit ourselves to indicating the entries in which we disagree, and refer the reader to~\Reff{Aebischer:2021raf} for the remaining entries of the matrix, as well as for the rotation matrix $\hat R$. 
The relevant part of the matrix $\Delta \hat{r}^{(1,0)}$, where this discrepancy arises, is its first $8\times18$ sub-block, denoted by~$\Delta \hat A^{(1,0)}$ in~\Reff{Aebischer:2021raf}.
We find
\eq{
\arraycolsep=4pt
\def\arraystretch{1.25}
\begin{aligned}
\Delta \hat{A}^{(1,0)} = \frac{1}{6} & \left(
\begin{array}{cccccccc}
    0 & 1 & 0 & -1 & 0 & \color{red}{-6} & 0 & 0
\end{array} \right)^T \\
& \times \left(
\begin{array}{cccccccccccccccccc}
    0 & 2 & 0 & 2 & \frac{-1}{N_c} & 1 & \frac{N_c-1}{N_c} & \frac{N_c-1}{N_c} & 0 & 2 & 0 & 2 & 0 & 2 & 0 & 2 & 0 & 2
\end{array} \right) \;.
\end{aligned}
}
where the discrepant terms result from the coefficient highlighted in {\color{red}red}.
Our final results for the ADM at $O(\alpha_s^2)$ can be read off the corresponding beta functions in the ancillary material (see~\Sec{sec:results}).

Knowing the ADM for the SM operators only (e.g. in the CMM basis) is not enough to produce any useful results in the JMS basis (since the JMS basis does not contain a similar~SM sub-block). Thus the known results for the ADM at $O(\alpha_s\alpha)$~\cite{Buras:1992zv,Ciuchini:1993vr,Ciuchini:1993ks,Bobeth:2003at,Huber:2005ig} cannot be used to obtain the corresponding information in the JMS basis.
We thus relegate the discussion of the corresponding results to later sections.

\subsection{ADMs from Flavor Symmetries}
\label{sec:Els Truquets}

We now follow a strategy similar to the one in~Section~6 of~\Reff{Buras:2000if} in order to derive the ADMs for the $d$-type $\Delta F=1.5$ sectors from the one for $\Delta F=1$.

We start by constructing a ``symmetrized'' basis for a 10-dimensional subsector of the $ud$-type $\Delta F=1$ sector (the subsector in~\Eq{eq:ADM DF=1}), which we call the ``$\pm$'' basis:
\begin{align*}
\Bigg\{ \; & 
\frac{1}{2}\bigg(\Op{ud}{LL}[V1][ijkl] + \Op{uddu}{LL}[V1][ijkl]\bigg), 
\frac{1}{2}\bigg(\Op{ud}{LR}[V1][ijkl] + \Op{uddu}{LR}[V1][ijkl]\bigg), 
\frac{1}{2}\bigg(\Op{ud}{LR}[V8][ijkl] + \Op{uddu}{LR}[V8][ijkl]\bigg), \\ 
& 
\frac{1}{2}\bigg(\Op{ud}{RR}[S1][ijkl] + \Op{uddu}{RR}[S1][ijkl]\bigg), 
\frac{1}{2}\bigg(\Op{ud}{RR}[S8][ijkl] + \Op{uddu}{RR}[S8][ijkl]\bigg), 
\\ 
& 
\frac{1}{2}\bigg(\Op{ud}{LL}[V1][ijkl] - \Op{uddu}{LL}[V1][ijkl]\bigg), 
\frac{1}{2}\bigg(\Op{ud}{LR}[V1][ijkl] - \Op{uddu}{LR}[V1][ijkl]\bigg), 
\frac{1}{2}\bigg(\Op{ud}{LR}[V8][ijkl] - \Op{uddu}{LR}[V8][ijkl]\bigg), \\
&
\frac{1}{2}\bigg(\Op{ud}{RR}[S1][ijkl] - \Op{uddu}{RR}[S1][ijkl]\bigg), 
\frac{1}{2}\bigg(\Op{ud}{RR}[S8][ijkl] - \Op{uddu}{RR}[S8][ijkl]\bigg) \Bigg\} \ ,
\end{align*}
where $\op_{uddu}^{V1,LL}$ is not a JMS operator but it is defined analogously.
This basis is ``symmetrized'' in the sense that, under the exchange $u_j\leftrightarrow d_l$, the first five `$+$' operators are symmetric and the last five `$-$' operators are antisymmetric.
We then calculate the $O(\alpha_s^2)$ ADM in this basis by means of a basis change from the ADM in the JMS basis given in~\Eq{eq:ADM DF=1}.

The key issue now is that, if we consider the set of operators obtained from the ``$\pm$'' basis by the replacements $u_i\to d_i$ and $u_j\to d_l$, then their ADM must be identical to the one just calculated, because of the $u_j\leftrightarrow d_l$ flavor symmetry.
But the ones arising from the `$+$' operators become,

\eq{
\Big\{ \Op{dd}{LL}[V][ijkj], \Op{dd}{LR}[V1][ijkj], \Op{dd}{LR}[V8][ijkj], \Op{dd}{RR}[S1][ijkj], \Op{dd}{RR}[S8][ijkj] \Big\} \ ,
\nonumber
}
which is a subset of the $d$-type $\Delta F=1.5$ sector. 
The ADM for this set is identical to that in~\Eq{eq:ADM DF=2}.
The ADMs at $O(\alpha_s^2)$ of all the RG-invariant sub-blocks in~\Sec{sec:DF=1.5 d-type} thus follow.

\subsection{Full ADM from Pole Coefficients}

Aside from those two-loop results for ADMs that are already known, some additional information is available concerning the separate contributions from the various two-loop diagrams~\cite{Buras:1989xd,Buras:1992tc,Buras:2000if}. This information is given in the form of ``tables of pole coefficients'', containing the $1/\epsilon^2$ and $1/\epsilon$ contributions from the sum of different two-loop diagrams and associated one-loop diagrams with counterterms. In a nutshell, the two-loop ADMs are written as
\eq{
\label{eq:TableRules}
\gamma_{ij}^{(m,n)} = -4 Z_{Q_i}^{(m,n,1)} \delta_{Q_iQ_j}
-4 \sum_{D\,\in\,\text{diags.}} C_{D,Q_iQ_j}^{(m,n)} T_{D,Q_iQ_j}^{(m,n;1)}
\ ,
}
where $T_{D,Q_iQ_j}^{(m,n;1)}$ represents a $1/\epsilon$ pole coefficient associated to a two-loop diagram of class~$D$, complemented with an appropriate color/charge factor $ C_{D,Q_iQ_j}^{(m,n)}$. 
In this section we will show how, knowing these coefficients $T_{D,Q_iQ_j}^{(2,0;1)}$ for certain $Q_i$, $Q_j$, is enough to reconstruct the full two-loop ADM in the LEFT.

We first clarify what the coefficients $T_{D,Q_iQ_j}^{(m,n;1)}$ are, and work out a representative example.
Comparing to~Eqs.~(\ref{eq:gamma20a})-(\ref{eq:gamma11a}) one finds that
\eqa{
\sum_{D\,\in\,\text{diags.}}\hspace{-2mm}
C_{D,Q_iQ_j}^{(2,0)} T_{D,Q_iQ_j}^{(2,0;1)} =&\ 
a^{(2,0;1)}_{Q_iQ_j} 
-a^{(1,0;1)}_{Q_iQ_k} a^{(1,0;0)}_{Q_kQ_j} 
-\frac12 a^{(1,0;1)}_{Q_iE_k} a^{(1,0;0)}_{E_kQ_j}\ ,
\\
\sum_{D\,\in\,\text{diags.}}\hspace{-2mm}
C_{D,Q_iQ_j}^{(0,2)} T_{D,Q_iQ_j}^{(0,2;1)} =&\ 
a^{(0,2;1)}_{Q_iQ_j} 
-a^{(0,1;1)}_{Q_iQ_k} a^{(0,1;0)}_{Q_kQ_j}
-\frac12 a^{(0,1;1)}_{Q_iE_k} a^{(0,1;0)}_{E_kQ_j}\ ,
\\
\sum_{D\,\in\,\text{diags.}}\hspace{-2mm}
C_{D,Q_iQ_j}^{(1,1)} T_{D,Q_iQ_j}^{(1,1;1)} =&\ 
a^{(1,1;1)}_{Q_iQ_j} 
-a^{(1,0;1)}_{Q_iQ_k} a^{(0,1;0)}_{Q_kQ_j} 
-a^{(0,1;1)}_{Q_iQ_k} a^{(1,0;0)}_{Q_kQ_j}
\nonumber\\
&\ -\frac12 \bigg[
a^{(1,0;1)}_{Q_iE_k} a^{(0,1;0)}_{E_kQ_j} 
+  a^{(0,1;1)}_{Q_iE_k} a^{(1,0;0)}_{E_kQ_j} 
\bigg]
\ .
}
How to group the different diagrammatic contributions into the classes $D$ is arbitrary, but it needs to be specified in order to understand the tables of pole coefficients in~\Refs{Buras:1989xd,Buras:1992tc,Ciuchini:1993vr,Buras:2000if}.
The general rule is that each type of two-loop diagram is grouped with the one-loop conterterms that renormalize its subdivergences.

For illustration, let us consider the $\op(\alpha_s^2)$ corrections to the $\Delta F = 1^{\bar q q}$ four-quark penguin operator $Q_1=(\bar s \gamma^\mu P_L d)(\bar u \gamma_\mu P_L u)$, as discussed in~\Reff{Buras:1989xd}, using the same numbering for the diagram types. For example:

\noindent
\hdashrule{\textwidth}{1pt}{1pt}
\\
{\bf Diagrams $D=4$:} The diagrams of this class correspond to the ladder diagrams on each fermion current. There are two such two-loop diagrams, giving the same contribution to the pole coefficient, which corresponds to the multiplicity factor $M=2$ in~\Reff{Buras:1989xd}. The color factor $C_F^2$, which is by definition removed from the pole coefficient, will be suppressed in the following. One finds,
\eqa{
&a^{(2,0;1)}_{Q_1Q_1} \bigg|_{D=4} =
\quad
\begin{tikzpicture}[baseline={(a2.base)}]
\begin{feynman}
\vertex (a1) at (0,0.05);
\vertex (a2) at (0,0) [label=right:$\scriptstyle\ Q_1$];
\vertex (v1) at (-0.5,0.55);
\vertex (v2) at (-0.3,0.35);
\vertex (v3) at (0.3,0.35);
\vertex (v4) at (0.5, 0.55);
\vertex (i1) at (-0.7,0.75);
\vertex (f1) at (0.7,0.75);
\vertex (i2) at (-0.7,-0.7);
\vertex (f2) at (0.7,-0.7);
\diagram{
(i1)--(a1)--(f1),
(i2)--(a2)--(f2),
(v1)--[photon](v4),
(v2)--[photon](v3)
};
\end{feynman}
\draw (1,-1) -- (1,1);
\end{tikzpicture}_{\ Q_1,1/\epsilon} 
+\hspace{5mm}
\begin{tikzpicture}[baseline={(a2.base)}]
\begin{feynman}
\vertex (a1) at (0,0.05);
\vertex (a2) at (0,0) [label=right:$\scriptstyle\ Q_1$];
\vertex (v1) at (-0.5,-0.5);
\vertex (v2) at (-0.3,-0.3);
\vertex (v3) at (0.3, -0.3);
\vertex (v4) at (0.5, -0.5);
\vertex (i1) at (-0.7,0.75);
\vertex (f1) at (0.7,0.75);
\vertex (i2) at (-0.7,-0.7);
\vertex (f2) at (0.7,-0.7);
\diagram{
(i1)--(a1)--(f1),
(i2)--(a2)--(f2),
(v1)--[photon](v4),
(v2)--[photon](v3)
};
\end{feynman}
\draw (1,-1) -- (1,1);
\end{tikzpicture}_{\ Q_1,1/\epsilon}
=  -\frac32 \ ,
\\[3mm]
&a^{(1,0;1)}_{Q_1Q_k} a^{(1,0;0)}_{Q_kQ_1} \bigg|_{D=4} =
\\
&\quad=
\begin{tikzpicture}[baseline={(a2.base)}]
\begin{feynman}
\vertex (a1) at (0,0.05);
\vertex (a2) at (0,0) [label=right:$\scriptstyle\ Q_1$];
\vertex (v1) at (-0.5,0.55);
\vertex (v4) at (0.5, 0.55);
\vertex (i1) at (-0.7,0.75);
\vertex (f1) at (0.7,0.75);
\vertex (i2) at (-0.7,-0.7);
\vertex (f2) at (0.7,-0.7);
\diagram{
(i1)--(a1)--(f1),
(i2)--(a2)--(f2),
(v1)--[photon](v4)
};
\end{feynman}
\draw (1,-1) -- (1,1);
\end{tikzpicture}_{\ Q_k,1/\epsilon} 
\hspace{-3mm}
\times
\begin{tikzpicture}[baseline={(a2.base)}]
\begin{feynman}
\vertex (a1) at (0,0.05);
\vertex (a2) at (0,0) [label=right:$\scriptstyle\ Q_k$];
\vertex (v1) at (-0.5,0.55);
\vertex (v4) at (0.5, 0.55);
\vertex (i1) at (-0.7,0.75);
\vertex (f1) at (0.7,0.75);
\vertex (i2) at (-0.7,-0.7);
\vertex (f2) at (0.7,-0.7);
\diagram{
(i1)--(a1)--(f1),
(i2)--(a2)--(f2),
(v1)--[photon](v4)
};
\end{feynman}
\draw (1,-1) -- (1,1);
\end{tikzpicture}_{\ Q_1,\text{finite}}
\hspace{-3mm}
+
\hspace{5mm}
\begin{tikzpicture}[baseline={(a2.base)}]
\begin{feynman}
\vertex (a1) at (0,0.05);
\vertex (a2) at (0,0) [label=right:$\scriptstyle\ Q_1$];
\vertex (v1) at (-0.5,-0.5);
\vertex (v4) at (0.5, -0.5);
\vertex (i1) at (-0.7,0.75);
\vertex (f1) at (0.7,0.75);
\vertex (i2) at (-0.7,-0.7);
\vertex (f2) at (0.7,-0.7);
\diagram{
(i1)--(a1)--(f1),
(i2)--(a2)--(f2),
(v1)--[photon](v4)
};
\end{feynman}
\draw (1,-1) -- (1,1);
\end{tikzpicture}_{\ Q_k,1/\epsilon} 
\hspace{-3mm}
\times
\begin{tikzpicture}[baseline={(a2.base)}]
\begin{feynman}
\vertex (a1) at (0,0.05);
\vertex (a2) at (0,0) [label=right:$\scriptstyle\ Q_k$];
\vertex (v1) at (-0.5,-0.5);
\vertex (v4) at (0.5, -0.5);
\vertex (i1) at (-0.7,0.75);
\vertex (f1) at (0.7,0.75);
\vertex (i2) at (-0.7,-0.7);
\vertex (f2) at (0.7,-0.7);
\diagram{
(i1)--(a1)--(f1),
(i2)--(a2)--(f2),
(v1)--[photon](v4)
};
\end{feynman}
\draw (1,-1) -- (1,1);
\end{tikzpicture}_{\ Q_1,\text{finite}}
\hspace{-3mm} 
= -4
\ ,
\nonumber\\[3mm]
&a^{(1,0;1)}_{Q_1E_k} a^{(1,0;0)}_{E_kQ_1} \bigg|_{D=4} =
\\
&\quad=
\begin{tikzpicture}[baseline={(a2.base)}]
\begin{feynman}
\vertex (a1) at (0,0.05);
\vertex (a2) at (0,0) [label=right:$\scriptstyle\ Q_1$];
\vertex (v1) at (-0.5,0.55);
\vertex (v4) at (0.5, 0.55);
\vertex (i1) at (-0.7,0.75);
\vertex (f1) at (0.7,0.75);
\vertex (i2) at (-0.7,-0.7);
\vertex (f2) at (0.7,-0.7);
\diagram{
(i1)--(a1)--(f1),
(i2)--(a2)--(f2),
(v1)--[photon](v4)
};
\end{feynman}
\draw (1,-1) -- (1,1);
\end{tikzpicture}_{\ E_k,1/\epsilon} 
\hspace{-3mm}
\times
\begin{tikzpicture}[baseline={(a2.base)}]
\begin{feynman}
\vertex (a1) at (0,0.05);
\vertex (a2) at (0,0) [label=right:$\scriptstyle\ E_k$];
\vertex (v1) at (-0.5,0.55);
\vertex (v4) at (0.5, 0.55);
\vertex (i1) at (-0.7,0.75);
\vertex (f1) at (0.7,0.75);
\vertex (i2) at (-0.7,-0.7);
\vertex (f2) at (0.7,-0.7);
\diagram{
(i1)--(a1)--(f1),
(i2)--(a2)--(f2),
(v1)--[photon](v4)
};
\end{feynman}
\draw (1,-1) -- (1,1);
\end{tikzpicture}_{\ Q_1,\text{finite}}
\hspace{-3mm}
+
\hspace{5mm}
\begin{tikzpicture}[baseline={(a2.base)}]
\begin{feynman}
\vertex (a1) at (0,0.05);
\vertex (a2) at (0,0) [label=right:$\scriptstyle\ Q_1$];
\vertex (v1) at (-0.5,-0.5);
\vertex (v4) at (0.5, -0.5);
\vertex (i1) at (-0.7,0.75);
\vertex (f1) at (0.7,0.75);
\vertex (i2) at (-0.7,-0.7);
\vertex (f2) at (0.7,-0.7);
\diagram{
(i1)--(a1)--(f1),
(i2)--(a2)--(f2),
(v1)--[photon](v4)
};
\end{feynman}
\draw (1,-1) -- (1,1);
\end{tikzpicture}_{\ E_k,1/\epsilon} 
\hspace{-3mm}
\times
\begin{tikzpicture}[baseline={(a2.base)}]
\begin{feynman}
\vertex (a1) at (0,0.05);
\vertex (a2) at (0,0) [label=right:$\scriptstyle\ Q_k$];
\vertex (v1) at (-0.5,-0.5);
\vertex (v4) at (0.5, -0.5);
\vertex (i1) at (-0.7,0.75);
\vertex (f1) at (0.7,0.75);
\vertex (i2) at (-0.7,-0.7);
\vertex (f2) at (0.7,-0.7);
\diagram{
(i1)--(a1)--(f1),
(i2)--(a2)--(f2),
(v1)--[photon](v4)
};
\end{feynman}
\draw (1,-1) -- (1,1);
\end{tikzpicture}_{\ Q_1,\text{finite}}
\hspace{-3mm} 
= 0
\nonumber\ ,
}
This gives for the contribution from diagrams $D=4$ to the $1/\epsilon$ pole coefficient:
\eq{
T_{4,Q_iQ_j}^{(2,0;1)} = -\frac32 - (-4) -\frac12 \cdot 0 = \frac52\ ,
}
which is consistent with the NDR scheme result in Table~3 of~\Reff{Buras:1989xd}
(see also~\Tab{tab:Tables CC} in~\App{app:Pole Tables}).
\\
\hdashrule{\textwidth}{1pt}{1pt}
\newline

\noindent
Several tables of pole coefficients for various sets of operators can be found in~\Reff{Buras:1989xd}~(BW), \Reff{Buras:1992tc}~(BJLW) and \Reff{Buras:2000if}~(BMU); all three references consider $O(\alpha_s^2)$ corrections. 
Following~\Eq{eq:TableRules}, it is thus straightforward to build the corresponding ADMs from the $1/\epsilon$ pole coefficients in the tables. For convenience, we have collected the various tables in~\App{app:Pole Tables}. 
In this section we show how one can use these pole coefficients to obtain the full set of two-loop ADMs to  $O(\alpha_s^2)$, $O(\alpha_s \alpha)$ and $O(\alpha^2)$ for all four-fermion operators in the LEFT.

We will distinguish between the contributions to the two-loop ADMs coming from current-current diagrams only, i.e. including only current-current diagrams in all of the matrices $\hat a$ in~\Eqs{eq:gamma20a}{eq:gamma11a}, and the rest of the contributions. We refer to them as {\it current-current} and {\it penguin} ADMs, respectively:
\eq{
\label{eq:ADM CC+P}
\hat \gamma^{(m,n)} = \hat \gamma^{(m,n)}_{cc} + \hat \gamma^{(m,n)}_p \;. 
}
By definition we include the term $Z_{Q_i}\delta_{Q_iQ_j}$ in $\hat\gamma_{cc}$.
While the case of current-current contributions is relatively straightforward, the case of penguin contributions is more involved due to the difficulties related to Dirac traces containing $\gamma_5$ in the NDR scheme.
When performing basis changes at NLO we will also split the matrix $\Delta \hat r ^{(m,n)}$ into current-current and penguin contributions, and define $\Delta\hat{r}_{cc}$ and $\Delta\hat{r}_{p}$ by:
\eqa{
\label{eq:NLO ADM Transformation CC}
\hat{\gamma}'^{(2,0)}_{cc} =& \hat{R} \hat{\gamma}^{(2,0)}_{cc} \hat{R}^{-1} + 4 Z_g^{(1,0;1)} \Delta\hat{r}^{(1,0)}_{cc} - \left[\Delta\hat{r}^{(1,0)}_{cc} , \hat{\gamma}'^{(1,0)}_{cc}\right] \;, \\
\label{eq:NLO ADM Transformation P}
\hat{\gamma}'^{(2,0)}_p =& \hat{R} \hat{\gamma}^{(2,0)}_p \hat{R}^{-1} + 4 Z_g^{(1,0;1)} \Delta\hat{r}^{(1,0)}_p - \left[\Delta\hat{r}^{(1,0)}_p , \hat{\gamma}'^{(1,0)}\right] - \left[\Delta\hat{r}^{(1,0)}_{cc} , \hat{\gamma}'^{(1,0)}_{p}\right] \;, 
}
with analogous expressions for $\hat \gamma^{(0,2)}$ and $\hat \gamma^{(1,1)}$. By definition, current-current ADMs transform under the change of basis independently of any penguin contribution at both one and two loops. Penguin ADMs at two loops transform with the full one-loop ADM in the commutator, and they also acquire an additional term (the last term in ~\Eq{eq:NLO ADM Transformation P}). 
This additional term however vanishes in all considered basis changes. This is due to the fact that the VLL, VLR and SLR subsectors in $\Delta\hat{r}^{(1,0)}_{cc}$ are zero in all our cases, while the other subsectors SRR and TRR have vainishing $\hat{\gamma}^{(1,0)}_{p}$. 
The same applies to $\hat{\gamma}^{(0,2)}_{p}$ and $\hat{\gamma}^{(1,1)}_{p}$.

\subsubsection{Current-Current Contributions}

Direct application of~\Eq{eq:TableRules}, with the 
pole coefficients given in~Tables~\ref{tab:Tables CC} and~\ref{tab:Tables CC2} and the appropriate color/charge factors, gives the ADM for a generic sector in the following basis
\eq{
\label{eq:Table Basis}
\begin{aligned}
& Q^{}_{1} = (\bar \psi_1 \gamma^\mu P_L \psi_2)(\bar \psi_3 \gamma_\mu P_L \psi_4) \;, & \quad 
& Q_{2} = (\bar \psi_1 \gamma^\mu P_L T^A \psi_2)(\bar \psi_3 \gamma_\mu P_L T^A\psi_4) \;, \\
& Q_{3} = (\bar \psi_1 \gamma^\mu P_L \psi_2)(\bar \psi_3 \gamma_\mu P_R \psi_4) \;, & \quad 
& Q_{4} = (\bar \psi_1 \gamma^\mu P_L T^A \psi_2)(\bar \psi_3 \gamma_\mu P_R T^A\psi_4) \;, \\
& Q_{5} = (\bar \psi_1 P_L \psi_2)(\bar \psi_3 P_R \psi_4) \;, & \quad 
& Q_{6} = (\bar \psi_1 P_L T^A \psi_2)(\bar \psi_3 P_R T^A\psi_4) \;, \\
& Q_{7} = (\bar \psi_1 P_R \psi_2)(\bar \psi_3 P_R \psi_4) \;, & \quad 
& Q_{8} = (\bar \psi_1 P_R T^A \psi_2)(\bar \psi_3 P_R T^A\psi_4) \;, \\
& Q_{9} = (\bar \psi_1 \sigma^{\mu\nu} P_R \psi_2)(\bar \psi_3 \sigma_{\mu\nu} P_R \psi_4) \;, & \quad 
& Q_{10} = (\bar \psi_1 \sigma^{\mu\nu} P_R T^A \psi_2)(\bar \psi_3 \sigma_{\mu\nu} P_R T^A\psi_4) \ ,
\end{aligned}
}
with $\psi_1,\psi_2,\psi_3,\psi_4$ being any four different fermion fields. Operators $Q_i$ in~\Eq{eq:Table Basis} for even $i$ are only present when all four fermion fields are quarks. 
From the generic ADM in this basis, one obtains the ADMs in the JMS basis for the different RG-invariant sectors as follows: 


\bigskip
\noindent
{\bf 1.} 
For the $\Delta F = 1$ sectors of type $ed$, $eu$, $\nu d$, $\nu u$, $\nu e$ and $\nu e d u$, as well as for the $\Delta L = 2$ sector, the generic ADM coincides with the ADM for each of these sectors in the JMS basis.


\bigskip
\noindent
{\bf 2.} 
The JMS basis for the $ud$-type $\Delta F = 1$ sector
is different from the one in~\Eq{eq:Table Basis}.
In this case, after obtaining the ADM from the pole coefficients in the basis $\{Q_i\}$, we perform a change of basis into the new basis 
$\{Q'_i\}$, given by $Q'_{1-6}=Q_{1-6}$ and 
\eq{
\label{eq:Table Basis Prime I}
\begin{aligned}
& Q'_{7} = (\bar \psi_1 P_R \psi_2)(\bar \psi_3 P_R \psi_4) \;, & \quad 
& Q'_{8} = (\bar \psi_1 P_R T^A \psi_2)(\bar \psi_3 P_R T^A\psi_4) \;, \\
& Q'_{9} = (\bar \psi_1 P_R \psi_4)(\bar \psi_3 P_R \psi_2) \;, & \quad 
& Q'_{10} = (\bar \psi_1 P_R T^A \psi_4)(\bar \psi_3 P_R T^A\psi_2) \ .
\end{aligned}
}
This change of basis corresponds essentially to the one worked out in~\Eqs{eq:R DF=1}{eq:Dr DF=1} for the $O(\alpha_s^2)$ contributions.


\bigskip
\noindent
{\bf 3.} 
For $\Delta F=2$  and $\Delta F=1.5$ sectors, in which $\psi_1 = \psi_3$ and/or $\psi_2 = \psi_4$, the basis in~\Eq{eq:Table Basis} is redundant.
We then follow the same symmetrization procedure as in Section~\ref{sec:Els Truquets} (as introduced in~\Reff{Buras:2000if}): we build the ADM in the redundant basis, and then perform a change of basis into a ``$\pm$'' basis. For example, in the case $\psi_2 = \psi_4$, we take
\begin{align}
& Q^\pm_{1} = \frac{1}{2} \Big[(\bar \psi_1 \gamma^\mu P_L \psi_2)(\bar \psi_3 \gamma_\mu P_L \psi_4) \pm (\bar \psi_1 \gamma^\mu P_L \psi_4)(\bar \psi_3 \gamma_\mu P_L \psi_2) \Big] \;, \nonumber \\ 
& Q^\pm_{2} = \frac{1}{2} \Big[(\bar \psi_1 \gamma^\mu P_L \psi_2)(\bar \psi_3 \gamma_\mu P_R \psi_4) \pm (\bar \psi_1 \gamma^\mu P_L \psi_4)(\bar \psi_3 \gamma_\mu P_R \psi_2) \Big] \;, \nonumber \\ 
& Q^\pm_{3} = \frac{1}{2} \Big[(\bar \psi_1 P_L \psi_2)(\bar \psi_3 P_R \psi_4) \pm (\bar \psi_1 P_L \psi_4)(\bar \psi_3 P_R \psi_2) \Big] \;, \label{eq:PM Basis}\\  
& Q^\pm_{4} = \frac{1}{2} \Big[(\bar \psi_1 P_R \psi_2)(\bar \psi_3 P_R \psi_4) \pm (\bar \psi_1 P_R \psi_4)(\bar \psi_3 P_R \psi_2) \Big] \;, \nonumber \\ 
& Q^\pm_{5} = \frac{1}{2} \Big[(\bar \psi_1 P_R T^A\psi_2)(\bar \psi_3 P_R T^A\psi_4) \pm (\bar \psi_1 P_R T^A\psi_4)(\bar \psi_3 P_R T^A\psi_2) \Big] \;. \nonumber 
\end{align}
The ADM for the operators $Q_i^+$ coincides then with the one in the JMS basis for the corresponding $\Delta F=2$  and $\Delta F=1.5$ sectors, as explained in more detail in Section~\ref{sec:Els Truquets}.


\bigskip
\noindent
{\bf 4.} 
The ADMs for the sectors $\Delta F = 1^{\bar f f}$ and $\Delta F=0$ have to be computed for $\psi_1=\psi_2$, $\psi_1=\psi_4$, $\psi_2=\psi_3$ and/or $\psi_2=\psi_4$. They receive contributions from penguin diagrams, which mix operators with different fermion content.
The splitting into current-current and penguin contributions mentioned above is now relevant, with the property that $\gamma_{cc}$ transforms independently from~$\gamma_{p}$, and can therefore be constructed separately. We thus split the sectors into subsectors of operators containing the same fermion fields. These subsectors can then be treated separately as in the previous cases. There is one exception, however, namely the subsectors $(\bar \psi_i\Gamma \psi_j)(\bar \psi_k\Gamma \psi_k)$ with $\psi=u,d,e,\nu$ and $k\ne \{i,j\}$.
In these exceptional cases the JMS basis has yet a different structure, and thus we perform a change of basis into a new basis 
$\{Q''_i\}$, given by $Q''_{3-10}=Q'_{3-10}$ and
\eq{
\label{eq:Table Basis Prime II}
\begin{aligned}
& Q''_{1} = (\bar \psi_1 \gamma^\mu P_L \psi_2)(\bar \psi_3 \gamma_\mu P_L \psi_4) \;, & \quad 
& Q''_{2} = (\bar \psi_1 \gamma^\mu P_L \psi_4)(\bar \psi_3 \gamma_\mu P_L \psi_2) \ ,
\end{aligned}
}
for quarks ($\psi = u,d$), and
\eq{
\label{eq:Table Basis Prime III}
\begin{aligned}
& Q''_{1} = \frac{1}{2}(\bar \psi_1 \gamma^\mu P_L \psi_2)(\bar \psi_3 \gamma_\mu P_L \psi_4) + \frac{1}{2}(\bar \psi_1 \gamma^\mu P_L \psi_4)(\bar \psi_3 \gamma_\mu P_L \psi_2)\;, 
\end{aligned}
}
for leptons  ($\psi = e,\nu$), as detailed at the end of~\Sec{sec:defLEFT}.
Note that the change of basis $\{Q'_i\}\to \{Q''_i\}$ is trivial, given that $\Delta \hat r^{(1,0)} = \Delta \hat r^{(0,1)} = 0$. This is the reason for the vanishing of the last term in~\Eq{eq:NLO ADM Transformation P}.


\bigskip
\noindent
{\bf 5.} For the baryon-number-violating (BNV) sectors, the tables of pole coefficients provide the ADMs in the following two bases, 
\eq{
\label{eq:BNV Basis}
\begin{aligned}
& B_{1} = \epsilon_{\alpha\beta\gamma}(q_1^{\alpha T}C P_L q^\beta_2)(q_3^{\gamma T}C P_R \ell) \;, & \; 
& 
\widetilde B_{1} = \epsilon_{\alpha\beta\gamma}(q_1^{\alpha T}C P_L q^\beta_2)(\bar \ell P_R q^\gamma_3) \;, 
\\
& B_{2} = \epsilon_{\alpha\beta\gamma}(q_1^{\alpha T}C P_R q^\beta_2)(\bar q^\gamma_3 P_R \ell) \;, & \; 
& 
\widetilde B_{2} = \epsilon_{\alpha\beta\gamma}(q_1^{\alpha T}C P_R q^\beta_2)(\bar \ell P_R q^\gamma_3) \;, 
\\
& B_{3} = \epsilon_{\alpha\beta\gamma}(q_1^{\alpha T}C \sigma^{\mu\nu}P_R q^\beta_2)(q_3^{\gamma T}C \sigma_{\mu\nu}P_R \ell) \;, & \; 
& 
\widetilde B_{3} = \epsilon_{\alpha\beta\gamma}(q_1^{\alpha T}C \sigma^{\mu\nu}P_R q^\beta_2)(\bar \ell \sigma_{\mu\nu}P_R q^\gamma_3) \;.
\end{aligned}
}
The operators $B_i$ correspond to $\Delta L = +1$ transitions, while $\widetilde B_i$ correspond to $\Delta L = -1$. 
Once again, these bases do not correspond to JMS operators and one needs to change them to the bases
$B'_{1,2}=B_{1,2}$ and $\widetilde B'_{1,2}=\widetilde B_{1,2}$, together with
\eq{
\label{eq:BNV LEFT Basis}
\begin{aligned}
& B'_{3} = \epsilon_{\alpha\beta\gamma}(q_1^{\alpha T}C P_R q_3^\beta)(q_2^{\gamma T}C P_R \ell) \;, & \; 
& 
\widetilde B'_{3} = \epsilon_{\alpha\beta\gamma}(q_1^{\alpha T}C P_R q_3^\beta)(\bar \ell P_R q^\gamma_2) \ . 
\end{aligned}
}
Note that, for $q_2 = q_3$, one has $B'_3 = B'_2$ and $\widetilde B'_3 = \widetilde B'_2$, and thus this basis becomes redundant. In such cases, one can symmetrize the two operators by changing them into a~``$\pm$'' basis (while leaving $B_1^{\scriptscriptstyle(\scriptstyle\prime\scriptscriptstyle)}$ and $\widetilde B_1^{\scriptscriptstyle(\scriptstyle\prime\scriptscriptstyle)}$ unchanged), in the style of~\Eq{eq:PM Basis} for the $\Delta F = 1.5$ and $\Delta F = 2$ sectors,
\eq{
\label{eq:BNV Sym Basis}
\begin{aligned}
& B_2^{\pm} = \frac12 \big(B_{2} \pm B'_{3} \big) \;, & \; 
& \widetilde B_2^{\pm} = \frac12 \big(\widetilde B_{2} \pm \widetilde B'_{3} \big) \;,
\end{aligned}
}
and then focus on the ``$+$'' operators to obtain the proper ADM.

\bigskip
\noindent
{\bf 6.} All current-current ADMs discussed here can be trivially extended to the sectors with chirally-flipped operators (related by $P_L \leftrightarrow P_R$). 

\bigskip
\noindent
The above mentioned manipulations provide the full current-current ADM in the LEFT.

\subsubsection{Penguin Contributions}
\label{sec:Penguin Reconstruction}

Penguin diagrams contribute to the ADMs of 
$\Delta F = 1^{\bar f f}$ and $\Delta F = 0$ sectors. These contributions are only non-zero for insertions of VLL, VLR and SRL operators (and their chirality-flipped counterparts). 

Although penguin contributions will introduce mixing among many sets of operators with different fermion content (i.e. different $\bar f f$ currents), we show in this section that the penguin ADMs can be reconstructed from a reduced set of small matrices, that we shall call ``penguin ADM seeds'', and which can be read off from a single table of pole coefficients (from \Tab{tab:Tables P} in this case, extracted from~\Reff{Buras:1992tc}).
A crucial ingredient in this operation is the property of flavor universality of anomalous dimensions~\cite{Morell:2024aml}.
For this purpose we will introduce the concepts of ``flavor-symmetric'' and ``flavor-decoupled'' operator bases. 

We start by defining some nomenclature: for a $\Delta F = 1^{\bar f f}$ operator, we call ``outer flavors'' those characterizing the net $\Delta F = 1$ transition, while the remaining ones are referred to as ``inner flavors''. Then:
\begin{itemize}
\item A basis $\{Q_i,E_i\}$ is said to be \textit{flavor-symmetric in flavors $f,f'$} if
\eq{
\{Q_i,E_i\}\big|_{f\leftrightarrow f'} = \{Q_i,E_i\}
\quad \text{(up\ to\ reordering).}
}
Note that this can only be true if $f,f'$ are inner flavors.

\item A basis is said to be \textit{flavor-decoupled} if, when decoupling any inner flavor $f$ from the EFT, the set of operators obtained by removing from this basis any terms containing the flavor $f$ defines a basis for the sector of the new EFT, whose penguin ADM is given by the old one with the appropriate rows and columns removed.

For an operator basis to be flavor decoupled it is sufficient that two conditions are simultaneously satisfied: 

\begin{enumerate}

\item Each operator contains no more than one inner flavor. This guarantees that all operators that are not removed completely in the procedure are left unaltered. 

\item When decomposing the penguin ADM into $\hat a$ matrices as in~\Eqs{eq:gamma10a}{eq:gamma11a}, each entry $(\hat\gamma_p)_{ij}$ depends only on flavors common to $Q_i$ and $Q_j$, whereas any contribution involving other flavors vanishes. This ensures that the ADM entries for all operators not eliminated by the decoupling (i.e. not featuring the decoupled field) retain exactly the same expression. At NLO this second condition reads:
\eq{
\label{eq:Flavor Decoupling NLO}
\left[ -4 \,p^{(2,0;1)}_{Q_iQ_j} + 4 \,p^{(1,0;1)}_{Q_iQ_k} p^{(1,0;0)}_{Q_kQ_j} + 2 \,p^{(1,0;1)}_{Q_iE_k} p^{(1,0;0)}_{E_kQ_j} \right]_{\text{featuring fields } \notin \, Q_i,Q_j} = 0 \; ,
}
where we used $\hat p \equiv \hat a_p$ for notational simplicity. Given that the matrices $\hat a_{cc}$ conserve flavor, the only products of one-loop $\hat{a}$ matrices involving fields that are not present in $Q_i$ or $Q_j$ are products of two $\hat a_p$ matrices.

\end{enumerate}

\end{itemize}

\begin{figure}
\vspace{-1.7cm}
\includegraphics[scale=0.75]{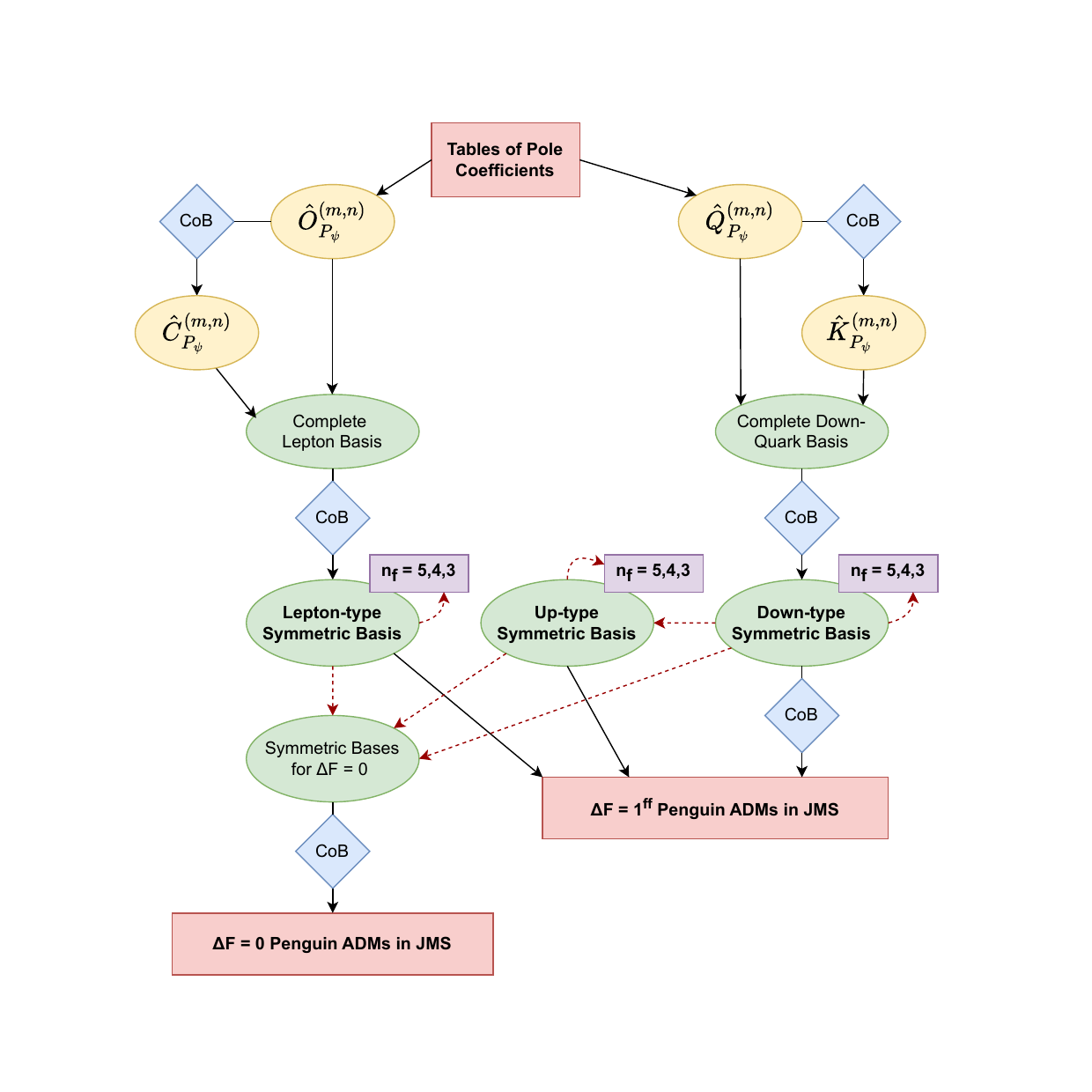}\vspace{-1.2cm}
\caption{Flowchart outlining the full procedure to obtain the two-loop ADMs starting from the tables of pole coefficients of Appendix \ref{app:Pole Tables}. In the figure, ``CoB'' is short for ``Change of Basis'', and the dashed arrows represent the steps which use surface-level flavor universality for the manipulation of the matrices.}
\label{fig:Flowchart}
\end{figure}

A flavor-decoupled basis that is also flavor-symmetric in any two lepton flavors $\ell,\ell'$ or quark flavors $q,q'$ implies a set of special properties for its ADM, which we will refer to as ``surface-level flavor universality''. Evidently, decoupling any flavor from the EFT will have a trivial effect on the penguin ADM. In addition, as dictated by flavor symmetry, for two operators $Q_i$ and $Q_j$ differing only in symmetric inner flavors one finds 
\begin{equation}
    \gamma_{ik} = \gamma_{jk}\,,\quad \gamma_{ki} = \gamma_{kj}\,,\quad \forall k\,,
\end{equation}
up to QED charges. Finally, extending the basis with an extra inner flavor will also have a trivial effect on the penguin ADM, mainly duplicating a set of rows and columns.\footnote{
Technically speaking, the process of extending the basis with extra flavors (e.g. $u_3$ in the LEFT) is more complicated. It amounts to extending the EFT with the extra flavor, thus defining a new EFT, and then building a $\Delta F = 1^{\bar f f}$ basis being flavor-decoupled and flavor-symmetric in all lepton/quark flavors, such that decoupling the extra flavor leads to the original basis. Then, $\hat \gamma'_p$ in the basis of the extended EFT will have the same entries as the original $\hat \gamma_p$ for all flavors common in both EFTs, due to flavor-decoupling. Moreover, the new rows and columns will simply be copies of some of the others, which follows from flavor symmetry.
}

\bigskip
\noindent
The complete procedure to obtain the full penguin ADM from the corresponding tables of pole coefficients is summarized in Figure~\ref{fig:Flowchart}, and consists of the following four steps:

\subsubsection*{1. Penguin ADM Seeds}

We start by generalizing the approach followed originally in~\Reff{Buras:1992tc} for the $\Delta F = 1^{\bar f f}$ SM operators in the BMU basis. 
We define two operator sets, closed under renormalization: 
the \textit{lepton penguin basis},
\begin{align}
\label{eq:Lepton Basis}
     & \op_1^{(\ell)} = (\bar\ell_i \gamma^\mu P_L \ell'_k)(\bar{\ell}'_k \gamma_\mu P_L \ell_j)\;, & & \op_2^{(\ell)} = (\bar\ell_i P_R \ell'_k)(\bar{\ell}'_k P_R \ell_j)\;, \nonumber \\[2mm]
     & \op_3^{(\ell)} = (\bar\ell_i \gamma^\mu P_L q_k)(\bar q_k \gamma_\mu P_L \ell_j)\;, & 
     & \op_4^{(\ell)} = (\bar\ell_i P_L q_k)(\bar q_k P_R \ell_j)\;, \nonumber \\[2mm]
     & \op_5^{(\ell)} = {\sum}_w (\bar\ell_i \gamma^\mu P_L \ell_j)(\bar e_w \gamma_\mu P_L e_w)\;, & & \op_6^{(\ell)} = {\sum}_w (\bar\ell_i \gamma^\mu P_L \ell_j)(\bar e_w \gamma_\mu P_R e_w)\;, \\[2mm]
     & \op_7^{(\ell)} = {\sum}_w (\bar\ell_i \gamma^\mu P_L \ell_j)(\bar u_w \gamma_\mu P_L u_w)\;, & 
     & \op_8^{(\ell)} = {\sum}_w (\bar\ell_i \gamma^\mu P_L \ell_j)(\bar u_w \gamma_\mu P_R u_w)\;, \nonumber \\[2mm]
     & \op_9^{(\ell)} = {\sum}_w (\bar\ell_i \gamma^\mu P_L \ell_j)(\bar d_w \gamma_\mu P_L d_w)\;, & 
     & \op_{10}^{(\ell)} = {\sum}_w (\bar\ell_i \gamma^\mu P_L \ell_j)(\bar d_w \gamma_\mu P_R d_w)\;, \nonumber 
\end{align}
and the \textit{down-quark penguin basis},
\begin{align}
\label{eq:Quark Basis}
     & \op_1^{(d)} = (\bar d^{\alpha}_i \gamma^\mu P_L \ell_k)(\bar\ell_k \gamma_\mu P_L d^{\alpha}_j)\;, & & 
     \op_2^{(d)} = (\bar d^{\alpha}_i P_R \ell_k)(\bar \ell_k P_L d^{\alpha}_j)\;, \nonumber \\[2mm]
     & \op_3^{(d)} = (\bar d^{\alpha}_i \gamma^\mu P_L q^{\alpha}_k)(\bar q^{\beta}_k \gamma_\mu P_L d^{\beta}_j)\;, & & 
     \op_4^{(d)} = (\bar d^{\alpha}_i \gamma^\mu P_L q^{\beta}_k)(\bar q^{\beta}_k \gamma_\mu P_L d^{\alpha}_j)\;, \nonumber\\[2mm]
     & \op_5^{(d)} = (\bar d^{\alpha}_i P_R q^{\alpha}_k)(\bar q^{\beta}_k P_L d^{\beta}_j)\;, & & 
     \op_6^{(d)} = (\bar d^{\alpha}_i P_R q^{\beta}_k)(\bar q^{\beta}_k P_L d^{\alpha}_j)\;, \nonumber \\[2mm]
     & \op_7^{(d)} = {\sum}_w (\bar d^{\alpha}_i \gamma^\mu P_L d^{\alpha}_j)(\bar e_w \gamma_\mu P_L e_w)\;, & & 
     \op_8^{(d)} = {\sum}_w (\bar d^{\alpha}_i \gamma^\mu P_L d^{\alpha}_j)(\bar e_w \gamma_\mu P_R e_w)\;,  \\[2mm]
     & \op_9^{(d)} = {\sum}_w (\bar d^{\alpha}_i \gamma^\mu P_L d^{\alpha}_j)(\bar u^{\beta}_w \gamma_\mu P_L u^{\beta}_w)\;, & & 
     \op_{10}^{(d)} = {\sum}_w (\bar d^{\alpha}_i \gamma^\mu P_L d^{\beta}_j)(\bar u^{\beta}_w \gamma_\mu P_L u^{\alpha}_w)\;, \nonumber \\[2mm]
     & \op_{11}^{(d)} = {\sum}_w (\bar d^{\alpha}_i \gamma^\mu P_L d^{\alpha}_j)(\bar u^{\beta}_w \gamma_\mu P_R u^{\beta}_w)\;, & & 
     \op_{12}^{(d)} = {\sum}_w (\bar d^{\alpha}_i \gamma^\mu P_L d^{\beta}_j)(\bar u^{\beta}_w \gamma_\mu P_R u^{\alpha}_w)\;, \nonumber \\[2mm]
     & \op_{13}^{(d)} = {\sum}_w (\bar d^{\alpha}_i \gamma^\mu P_L d^{\alpha}_j)(\bar d^{\beta}_w \gamma_\mu P_L d^{\beta}_w)\;, & & 
     \op_{14}^{(d)} = {\sum}_w (\bar d^{\alpha}_i \gamma^\mu P_L d^{\beta}_j)(\bar d^{\beta}_w \gamma_\mu P_L d^{\alpha}_w)\;, \nonumber \\[2mm]
     & \op_{15}^{(d)} = {\sum}_w (\bar d^{\alpha}_i \gamma^\mu P_L d^{\alpha}_j)(\bar d^{\beta}_w \gamma_\mu P_R d^{\beta}_w)\;, & & 
     \op_{16}^{(d)} = {\sum}_w (\bar d^{\alpha}_i \gamma^\mu P_L d^{\beta}_j)(\bar d^{\beta}_w \gamma_\mu P_R d^{\alpha}_w)\;,\nonumber
\end{align}
where $\ell(\ell')$ and $q$ are generic lepton and quark fields, respectively, and $\alpha,\beta$ are $SU(3)_c$ indices. The index $w$ runs over the corresponding flavors, and $k$ is a flavor index different from~$i,j$. 
Two comments are in order. First, the lepton and down-quark penguin bases are not complete bases, but specific bases for certain sub-blocks closed under renormalization. Second, there is one particular lepton and one down-quark penguin basis for each choice of $\ell_k= e_k,\nu_k$ and  $q_k= d_k,u_k$.

The advantage of these bases is that the tables of pole coefficients directly provide the contributions to the penguin ADMs from each of the $\op_{1-4}^{(\ell)}$ and $\op_{1-6}^{(d)}$ operators via~\Eq{eq:TableRules}.
Let us focus, for simplicity, on the simpler case of $O(\alpha_s^2)$~contributions to the $10 \times 10$ penguin~ADM in the lepton penguin basis (for arbitrary flavor indices), which has the following structure:
\eq{
\label{eq:Open Penguin ADM Seeds}
\hat\gamma_p^{(2,0)}\Big|_{\text{Lepton Basis}} = 
\arraycolsep=5pt
\def\arraystretch{1.25}
\begin{pmatrix}
    0_{4\times6} & \hat O_P & \hat O_P \\
    0_{4\times6} &\hat M_C \, \hat C_P & \hat M_C \, \hat C_P \\
    0_{2\times6} &\hat N_C \, \hat C_P & \hat N_C \, \hat C_P \\
\end{pmatrix} \;, \quad 
\begin{array}{l}
    \hat M_C = \text{diag}(n_e,n_e,n_u,n_u) \;, \\[2mm]
    \hat N_C = n_d
    \begin{pmatrix}
        0_{2\times2} & \hat{\mathbb{1}}_{2\times2}
    \end{pmatrix} \;, \\
\end{array}
}
where $\hat{\mathbb{1}}_{2\times2}$ is the two-by-two identity matrix. We call the $4\times2$ matrices $\hat O_P$ and $\hat C_P$ the {\it ``penguin ADM seeds''}. The matrix $\hat O_P$ is obtained directly from the tables. However, $\hat C_P$ cannot be directly computed in the NDR scheme because it requires the evaluation of ``closed-penguin'' two-loop diagrams (i.e. penguins with insertions of operators with flavor indices $ijkk$) which contain traces that involve $\gamma_5$. This is not the case for $\hat O_P$, which arises from ``open-penguin'' diagrams (with insertions of operators with flavor indices $ikkj$) alone.

The closed-penguin seed $\hat C_P$ can be obtained through a change of basis at NLO into an alternative basis $\widetilde\op_i$ where all current-current operators $\op_{1-4}^{(\ell)}$, which give rise to open-penguin diagrams, are replaced by their closed-penguin counterparts: 
\eq{
\label{eq:Alternate Lepton Basis}
\begin{aligned}
    &\widetilde\op_1^{\,(\ell)} = (\bar\ell_i \gamma^\mu P_L \ell_j)(\bar\ell_k \gamma_\mu P_L \ell_k)\;, & & \widetilde\op_2^{\,(\ell)} = (\bar\ell_i \gamma^\mu P_L \ell_j)(\bar\ell_k \gamma_\mu P_R \ell_k)\;, \\[2mm] 
    &\widetilde\op_3^{\,(\ell)} = (\bar\ell_i \gamma^\mu P_L \ell_j)(\bar q^{\alpha}_k \gamma_\mu P_L q^{\alpha}_k)\;, & & \widetilde\op_4^{\,(\ell)} = (\bar\ell_i \gamma^\mu P_L \ell_j)(\bar q^{\alpha}_k \gamma_\mu P_R q^{\alpha}_k) \;,
\end{aligned} 
}
and $\widetilde\op_{5-10}^{\,(\ell)} = \op_{5-10}^{\,(\ell)}$. The penguin ADM in this alternative basis has the form
\eq{
\label{eq:Closed Penguin ADM Seeds}
\hat\gamma_p^{(2,0)}\Big|_{\text{Alt. Lepton Basis}} = 
\arraycolsep=5pt
\def\arraystretch{1.25}
\begin{pmatrix}
    0_{4\times6} & \hat C_P & \hat C_P \\
    0_{4\times6} &\hat M_C \, \hat C_P & \hat M_C \, \hat C_P \\
    0_{2\times6} &\hat N_C \, \hat C_P & \hat N_C \, \hat C_P \\
\end{pmatrix} \;.
}
Comparing this structure with the result of the change of basis completely determines $\hat C_P$ in terms of $\hat O_P$. 
Once both penguin seeds $\hat O_P, \hat C_P$ are known, the rest of the ADM can be trivially assembled.

This procedure can be analogously followed in the case of the $O(\alpha_s\alpha)$ and $O(\alpha^2)$ ADMs to obtain the corresponding penguin ADM seeds. The main difference in these cases is that the ADMs in~Eqs.~(\ref{eq:Open Penguin ADM Seeds}) and~(\ref{eq:Closed Penguin ADM Seeds}) will be less sparse. Their form resembles more the general case displayed in~\App{app:ADM Seeds}.

The seeds for the down-quark penguin basis are obtained completely analogously, albeit in a slightly larger basis. The penguin seeds will be contained in $6\times2$ and $6\times4$ matrices in this case, and the alternative penguin basis will change the first 6 operators in~\Eq{eq:Quark Basis} into
\eq{
\label{eq:Alternate Quark Basis}
\begin{aligned}
    &\widetilde\op_1^{\,\text{Quark}} = (\bar d^{\alpha}_i \gamma^\mu P_L d^{\alpha}_j)(\bar\ell_k \gamma_\mu P_L \ell_k)\;, & & \widetilde\op_2^{\,\text{Quark}} = (\bar d^{\alpha}_i \gamma^\mu P_L d^{\alpha}_j)(\bar\ell_k \gamma_\mu P_R \ell_k)\;, \\[2mm] 
    &\widetilde\op_3^{\,\text{Quark}} = (\bar d^{\alpha}_i \gamma^\mu P_L d^{\alpha}_j)(\bar q^{\beta}_k \gamma_\mu P_L q^{\beta}_k)\;, & & \widetilde\op_4^{\,\text{Quark}} = (\bar d^{\alpha}_i \gamma^\mu P_L d^{\beta}_j)(\bar q^{\beta}_k \gamma_\mu P_L q^{\alpha}_k) \;,\\[2mm]
    &\widetilde\op_5^{\,\text{Quark}} = (\bar d^{\alpha}_i \gamma^\mu P_L d^{\alpha}_j)(\bar q^{\beta}_k \gamma_\mu P_R q^{\beta}_k)\;, & & \widetilde\op_6^{\,\text{Quark}} = (\bar d^{\alpha}_i \gamma^\mu P_L d^{\beta}_j)(\bar q^{\beta}_k \gamma_\mu P_R q^{\alpha}_k) \;.
\end{aligned} 
}
The ADMs for both the lepton and down-quark penguin bases, their alternative versions and the corresponding penguin seeds are discussed in~\App{app:ADM Seeds}.

\subsubsection*{2. Completion of the Penguin Bases}
\label{sec:Penguin Basis Completion}

The next step in the procedure consists of completing the lepton and down-quark penguin bases in order to cover all possible operators in the penguin sub-blocks in Sections~\ref{sec:DF=1qq d-type} and~\ref{sec:DF=1qq e-type}. The procedure follows~\Reff{Buras:2000if} for the $d$-type basis, where it is shown how all penguin contributions can be expressed trivially in terms of the penguin seeds obtained above, as long as the basis is built appropriately~\cite{Morell:2024aml}.
The basis is completed with all the different copies of the current-current operators, $\op_{1-4}^{(\ell)}$ and $\op_{1-6}^{(d)}$, one for each choice of $\ell_k$ and $q_k$, and then finally including the operators
\eq{
\label{eq:Lepton Basis Completion}
\begin{aligned}
    &\op_{11}^{\,(\ell)} = (\bar\ell_i \gamma^\mu P_L \ell_j)(\bar\ell_j \gamma_\mu P_L \ell_j) + (\bar\ell_i \gamma^\mu P_L \ell_j)(\bar \ell^{\alpha}_i \gamma_\mu P_L \ell^{\alpha}_i) \;, \\[2mm] 
    &\op_{12}^{\,(\ell)} = (\bar\ell_i \gamma^\mu P_L \ell_j)(\bar\ell_j \gamma_\mu P_R \ell_j) + (\bar\ell_i \gamma^\mu P_L \ell_j)(\bar\ell_i \gamma_\mu P_R \ell_i) \;,
\end{aligned} 
}
for the lepton penguin basis, and
\eq{
\label{eq:Quark Basis Completion}
\begin{aligned}
    &\op_{17}^{\,(d)} = \frac{1}{2} \left[ (\bar d^{\alpha}_i \gamma^\mu P_L d^{\alpha}_j)(\bar d^{\beta}_j \gamma_\mu P_L d^{\beta}_j) + (\bar d^{\alpha}_i \gamma^\mu P_L d^{\beta}_j)(\bar d^{\beta}_j \gamma_\mu P_L d^{\alpha}_j) \right. \\[2mm] 
    & \hspace{2cm} \left. + (\bar d^{\alpha}_i \gamma^\mu P_L d^{\alpha}_j)(\bar d^{\beta}_i \gamma_\mu P_L d^{\beta}_i) + (\bar d^{\alpha}_i \gamma^\mu P_L d^{\beta}_j)(\bar d^{\beta}_i \gamma_\mu P_L d^{\alpha}_i)\right] \;, \\[2mm] 
    &\op_{18}^{\,(d)} = (\bar d^{\alpha}_i \gamma^\mu P_L d^{\alpha}_j)(\bar d^{\beta}_j \gamma_\mu P_R d^{\beta}_j) + (\bar d^{\alpha}_i \gamma^\mu P_L d^{\alpha}_j)(\bar d^{\beta}_i \gamma_\mu P_R d^{\beta}_i)\;, \\[2mm]
    &\op_{19}^{\,(d)} = (\bar d^{\alpha}_i \gamma^\mu P_L d^{\beta}_j)(\bar d^{\beta}_j \gamma_\mu P_R d^{\alpha}_j) + (\bar d^{\alpha}_i \gamma^\mu P_L d^{\beta}_j)(\bar d^{\beta}_i \gamma_\mu P_R d^{\alpha}_i)\;,  
\end{aligned} 
}
for the down-quark penguin basis. These extra operators contribute to the penguin ADM through $\hat O_P$ and $\hat C_P$, since they generate both open and closed penguin diagrams. The contribution from $\op_{17}^{\,(d)}$ is slightly more involved, and can be obtained through a tree-level change of basis similar to the ``$\pm$'' symmetrization of~\Eq{eq:PM Basis}. Using the notation of~\App{app:ADM Seeds} the contributions of these additional operators are given by
\begin{align}
\arraycolsep=5pt
\def\arraystretch{1.25}
\left(
\begin{array}{c}
    {\gamma}_{\op_{11}^{\,(\ell)},\op_{5-6}^{\,(\ell)}} \\
    {\gamma}_{\op_{12}^{\,(\ell)},\op_{5-6}^{\,(\ell)}} 
\end{array} \right)_p &= \left(
\arraycolsep=5pt
\def\arraystretch{1.25}
\begin{array}{cccc}
    2 & 0 & 0 & 0 \\
    0 & 2 & 0 & 0
\end{array} \right) \times \left[\hat O_{P_e} + \hat C_{P_e}\right]_{Q_{\ell} \to Q_f} \;, \\
\arraycolsep=5pt
\def\arraystretch{1.25}
\left(
\begin{array}{c}
    {\gamma}_{\op_{17}^{\,(d)},\op_{7-8}^{\,(d)}} \\
    {\gamma}_{\op_{18}^{\,(d)},\op_{7-8}^{\,(d)}} \\
    {\gamma}_{\op_{19}^{\,(d)},\op_{7-8}^{\,(d)}} 
\end{array} \right)_p &= \left(
\arraycolsep=5pt
\def\arraystretch{1.25}
\begin{array}{cccccc}
    0 & 0 & 1 & 1 & 0 & 0 \\
    0 & 0 & 0 & 0 & 2 & 0 \\
    0 & 0 & 0 & 0 & 0 & 2 
\end{array} \right) \times \left[\hat Q_{P_e} + \hat K_{P_e}\right]_{Q_{q} \to Q_f} \;,
\end{align}
with identical expressions for their mixing into $\op_{7-8}^{\,(\ell)}$ and $\op_{9-12}^{\,(d)}$ in combination with the $P_u$ seeds, and into $\op_{9-10}^{\,(\ell)}$ and $\op_{13-16}^{\,(d)}$ combined with the $P_d$ seeds. Again, these are the only non-vanishing entries of the penguin ADM in the lepton and down-quark penguin bases.

\subsubsection*{3. Reconstruction of $\Delta F = 1^{\bar f f}$ in the JMS basis}
\label{sec:Penguin Reconstruction DF=1}

We now define a new set of four flavor-symmetric and flavor-decoupled bases:
\begin{align}
    & \begin{aligned}
    \label{eq:e-Sym Basis}
    \left\{ \mathcal{S}_i^{(e)} \right\} = \bigg\{ & \Op{\vphantom{d}ed}{LL}[V][ij11], \Op{\vphantom{d}ed}{LL}[V][ij22], \Op{\vphantom{d}ed}{LL}[V][ij33], \Op{\vphantom{d}eu}{LL}[V][ij11], \Op{\vphantom{d}eu}{LL}[V][ij22], \Op{\vphantom{d}\nu e}{LL}[V][11ij], \Op{\vphantom{d}\nu e}{LL}[V][22ij], \Op{\vphantom{d}\nu e}{LL}[V][33ij], \\[-0.25mm]
    & {\color{red}\Sym{\vphantom{d}ee}{LL}[V][ijkk]}, \Op{\vphantom{d}ee}{LL}[V][ijjj], \Op{\vphantom{d}ee}{LL}[V][ijii], \Op{\vphantom{d}ed}{LR}[V][ij11], \Op{\vphantom{d}ed}{LR}[V][ij22], \Op{\vphantom{d}ed}{LR}[V][ij33], \Op{\vphantom{d}eu}{LR}[V][ij11], \Op{\vphantom{d}eu}{LR}[V][ij22], \\[-2mm]
    & \Op{\vphantom{d}ee}{LR}[V][ijkk], \Op{\vphantom{d}ee}{LR}[V][ijjj], \Op{\vphantom{d}ee}{LR}[V][ijii] \bigg\}\;,
    \end{aligned}  \\
    & \begin{aligned}
    \label{eq:nu-Sym Basis}
    \left\{ \mathcal{S}_i^{(\nu)} \right\} = \bigg\{ & \Op{\vphantom{d}\nu e}{LL}[V][ij11], \Op{\vphantom{d}\nu e}{LL}[V][ij22], \Op{\vphantom{d}\nu e}{LL}[V][ij33], \Op{\vphantom{d}\nu d}{LL}[V][ij11], \Op{\vphantom{d}\nu d}{LL}[V][ij22], \Op{\vphantom{d}\nu d}{LL}[V][ij33], \Op{\vphantom{d}\nu u}{LL}[V][ij11], \Op{\vphantom{d}\nu u}{LL}[V][ij22], \\[-0.5mm] 
    & {\color{red}\Sym{\vphantom{d}\nu \nu}{LL}[V][ijkk]}, \Op{\vphantom{d}\nu \nu}{LL}[V][ijjj], \Op{\vphantom{d}\nu \nu}{LL}[V][ijii], \Op{\vphantom{d}\nu e}{LR}[V][ij11], \Op{\vphantom{d}\nu e}{LR}[V][ij22], \Op{\vphantom{d}\nu e}{LR}[V][ij33], \Op{\vphantom{d}\nu d}{LR}[V][ij11], \Op{\vphantom{d}\nu d}{LR}[V][ij22], \\[-2mm]
    & \Op{\vphantom{d}\nu d}{LR}[V][ij33], \Op{\vphantom{d}\nu u}{LR}[V][ij11], \Op{\vphantom{d}\nu u}{LR}[V][ij22] \bigg\}\;,
    \end{aligned}  \\
    & \begin{aligned}
    \label{eq:u-Sym Basis}
    \left\{ \mathcal{S}_i^{(u)} \right\} =\bigg\{ & \Op{\vphantom{d}eu}{LL}[V][11ij], \Op{\vphantom{d}eu}{LL}[V][22ij], \Op{\vphantom{d}eu}{LL}[V][33ij],  \Op{\vphantom{d}\nu u}{LL}[V][11ij], \Op{\vphantom{d}\nu u}{LL}[V][22ij], \Op{\vphantom{d}\nu u}{LL}[V][33ij], \Op{\vphantom{d}ud}{LL}[V1][ij11], \Op{\vphantom{d}ud}{LL}[V8][ij11], \\[-0.25mm]
    & \Op{\vphantom{d}ud}{LL}[V1][ij22], \Op{\vphantom{d}ud}{LL}[V8][ij22], \Op{\vphantom{d}ud}{LL}[V1][ij33], \Op{\vphantom{d}ud}{LL}[V8][ij33], \Op{\vphantom{d}uu}{LL}[V][ijjj], \Op{\vphantom{d}uu}{LL}[V][ijii], \Op{\vphantom{d}ue}{LR}[V][ij11], \\
    & \Op{\vphantom{d}ue}{LR}[V][ij22], \Op{\vphantom{d}ue}{LR}[V][ij33], \Op{\vphantom{d}ud}{LR}[V1][ij11], \Op{\vphantom{d}ud}{LR}[V8][ij11], \Op{\vphantom{d}ud}{LR}[V1][ij22], \Op{\vphantom{d}ud}{LR}[V8][ij22], \Op{\vphantom{d}ud}{LR}[V1][ij33], \\[-2mm] 
    & \Op{\vphantom{d}ud}{LR}[V8][ij33], \Op{\vphantom{d}uu}{LR}[V1][ijjj], \Op{\vphantom{d}uu}{LR}[V8][ijjj], \Op{\vphantom{d}uu}{LR}[V1][ijii], \Op{\vphantom{d}uu}{LR}[V8][ijii] \bigg\}\;,
    \end{aligned}  \\
    & \begin{aligned}
    \label{eq:d-Sym Basis}
    \left\{ \mathcal{S}_i^{(d)} \right\} = \bigg\{ & \Op{\vphantom{d}ed}{LL}[V][11ij], \Op{\vphantom{d}ed}{LL}[V][22ij], \Op{\vphantom{d}ed}{LL}[V][33ij],  \Op{\vphantom{d}\nu d}{LL}[V][11ij], \Op{\vphantom{d}\nu d}{LL}[V][22ij], \Op{\vphantom{d}\nu d}{LL}[V][33ij], \Op{\vphantom{d}ud}{LL}[V1][11ij], \Op{\vphantom{d}ud}{LL}[V8][11ij], \\[-0.25mm] 
    & \Op{\vphantom{d}ud}{LL}[V1][22ij], \Op{\vphantom{d}ud}{LL}[V8][22ij], \Op{\vphantom{d}dd}{LL}[V][ijkk],
    {\color{red}\Sym{\vphantom{d}dd}{LL}[V8][ijkk]},
    \Op{\vphantom{d}dd}{LL}[V][ijjj], \Op{\vphantom{d}dd}{LL}[V][ijii], \Op{\vphantom{d}de}{LR}[V][ij11], \\ 
    & \Op{\vphantom{d}de}{LR}[V][ij22], \Op{\vphantom{d}de}{LR}[V][ij33], \Op{\vphantom{d}du}{LR}[V1][ij11], \Op{\vphantom{d}du}{LR}[V8][ij11], \Op{\vphantom{d}du}{LR}[V1][ij22], \Op{\vphantom{d}du}{LR}[V8][ij22], \Op{\vphantom{d}dd}{LR}[V1][ijkk], \\[-2mm]
    & \Op{\vphantom{d}dd}{LR}[V8][ijkk], \Op{\vphantom{d}dd}{LR}[V1][ijjj], \Op{\vphantom{d}dd}{LR}[V8][ijjj], \Op{\vphantom{d}dd}{LR}[V1][ijii], \Op{\vphantom{d}dd}{LR}[V8][ijii] \bigg\} \;,
    \end{aligned}
\end{align}
where the basis $\big\{ \mathcal{S}_i^{(u)} \big\}$ is built from operators in the JMS basis, while 
$\big\{ \mathcal{S}_i^{(e,\nu,d)} \big\}$ contain each one operator (highlighted in {\color{red}red} above) which is not in the JMS basis\footnote{For the $e$- and $\nu$-type bases, the mismatch corresponds to the discussion at the end of \Sec{sec:defLEFT}.}:
\eq{
\Sym{\vphantom{d}\ell\ell}{LL}[V][ijkk] = (\bar \ell_i \gamma^\mu P_L \ell_j)(\bar \ell_k \gamma_\mu P_L \ell_k) \;, \qquad
\Sym{\vphantom{d}dd}{LL}[V8][ijkk] = (\bar d_i \gamma^\mu P_L T^A d_j)(\bar d_k \gamma_\mu P_L T^A d_k) \;.
}
This extra operator is necessary in order to make the basis $\big\{ \mathcal{S}_i^{(d)} \big\}$ flavor-symmetric and flavor-decoupled.
We call $\big\{ \mathcal{S}_i^{(f)} \big\}$ the {\it $f$-type symmetric basis}.

Except for the $u$-type case, the two-loop penguin ADMs in these symmetric bases can be obtained by performing a change of basis at NLO from the completed lepton/down-quark bases. 
The penguin ADMs for the $u$-type as well as the penguin ADMs for the LEFT(4) and the LEFT(3) can be obtained through surface-level flavor universality, as we discuss below.

\begin{figure}
    \centering
    \begin{subfigure}{0.265\textwidth}
    \hspace{0.27cm}
    \includegraphics[valign=c, scale=0.15]{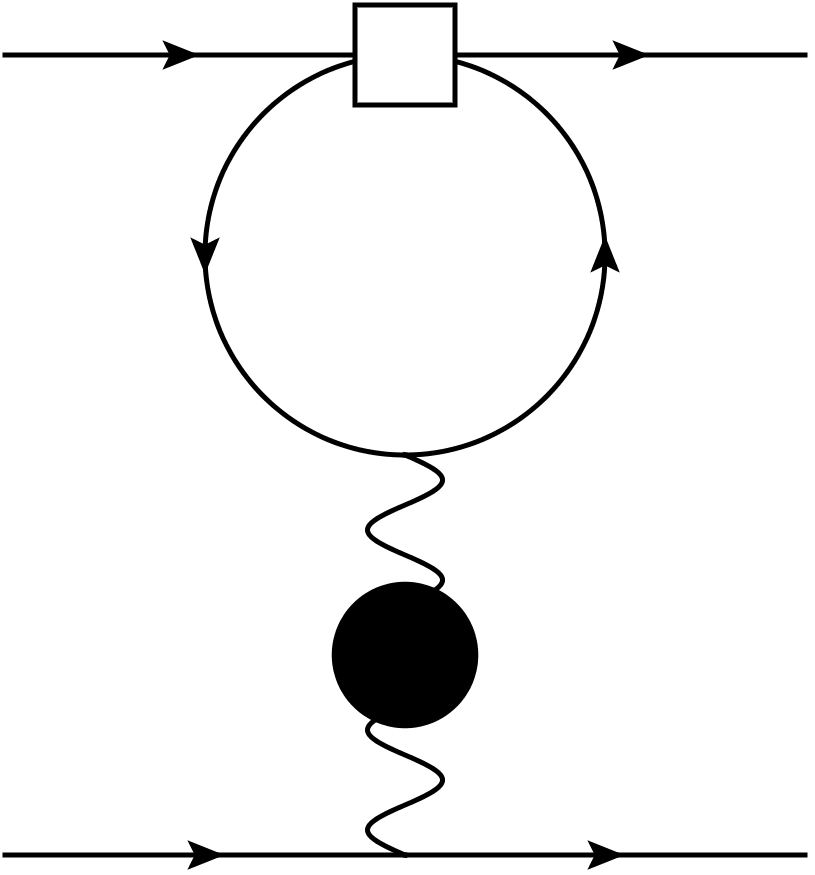}
    \caption{}
    \end{subfigure}
    \hspace{1cm}
    \begin{subfigure}{0.265\textwidth}
    \hspace{0.27cm}
    \includegraphics[valign=c, scale=0.15]{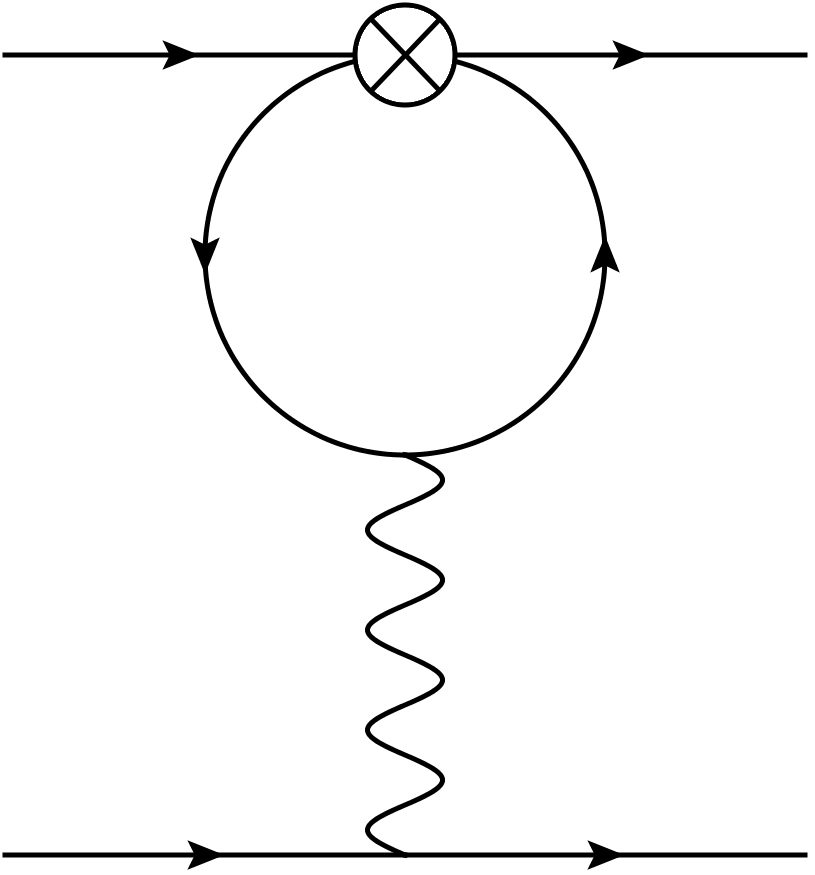}
    \caption{}
    \end{subfigure}
    \caption{Diagrams contributing to the cancellation given in~\Eq{eq:Flavor Decoupling NLO}, with~(a) contributing to the first term and~(b) contributing to the second term. The square represents an insertion of $Q_i$, the black blob represents the gluon/photon self energy, and the crossed circle is an EFT counterterm corresponding to $p^{(1,0;0)}_{Q_kQ_j}$.}
    \label{fig:Flavor Diagrams}
\end{figure}

Before going on, however, it is necessary to prove that these symmetric bases are indeed flavor-decoupled, i.e. that~\Eq{eq:Flavor Decoupling NLO} is satisfied: 
\eq{
\left[ -4 \,p^{(2,0;1)}_{Q_iQ_j} + 4 \,p^{(1,0;1)}_{Q_iQ_k} p^{(1,0;0)}_{Q_kQ_j} + 2 \,p^{(1,0;1)}_{Q_iE_k} p^{(1,0;0)}_{E_kQ_j} \right]_{\text{featuring flavors } \notin \, Q_i,Q_j} = 0 \; . 
\nonumber
}
There is only one type of two-loop penguin diagram that can contribute to the first term in this equation, see Fig.~\ref{fig:Flavor Diagrams}~(a), and the second term corresponds also to a single EFT counterterm diagram, see Fig.~\ref{fig:Flavor Diagrams}~(b).
It is simple to check that these terms lead to a complete cancellation~\cite{Buras:1992tc}. 
The third term is zero in all four bases due to their closed-penguin structure.
With this,~\Eq{eq:Flavor Decoupling NLO} is satisfied and thus all symmetric bases realize surface-level flavor universality.

Let us now discuss the derivation of the ADM for the $u$-type operators. We start with the $d$-type symmetric basis and replace the flavor $d_k$ by an additional $u$-type quark (i.e. a fictitious $u_3$). The flavors $d_i$ and $d_j$ are however left unaltered. Due to the basis being flavor-decoupled, the alterations in the resulting ADM are easy to track: The four columns/rows for the operators containing $d_k$ are replaced by four new ones corresponding to $u_3$. This modified basis is functionally equivalent to the up-type symmetric basis up to a relabeling of field names, $u \leftrightarrow d$. Hence, when replacing charges in the ADM, $Q_d \leftrightarrow Q_u$, one finds the desired ADM in the up-quark symmetric basis.

The two-loop ADMs in the symmetric bases, as obtained thus far, are only valid for $n_q=5$, given that the structure of the basis itself depends on the number of active fermions. The symmetric bases also allow for a trivial transformation of the ADMs when decoupling heavy fermions from the LEFT. As already discussed, decoupling a flavor in a symmetric basis impacts the ADM in a trivial way, namely by removing the corresponding rows and columns. The penguin ADMs for the $\Delta F=1^{\bar f f}$ sectors in the LEFT(4) and the LEFT(3) follow then quite straightforwardly.

In each case, once the full penguin ADMs have been obtained to NLO in the symmetric bases, one can perform a final change of basis to the JMS basis, and thus obtain the final result. 
This is not needed in the case of the up-type symmetric basis, however, given that it corresponds already to its JMS counterpart.
Finally, the ADMs for the chirally-flipped sectors are identical to the ones derived above.

\subsubsection*{4. Reconstruction of $\Delta F = 0$ in the JMS basis}
\label{sec:Penguin Reconstruction DF=0}

\begin{figure}
    \centering
    \includegraphics[valign=c, scale=0.15, angle=90]{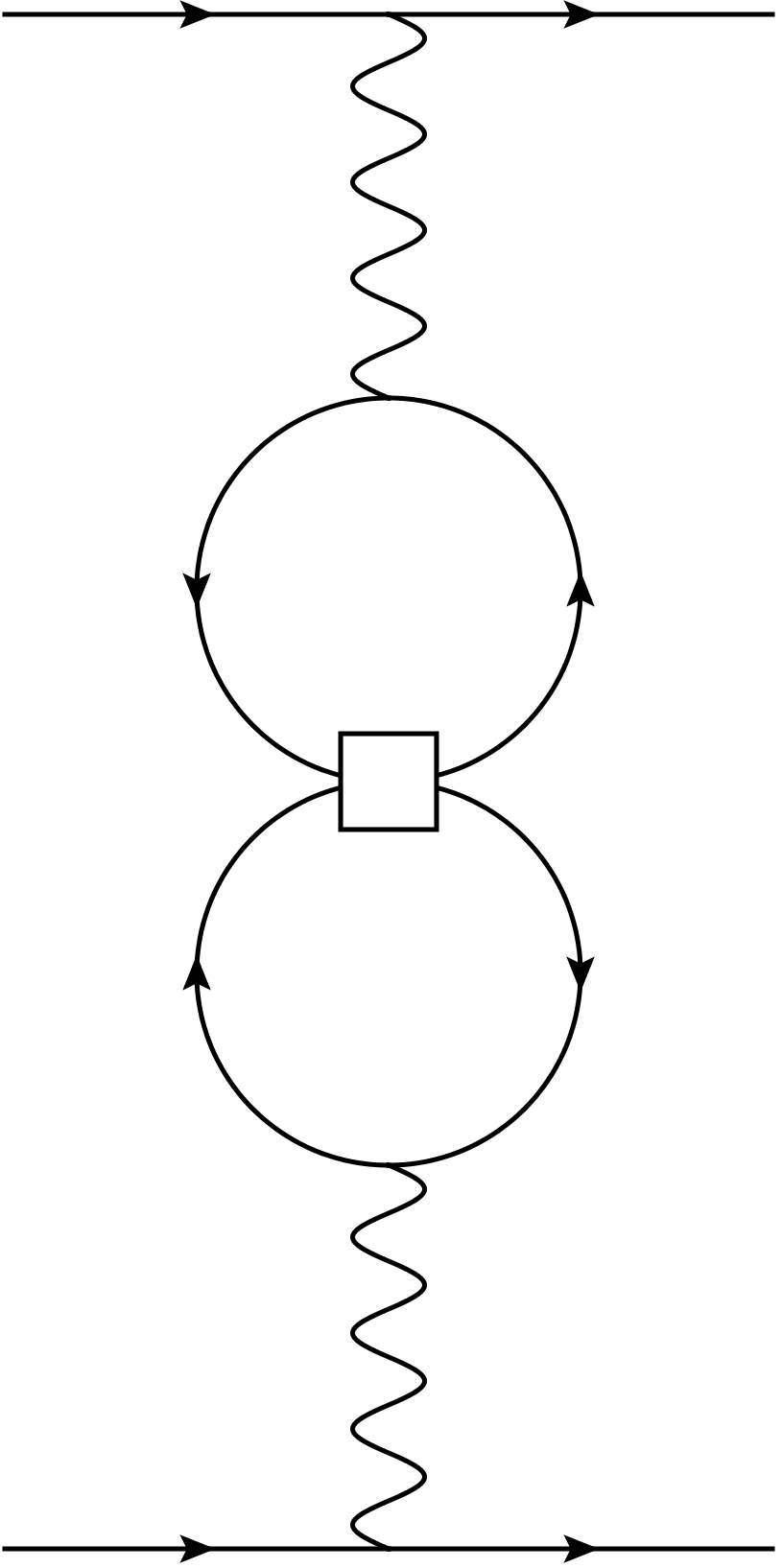}
    \caption{Class of diagrams exclusive to operators in the $\Delta F = 0$ sector. Their contribution to the penguin ADM vanishes in any flavor-decoupled basis, in exactly the same way as in the diagrams in Fig.~\ref{fig:Flavor Diagrams}.}
    \label{fig:DF0 Diagram}
\end{figure}

In the $\Delta F = 0$ sector the operators have a flavor structure of the form $\op_i = \bar\psi_1\Gamma\psi_1\,\bar\psi_2\Gamma'\psi_2$ or $\widetilde \op_i =\bar\psi_1\Gamma\psi_2\,\bar\psi_2\Gamma'\psi_1$, possibly with $\psi_1=\psi_2$.
Thus both fields $\psi_1,\psi_2$ play the role of ``inner flavors'' and can become part of fermion loops.
This means that the beta function $\dot L_{iikk}$
receives separate contributions proportional to $L_{iijj}$ and to $L_{jjkk}$, for any $j$.
These two contributions can be independently associated to the corresponding $\Delta F=1^{\bar f f}$ operators with inner flavor $j$.
Schematically we write:
\eq{
\begin{aligned}
\label{eq:DF=0 Reconstruction}
\left[\dot{L}_{iikk}\right]_{p} =& \left.\left[\dot{L}_{i4kk}\right]_{p}\right|_{4 \to i} + \left.\left[\dot{L}_{iik4}\right]_{p}\right|_{4 \to k} \ , \\
\left[\dot{L}_{ikki}\right]_{p} =&\left.\left[\dot{L}_{ikk4}\right]_{p}\right|_{4 \to i} + \left.\left[\dot{L}_{i4ki}\right]_{p}\right|_{4 \to k} \ , 
\end{aligned}
}
where the index `4' is an artificial flavor introduced for the purpose of the method, as explained below. However, it is crucial to note that the straightforward combination in the RHS of this equation only accounts for all $\Delta F = 0$ penguin contributions to the beta-function when working in a flavor-decoupled basis, both in $\Delta F = 1^{\bar f f}$ and $\Delta F = 0$. This is most easily seen at the level of Feynman diagrams, in an expansion of the ADM like that of~\Eqs{eq:gamma20a}{eq:gamma11a}, where one can see that it is only in such bases that the contributions to the $\Delta F = 0$ ADM from the class of diagrams in Fig.~\ref{fig:DF0 Diagram} vanish completely.

To apply~\Eq{eq:DF=0 Reconstruction}, our starting point is to define a set of $\Delta F = 1^{\bar f f}$ flavor-decoupled and fully flavor-symmetric bases, analogous to the symmetric bases in~\Eqs{eq:e-Sym Basis}{eq:d-Sym Basis}. Each of these bases is defined in a separate theory with one extra fermion, which we arbitrarily call $e_4,\nu_4,u_4,d_4$, respectively, and refer to the $\Delta F = 1^{\bar f f}$ sector relative to the corresponding $4\to i$ transition. 
These bases are:
\begin{align}
    & \begin{aligned}
    \label{eq:e-Sym Basis 2}
    \left\{ {\mathcal{S}'}_i^{(e)} \right\} = \bigg\{ & \Op{\vphantom{d}ed}{LL}[V][i411], \Op{\vphantom{d}ed}{LL}[V][i422], \Op{\vphantom{d}ed}{LL}[V][i433], \Op{\vphantom{d}eu}{LL}[V][i411], \Op{\vphantom{d}eu}{LL}[V][i422], \Op{\vphantom{d}\nu e}{LL}[V][11i4], \Op{\vphantom{d}\nu e}{LL}[V][22i4], \Op{\vphantom{d}\nu e}{LL}[V][33i4], \\[-0.25mm]
    & {\color{red}\Sym{\vphantom{d}ee}{LL}[V][i4kk]}, {\color{red}\Sym{\vphantom{d}ee}{LL}[V][i4ll]}, \Op{\vphantom{d}ee}{LL}[V][i444], \Op{\vphantom{d}ee}{LL}[V][i4ii], \Op{\vphantom{d}ed}{LR}[V][i411], \Op{\vphantom{d}ed}{LR}[V][i422], \Op{\vphantom{d}ed}{LR}[V][i433], \Op{\vphantom{d}eu}{LR}[V][i411], \\[-2mm]
    & \Op{\vphantom{d}eu}{LR}[V][i422], \Op{\vphantom{d}ee}{LR}[V][i4kk], \Op{\vphantom{d}ee}{LR}[V][i4ll], \Op{\vphantom{d}ee}{LR}[V][i444], \Op{\vphantom{d}ee}{LR}[V][i4ii] \bigg\} \;,
    \end{aligned} \\
    & \begin{aligned}
    \label{eq:nu-Sym Basis 2}
    \left\{ {\mathcal{S}'}_i^{(\nu)} \right\} = \bigg\{ & \Op{\vphantom{d}\nu e}{LL}[V][i411], \Op{\vphantom{d}\nu e}{LL}[V][i422], \Op{\vphantom{d}\nu e}{LL}[V][i433], \Op{\vphantom{d}\nu d}{LL}[V][i411], \Op{\vphantom{d}\nu d}{LL}[V][i422], \Op{\vphantom{d}\nu d}{LL}[V][i433], \Op{\vphantom{d}\nu u}{LL}[V][i411], \Op{\vphantom{d}\nu u}{LL}[V][i422], \\[-0.5mm] 
    & {\color{red}\Sym{\vphantom{d}\nu \nu}{LL}[V][i4kk]}, {\color{red}\Sym{\vphantom{d}\nu \nu}{LL}[V][i4ll]}, \Op{\vphantom{d}\nu \nu}{LL}[V][i444], \Op{\vphantom{d}\nu \nu}{LL}[V][i4ii], \Op{\vphantom{d}\nu e}{LR}[V][i411], \Op{\vphantom{d}\nu e}{LR}[V][i422], \Op{\vphantom{d}\nu e}{LR}[V][i433], \Op{\vphantom{d}\nu d}{LR}[V][i411],\\[-2mm]
    &  \Op{\vphantom{d}\nu d}{LR}[V][i422], \Op{\vphantom{d}\nu d}{LR}[V][i433], \Op{\vphantom{d}\nu u}{LR}[V][i411], \Op{\vphantom{d}\nu u}{LR}[V][i422] \bigg\} \;,
    \end{aligned} \\
    & \begin{aligned}
    \label{eq:u-Sym Basis 2}
    \left\{ {\mathcal{S}'}_i^{(u)} \right\} =\bigg\{ & \Op{\vphantom{d}eu}{LL}[V][11i4], \Op{\vphantom{d}eu}{LL}[V][22i4], \Op{\vphantom{d}eu}{LL}[V][33i4],  \Op{\vphantom{d}\nu u}{LL}[V][11i4], \Op{\vphantom{d}\nu u}{LL}[V][22i4], \Op{\vphantom{d}\nu u}{LL}[V][33i4], \Op{\vphantom{d}ud}{LL}[V1][i411], \Op{\vphantom{d}ud}{LL}[V8][i411], \\[-0.25mm]
    & \Op{\vphantom{d}ud}{LL}[V1][i422], \Op{\vphantom{d}ud}{LL}[V8][i422], \Op{\vphantom{d}ud}{LL}[V1][i433], \Op{\vphantom{d}ud}{LL}[V8][i433], \Op{\vphantom{d}uu}{LL}[V][i4ll], {\color{red}\Sym{\vphantom{d}uu}{LL}[V8][i4ll]}, \Op{\vphantom{d}uu}{LL}[V][i444], \Op{\vphantom{d}uu}{LL}[V][i4ii], \\
    & \Op{\vphantom{d}ue}{LR}[V][i411], \Op{\vphantom{d}ue}{LR}[V][i422], \Op{\vphantom{d}ue}{LR}[V][i433], \Op{\vphantom{d}ud}{LR}[V1][i411], \Op{\vphantom{d}ud}{LR}[V8][i411], \Op{\vphantom{d}ud}{LR}[V1][i422], \Op{\vphantom{d}ud}{LR}[V8][i422], \Op{\vphantom{d}ud}{LR}[V1][i433], \\[-2mm] 
    & \Op{\vphantom{d}ud}{LR}[V8][i433], \Op{\vphantom{d}uu}{LR}[V1][i4ll], \Op{\vphantom{d}uu}{LR}[V8][i4ll], \Op{\vphantom{d}uu}{LR}[V1][i444], \Op{\vphantom{d}uu}{LR}[V8][i444], \Op{\vphantom{d}uu}{LR}[V1][i4ii], \Op{\vphantom{d}uu}{LR}[V8][i4ii] \bigg\} \;,
    \end{aligned} \\
    & \begin{aligned}
    \label{eq:d-Sym Basis 2}
    \left\{ {\mathcal{S}'}_i^{(d)} \right\} = \bigg\{ & \Op{\vphantom{d}ed}{LL}[V][11i4], \Op{\vphantom{d}ed}{LL}[V][22i4], \Op{\vphantom{d}ed}{LL}[V][33i4],  \Op{\vphantom{d}\nu d}{LL}[V][11i4], \Op{\vphantom{d}\nu d}{LL}[V][22i4], \Op{\vphantom{d}\nu d}{LL}[V][33i4], \Op{\vphantom{d}ud}{LL}[V1][11i4], \Op{\vphantom{d}ud}{LL}[V8][11i4], \\[-0.25mm] 
    & \Op{\vphantom{d}ud}{LL}[V1][22i4], \Op{\vphantom{d}ud}{LL}[V8][22i4], \Op{\vphantom{d}dd}{LL}[V][i4kk],  {\color{red}\Sym{\vphantom{d}dd}{LL}[V8][i4kk]}, \Op{\vphantom{d}dd}{LL}[V][i4ll], {\color{red}\Sym{\vphantom{d}dd}{LL}[V8][i4ll]}, \Op{\vphantom{d}dd}{LL}[V][i444], \Op{\vphantom{d}dd}{LL}[V][i4ii], \\ 
    & \Op{\vphantom{d}de}{LR}[V][i411], \Op{\vphantom{d}de}{LR}[V][i422], \Op{\vphantom{d}de}{LR}[V][i433], \Op{\vphantom{d}du}{LR}[V1][i411], \Op{\vphantom{d}du}{LR}[V8][i411], \Op{\vphantom{d}du}{LR}[V1][i422], \Op{\vphantom{d}du}{LR}[V8][i422],  \\[-2mm]
    & \Op{\vphantom{d}dd}{LR}[V1][i4kk], \Op{\vphantom{d}dd}{LR}[V8][i4kk], \Op{\vphantom{d}dd}{LR}[V1][i444], \Op{\vphantom{d}dd}{LR}[V8][i444], \Op{\vphantom{d}dd}{LR}[V1][i4ii], \Op{\vphantom{d}dd}{LR}[V8][i4ii] \bigg\} \;,
    \end{aligned}
\end{align}
where $k\ne l$ are flavor indices  both different from $i,4$. Obtaining the penguin ADMs for these bases from the ones in the previous section is a straightforward task given their surface-level flavor universality: again, the basis needs to be extended, this time including the extra flavor.
 
In order to avoid what will result in double-counting related to Wilson coefficients of the type $L_{iiii}$, we now put to zero the Wilson coefficients $L_{i444}$. These contributions are already accounted for by $L_{i4ii}$.
Finally, in order to compensate for the symmetry factor of~2 in the tree-level matrix element of operators $\op^{V,LL}_{iiii}$, we substitute $L^{V,LL}_{i4ii}\to 2 L^{V,LL}_{i4ii}$ everywhere.
The resulting beta functions are now ready to be introduced in~\Eq{eq:DF=0 Reconstruction}. This provides the full two-loop penguin ADM for a particular $\Delta F = 0$ basis (specified below). A final change to the JMS basis provides the final result.

The aforementioned $\Delta F = 0$ basis
is identical to the JMS basis, except for eight operators in the quark sectors:
\begin{gather*}
\Sym{\vphantom{d}dd}{LL}[V8][1122]\,,\; \Sym{\vphantom{d}dd}{LL}[V8][1133]\,,\; \Sym{\vphantom{d}dd}{LL}[V8][2233]\,,\; \Sym{\vphantom{d}uu}{LL}[V8][1122]\,,\;
\Sym{\vphantom{d}dd}{RR}[V8][1122]\,,\; \Sym{\vphantom{d}dd}{RR}[V8][1133]\,,\; \Sym{\vphantom{d}dd}{RR}[V8][2233]\,,\; \Sym{\vphantom{d}uu}{RR}[V8][1122]\,,
\end{gather*}
which under the change of basis will lead to the JMS operators
\begin{align*}
\Op{\vphantom{d}dd}{LL}[V][1221]\,,\; \Op{\vphantom{d}dd}{LL}[V][1331]\,,\; \Op{\vphantom{d}dd}{LL}[V][2332]\,,\; \Op{\vphantom{d}uu}{LL}[V][1221]\,,\;
\Op{\vphantom{d}dd}{RR}[V][1221]\,,\; \Op{\vphantom{d}dd}{RR}[V][1331]\,,\; \Op{\vphantom{d}dd}{RR}[V][2332]\,,\; \Op{\vphantom{d}uu}{RR}[V][1221]\,.
\end{align*}
In the lepton sectors, it will be the nine operators
\begin{align*}
\Sym{\vphantom{d}ee}{LL}[V][1122]\,,\; \Sym{\vphantom{d}ee}{LL}[V][1133]\,,\; \Sym{\vphantom{d}ee}{LL}[V][2233]\,,\; 
\Sym{\vphantom{d}\nu\nu}{LL}[V][1122]\,,\; \Sym{\vphantom{d}\nu\nu}{LL}[V][1133]\,,\; \Sym{\vphantom{d}\nu\nu}{LL}[V][2233]\,,\;
\Sym{\vphantom{d}ee}{RR}[V][1122]\,,\; \Sym{\vphantom{d}ee}{RR}[V][1133]\,,\; \Sym{\vphantom{d}ee}{RR}[V][2233]\,,\;
\end{align*}
that will transform in the change of basis into the symmetrized operators in Eqs.~(\ref{eq:SymRules-ee}) and~(\ref{eq:SymRules-nunu}), namely through
\begin{gather*}
\Sym{\vphantom{d}ee}{LL}[V][iijj] \to \frac{1}{2}\Big(\Op{\vphantom{d}ee}{LL}[V][iijj] + \Op{\vphantom{d}ee}{LL}[V][ijji]\Big)\,,\quad
\Sym{\vphantom{d}\nu\nu}{LL}[V][iijj] \to \frac{1}{2}\Big(\Op{\vphantom{d}\nu\nu}{LL}[V][iijj] + \Op{\vphantom{d}\nu\nu}{LL}[V][ijji]\Big)\,,\\
\Sym{\vphantom{d}ee}{RR}[V][iijj] \to \frac{1}{2}\Big(\Op{\vphantom{d}ee}{RR}[V][iijj] + \Op{\vphantom{d}ee}{RR}[V][ijji]\Big)\,.
\end{gather*}

The discussion above only holds for the $\Delta F = 0$ ADM with $n_q = 5$. To obtain the ADMs for $n_q < 5$ one can use the fact that the aforementioned $\Delta F = 0$ basis is flavor-decoupled and flavor symmetric in any two flavors of the same type $e$, $\nu$ or $q$ ($u$ or $d$). The ADM in this basis is therefore surface-level flavor universal, and thus simply removing the Wilson coefficients containing the heavy fields is enough to properly implement the decoupling. 
The ADM for $n_q < 5$ in the JMS basis follows after the appropriate change of basis.

\subsection{Results: Beta-Functions and \dsix Ancillaries}
\label{sec:results}

The complete four-fermion ADMs at NLO are presented in an ancillary pdf file attached to the arXiv submission of this article, where the beta-functions $\dot L_i$ are given for the three different cases LEFT(5), LEFT(4) and LEFT(3). 
The full set of LO and NLO results are also implemented in \texttt{Mathematica} format in \dsix (available at~\href{https://github.com/DsixTools/DsixTools/tree/Two-loop-LEFT}{github}).

We have checked that our results agree with those available in the literature.
For the comparison the three methods described above were employed, namely performing direct NLO basis changes, using flavor symmetrization to relate ADMs of disconnected RG-invariant sectors, as well as building the ADMs from the available tables of pole coefficients. 
All three methods produce identical results in all available cases. To the best of our knowledge all NLO corrections to the BNV sector, part of the $\Delta F = 1^{\bar f f}$ sector at order $O(\alpha_s \alpha)$, most NLO ADMs for the $\Delta F = 0$ sector, and all four-fermion sectors at order $O(\alpha^2)$ are new. As a further consistency check we have tested all ADMs under flavor-symmetry transformations as discussed in~\Reff{Morell:2024aml}, consisting in the renaming of two inner flavors (e.g. $u \leftrightarrow d$) followed by the change of basis needed to recover the initial basis. Such transformations must leave the ADM unchanged $-$up to the same swapping of QED charges in the ADM$-$, as long as the swapped flavors are in the same representation of $SU(3)_c$. We have checked that all ADMs satisfy this condition.

Recent results by~\Reff{Naterop:2025lzc} on the two-loop beta functions for the BNV sectors, posterior to the online publication of v1 of this paper, have helped us identify a few bugs that affected our code for the writing of the ADMs in the ancillaries. Our results agree with that reference in all cases, up to a change of evanescent scheme.

\subsubsection{Renormalization of Vector and Axial Currents}

As an additional cross-check, we have also studied the compatibility of our beta functions with known results on the renormalization of vector and axial currents. The operators most suitable for such comparison are those containing a neutrino current, $\mathcal{O} \sim (\bar\nu_i \Gamma_1 \nu_j)(\psi_k \Gamma_2 \psi_l)$, given that neutrinos are sterile in the LEFT, and thus all contributions to the ADM arise from the renormalization of the other fermionic current, $J \sim \psi_k \Gamma_2 \psi_l$. 
This includes operators in the $\nu d$-, $\nu u$- and $\nu e$-type $\Delta F = 1$ sectors; the $\nu$-type $\Delta F = 1^{\bar f f}$ sector; and some operators within the $\Delta F = 0$ sector. 

As a first step, we have checked that our results agree with the well-known non-renormalization of flavor-changing vector and axial currents (e.g. see~\cite{Peskin:1995ev,Manohar:2000dt}):
\eq{
\dlwc{\nu \psi}{LL}[V][ijkl] = \dlwc{\nu \psi}{LR}[V][ijkl] = 0 \;,
}
for $\psi = d,u,e$ and any flavor indices $i,j,k\neq l$.
However, the flavor-conserving vector and axial currents, $\bar\psi_k \gamma^\mu (\gamma_5) \psi_k$, get nonzero running from penguin-type contributions. Up to two-loop order, the corresponding effects in our beta functions are:
\eqa{
\dlwc{\nu \psi}{LL}[V][ijkk] &= \sum_{f=d,u,e}\sum_w \left[ \left(V_{f\psi} - A_{f\psi} \right) \lwc{\nu f}{LL}[V][ijww] +  \left(V_{f\psi} + A_{f\psi} \right) \lwc{\nu f}{LR}[V][ijww] \right] \;,
\\[2mm]
\dlwc{\nu \psi}{LR}[V][ijkk] &= \sum_{f=d,u,e}\sum_w \left[ \left(V_{f\psi} + A_{f\psi} \right) \lwc{\nu f}{LL}[V][ijww] +  \left(V_{f\psi} - A_{f\psi} \right) \lwc{\nu f}{LR}[V][ijww] \right] \;,
}
again for $\psi = d,u,e$ and any flavor indices $i,j,k$. The coefficients $V_{f\psi}$ and $A_{f\psi}$ are generated through two different types of penguin contributions. As noted in~\Reff{Collins:2005nj}, the coefficients $V_{f\psi}$ emerge from the renormalization of the vector current through a contribution that mimics the photon self-energy, starting at $O(\alpha)$. Our results for these coefficients are:
\begin{equation}
    V_{f\psi} = e^2 \, \frac{4\,Q_f Q_\psi}{3} \times \left\{
    \arraycolsep=3pt
    \def\arraystretch{1.5}
    \begin{array}{ccl}
        N_c \left(1 + 3 C_F \frac{\alpha_s}{4\pi} + 3 Q_f^2 \frac{\alpha}{4\pi} \right) & \text{if} & f = u,d  \\
         \left(1 + 3 Q_f^2 \frac{\alpha}{4\pi}  \right) & \text{if} & f = e
    \end{array}\right. \;.
\end{equation}
Apart from agreeing with the one-loop results in~\Reff{Collins:2005nj}, these coefficients reproduce the known two-loop $1/\epsilon$ poles in the photon self-energy (see e.g.~\cite{Schnubel:2025ejl}). At the same time, as shown, for instance, in~Refs.~\cite{Larin:1993tq,Bernreuther:2005rw,Fael:2023zqr}, the coefficients $A_{f\psi}$ emerge from the contribution to the ADM of the ABJ triangle anomaly, both in the case of two gluons, at $O(\alpha_s^2)$, and two photons, at $O(\alpha^2)$,
\begin{equation}
    A_{f\psi} = \left\{
    \arraycolsep=3pt
    \def\arraystretch{1.5}
    \begin{array}{ccl}
        6 C_F \alpha_s^2 + 12 N_c Q_f^2 Q_\psi^2 \alpha^2 & \text{if} & f,\psi = u,d  \\
        12 N_c Q_f^2 Q_\psi^2 \alpha^2 & \text{if} & f = u,d \;\,\text{and}\;\, \psi = e  \\
        12 Q_f^2 Q_\psi^2 \alpha^2 & \text{if} & f = e
    \end{array}\right. \;.
\end{equation}
These results agree with~Refs.~\cite{Larin:1993tq,Bernreuther:2005rw,Fael:2023zqr}, and thus properly take into account the anomaly contributions. Let us remark here that, as in all closed-penguin ADMs, we obtain these results without encountering any problematic Dirac traces in the process.

\section{Numerical Analysis}
\label{sec:numerics}

Having obtained the full set of four-fermion beta functions at NLO, it is interesting to showcase through a few examples the numerical impact of these corrections. We thus present here a numerical comparison between the two-loop RGE and its one-loop counterpart, as implemented in the previous version of \dsix~\cite{Fuentes-Martin:2020zaz}. We limit ourselves to a few illustrative examples for which we compute the running from the electroweak scale ($\mu = M_Z$) down to the bottom mass scale ($\mu = m_b$). The solution to the RGEs depend on the initial conditions at the electroweak scale, namely, the matching conditions from the underlying full theory (e.g. the SMEFT). For our purposes we will simply set the initial conditions to be arbitrary real numbers in the range $\left(-10^{-5},10^{-5}\right)\GeV^{-2}$.

\begin{figure}[tb]
    \centering
    \includegraphics[scale=0.95]{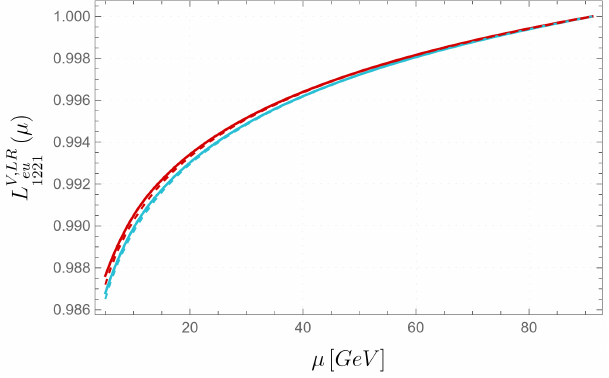}
    \caption{Running of $L_{eu}^{V,LR}(\mu)$ from the EW scale $\mu = M_Z$ down to the bottom mass scale $\mu = m_b$ with LL ({\color{blue-plot}blue}) and NLL ({\color{red-plot}red}) resummation and the corresponding LO ({\color{blue-plot}blue-dashed}) and NLO ({\color{red-plot}red-dashed}) fixed-order approximations. The Wilson coefficient is shown normalized to 1 at the matching scale $\mu = M_Z$, that is $L_{eu}^{V,LR}(\mu)/L_{eu}^{V,LR}(M_Z)$.} 
    \label{fig:RGE eu-SRR}
\end{figure}

We begin by studying the running of a VLR sub-block in the $eu$-type $\Delta F = 1$ sector, with flavors $(\bar e\mu)(\bar c u)$. This sub-block consists of a single operator, with its beta function given by
\eq{
\dlwc{eu}{LR}[V][1221]  =  8 e^2 \lwc{eu}{LR}[V][1221] -32 \alpha_s \alpha \lwc{eu}{LR}[V][1221] -\frac{4960}{27} \alpha^2 \lwc{eu}{LR}[V][1221]\ .
}
Thus the next-to-leading-log (NLL) running is an $\alpha_s$ correction to the leading-log (LL) one, and no mixing is present.
The running with LL and NLL resummation, together with the corresponding LO and NLO fixed-order approximations without resummation, are shown in~Figure~\ref{fig:RGE eu-SRR}. 

\begin{figure}[tb]
    \centering
    \includegraphics[scale=1.05]{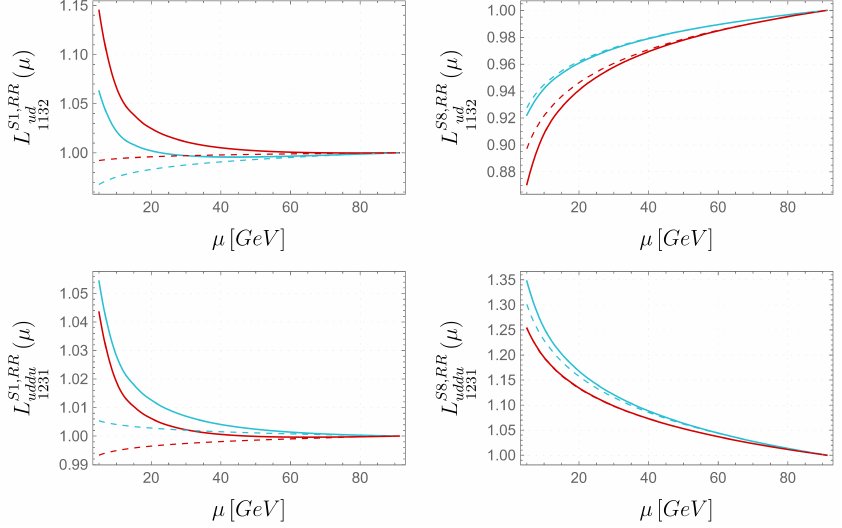}
    \caption{Running from the EW scale $\mu = M_Z$ down to the bottom mass scale $\mu = m_b$ with LL ({\color{blue-plot}blue}) and NLL ({\color{red-plot}red}) resummation, and the corresponding LO ({\color{blue-plot}blue-dashed}) and NLO ({\color{red-plot}red-dashed}) fixed-order approximations. The Wilson coefficients are shown normalized to~1 at the matching scale $\mu = M_Z$.} 
    \label{fig:RGE ud-SRR}
\end{figure}

A more interesting case to study is the running of an SRR sub-block in the $ud$-type $\Delta F = 1$ sector. This case features operator mixing, which can lead to greater effects in the running. As an example, the two-loop beta function of the first coefficient in the sub-block is given by
\eqa{
\dlwc{ud}{RR}[S1][ijkl] =
& 
\ g_s^2\, \Bigg[\frac{64}{9} \lwc{uddu}{RR}[S1][ilkj] +\frac{112}{27} \lwc{uddu}{RR}[S8][ilkj] -16 \lwc{ud}{RR}[S1][ijkl] +\frac{16}{9} \lwc{ud}{RR}[S8][ijkl] \Bigg] 
\nonumber\\ 
& 
+e^2\, \Bigg[\frac{4}{27} \lwc{uddu}{RR}[S1][ilkj] +\frac{16}{81} \lwc{uddu}{RR}[S8][ilkj] -\frac{46}{9} \lwc{ud}{RR}[S1][ijkl] \Bigg] 
\nonumber\\ 
& 
+\alpha_s^2\,  \Bigg[\frac{9472}{81} \lwc{uddu}{RR}[S1][ilkj] +\frac{16576}{243} \lwc{uddu}{RR}[S8][ilkj] 
-\frac{2120}{9} \lwc{ud}{RR}[S1][ijkl] +\frac{928}{81} \lwc{ud}{RR}[S8][ijkl] \Bigg]
\nonumber\\ 
&
+\alpha_s \alpha\, \Bigg[\frac{1888}{81} \lwc{uddu}{RR}[S1][ilkj] +\frac{3808}{243} \lwc{uddu}{RR}[S8][ilkj] +\frac{136}{27} \lwc{ud}{RR}[S1][ijkl] 
+\frac{224}{81} \lwc{ud}{RR}[S8][ijkl] \Bigg] 
\nonumber\\ 
&
+\alpha^2 \,\Bigg[\frac{98}{729} \lwc{uddu}{RR}[S1][ilkj] +\frac{392}{2187} \lwc{uddu}{RR}[S8][ilkj] +\frac{6391}{243} \lwc{ud}{RR}[S1][ijkl] \Bigg] \ .
}
Although the ADM of this sector already includes mixing at the LO, notable effects can still emerge depending on the initial conditions. We assume the following matching conditions at the EW scale:
\eq{
\begin{aligned}
&\lwc{ud}{RR}[S1][1132](M_Z) = 7.3 \cdot 10^{-7}\GeV^{-2}, &\lwc{ud}{RR}[S8][1132](M_Z) = 6.2 \cdot 10^{-6}\GeV^{-2}, \\
&\lwc{uddu}{RR}[S1][1231](M_Z) = 1.7 \cdot 10^{-6}\GeV^{-2} , &\lwc{uddu}{RR}[S8][1231](M_Z) = -2.4 \cdot 10^{-6}\GeV^{-2}.
\end{aligned}
}
Their running with LL and NLL resummation, together with the corresponding LO and NLO fixed-order approximations are depicted in Figure~\ref{fig:RGE ud-SRR}. Up to a scale of $\sim 40\GeV$ all four results are compatible within $\lesssim 5\%$. Below that scale, however, the differences start accumulating quite remarkably. It is interesting to note that the effects become quite sizable in going from LL to NLL running. 

Finally, we study the running of four operators affected by penguin-type mixing: two corresponding to $\Delta F = 1$ transitions, and two being part of the large vector sub-block in $\Delta F = 0$. 
The ADMs receiving penguin contributions are much less sparse than those affected only by current-current contributions, and this effect is enhanced at two loops with respect to the one-loop case. As such, the corrections brought upon by the NLL running can be more prominent in this case, although the picture is highly dependent on the particular matching conditions at hand, with high variability between sets of initial conditions. Again, we choose a set of randomized initial conditions for all Wilson coefficients, drawn uniformly form the interval $(-10^{-5},10^{-5})\GeV^{-2}$. 

\begin{figure}[tb]
    \centering
    \includegraphics[scale=1.05]{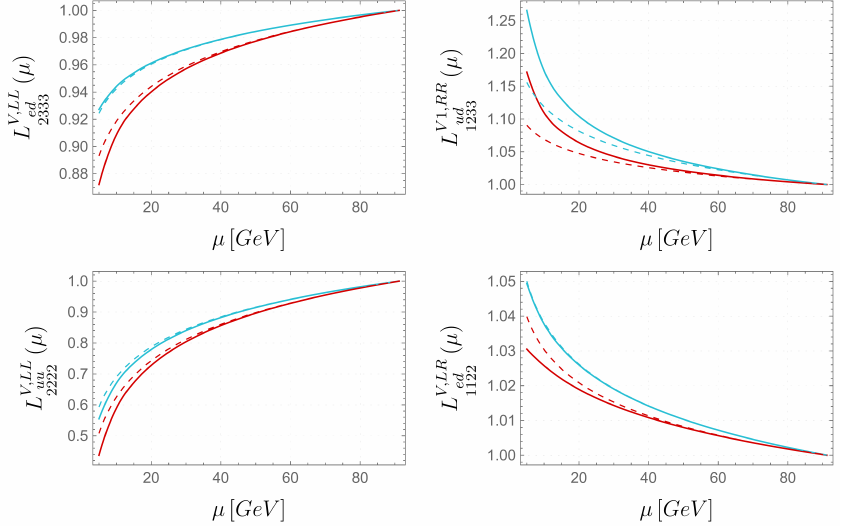}
    \caption{Running from the EW scale $\mu = M_Z$ down to the bottom mass scale $\mu = m_b$ with LL ({\color{blue-plot}blue}) and NLL ({\color{red-plot}red}) resummation, and the corresponding LO ({\color{blue-plot}blue-dashed}) and NLO ({\color{red-plot}red-dashed}) fixed-order approximations. The Wilson coefficients are shown normalized to~1 at the matching scale $\mu = M_Z$.} 
    \label{fig:RGE Penguin}
\end{figure}

The resulting running, as depicted in Figure~\ref{fig:RGE Penguin}, shows a varied picture. For many of the Wilson coefficients the NLL corrections are small, as for example for \pkg{LedVLR[1,1,2,2]} (in \dsix notation). Some operators, however, show large differences in their LL and NLL running. For instance, the running of \pkg{LedVLL[1,2,3,3]} and \pkg{LedVLL[2,3,3,3]} diverges notably, even though the NLL effects remain moderate ($<10\%$) compared to LL. In the case of \pkg{LuuV1LL[2,2,2,2]} the difference between the running at LL and NLL amounts to a $20\%$ effect. 
%

\section{Conclusions}
\label{sec:conclusions}

In this article we have derived the complete set of two-loop four-fermion ADMs in the LEFT in the NDR scheme. For the derivation we have adopted three different approaches which are based on results available in the literature. The first approach consists of transforming ADMs already known from the literature into the JMS basis via NLO basis changes, which allow to obtain parts of the $\Delta F = 2$, $\Delta F = 1$ and $\Delta F = 1^{\bar f f}$ sectors. The second method is based on exploiting flavor-symmetric properties of ADMs, which allows to derive the ADMs for the sectors $\Delta F = 2$, $\Delta F = 1.5$ and $\Delta F = 1^{\bar f f}$. These two approaches provide the ADMs in the JMS basis only to $O(\alpha_s^2)$. 
The third approach employs ``tables of pole coefficients'', available from the literature, from which the ADMs can be directly derived. This method provides the complete NLO four-fermion ADM in the JMS basis, including $O(\alpha_s^2)$, $O(\alpha_s \alpha)$ and $O(\alpha^2)$ corrections. 
With this method we have also verified the results from the two previous approaches. 
Regarding penguin contributions, this method seems necessary in order to tackle the issue of traces involving $\gamma_5$ in the NDR scheme. 
But it also illustrates that arguments based on flavor universality together with the proper treatment of evanescent operators are useful when deriving ADMs and provide powerful checks. 
Finally, as another consistency check we have recomputed several parts of the ADM via direct calculations, finding full agreement.

Many of our results are new. To the best of our knowledge the complete BNV sector, parts of the $\Delta F = 1^{\bar f f}$ sector at order $O(\alpha_s \alpha)$, most of the $\Delta F = 0$ sector, as well as all four-fermion sectors at order $O(\alpha^2)$ have never been presented before. We have also provided an exhaustive list of all four-fermion flavor sectors in the LEFT, where each sector is further divided into RG-invariant sub-blocks.

The results for the complete two-loop ADM are provided in an ancillary file, together with an implementation in \dsix format~\cite{Celis:2017hod,Fuentes-Martin:2020zaz}. We plan to provide an analogous implementation for \texttt{wilson}~\cite{Aebischer:2018bkb} in the near future.

Finally, we have performed a numerical study of the NLO RGE effects in comparison to the running at LO. The two-loop corrections can be sizable, especially in the $\Delta F = 1^{\bar f f}$ and $\Delta F = 0$ sectors, where penguin diagrams mix a large number of operators at NLO.


\acknowledgments

We thank Luca Naterop and Peter Stoffer, as well as Lukas Born and Anders Eller Thomsen, for cross-checks helpful in fixing bugs affecting some of the results in the first version of this paper.
The work of J.A. is supported by the European Union’s Horizon 2020 research and innovation program under the Marie Skłodowska-Curie grant agreement No.~101145975-EFT-NLO.
P.M. acknowledges funding from the Spanish MCIN/AEI/10.13039/501100011033: grant PRE2022-103999 funded by MCIN/AEI/10.13039/501100011033 and by "ESF Investing in your future", grant CEX2019-000918-M through the “Unit of Excellence Mar\'ia de Maeztu 2020-2023” award to the Institute of Cosmos Sciences.
M.P acknowledges funding by the European Research Council (ERC) under the European Union’s Horizon 2020 research and innovation programme under grant agreement 833280 (FLAY), and by the Swiss National Science Foundation (SNF) under contract 200020-204428.
J.V. acknowledges funding from grant 2021-SGR-249 (Generalitat de Catalunya), and from the Spanish MCIN/AEI/10.13039/501100011033 through the following grants: grant CNS2022-135262 funded by the “European Union NextGenerationEU/PRTR”, grant CEX2019-000918-M through the “Unit of Excellence Mar\'ia de Maeztu 2020-2023” award to the Institute of Cosmos Sciences, and grant PID2022-136224NB-C21.


\appendix

\newpage

\section{Evanescent Operator Basis}
\label{app:EV Basis}

The set of evanescent operators used to complete the JMS operator basis in the NDR scheme corresponds to $a_{\text{ev}}, b_{\text{ev}}, c_{\text{ev}}, \cdots =1$ in~\cite{Dekens:2019ept}, which is equivalent to the method of ``greek projections''~\cite{Tracas:1982gp,Buras:2000if}. This choice also matches the evanescent basis used in~\cite{Buras:1992tc,Buras:1992zv,Ciuchini:1993vr}. For simplicity, we refrain from displaying the full form of the evanescent basis, but instead present the general structures built out of generic fermion fields $\psi_{i,j,k,l}$. The evanescent structures read:
\begin{align}
E^{V,LL}_{1} \;\sim\quad & (\bar{\psi}_i\gamma^{\mu}P_L\psi_l)(\bar{\psi}_k\gamma_{\mu}P_L\psi_j) - (\bar{\psi}_i\gamma^{\mu}P_L\psi_j)(\bar{\psi}_k\gamma_{\mu}P_L\psi_l) \;, 
\\[1mm] 
E^{V,LL}_{2} \;\sim\quad & (\bar{\psi}_i\gamma^{\mu}\gamma^{\nu}\gamma^{\rho}P_L\psi_j)(\bar{\psi}_k\gamma_{\mu}\gamma_{\nu}\gamma_{\rho}P_L\psi_l) - (16-4\epsilon)(\bar{\psi}_i\gamma^{\mu}P_L\psi_j)(\bar{\psi}_k\gamma_{\mu}P_L\psi_l) \;, 
\\[0.5mm] 
E^{V,LL}_{3} \;\sim\quad & (\bar{\psi}_i\gamma^{\mu}\gamma^{\nu}\gamma^{\rho}\gamma^{\sigma}\gamma^{\lambda}P_L\psi_j)(\bar{\psi}_k\gamma_{\mu}\gamma_{\nu}\gamma_{\rho}\gamma_{\sigma}\gamma_{\lambda}P_L\psi_l) 
\nonumber\\
& \hspace{4cm} - (256-224\epsilon)(\bar{\psi}_i\gamma^{\mu}P_L\psi_j)(\bar{\psi}_k\gamma_{\mu}P_L\psi_l) \;, 
\\[0.5mm]  
E^{V,LR}_{1} \;\sim\quad & 2(\bar{\psi}_iP_R\psi_l)(\bar{\psi}_kP_L\psi_j) + (\bar{\psi}_i\gamma^{\mu}P_L\psi_j)(\bar{\psi}_k\gamma_{\mu}P_R\psi_l) \;, 
\\[0.5mm]
E^{V,LR}_{2} \;\sim\quad & (\bar{\psi}_i\gamma^{\mu}\gamma^{\nu}\gamma^{\rho}P_L\psi_j)(\bar{\psi}_k\gamma_{\mu}\gamma_{\nu}\gamma_{\rho}P_R\psi_l) - (4+4\epsilon)(\bar{\psi}_i\gamma^{\mu}P_L\psi_j)(\bar{\psi}_k\gamma_{\mu}P_R\psi_l) \;, \\ 
E^{V,LR}_{3} \;\sim\quad & (\bar{\psi}_i\gamma^{\mu}\gamma^{\nu}\gamma^{\rho}\gamma^{\sigma}\gamma^{\lambda}P_L\psi_j)(\bar{\psi}_k\gamma_{\mu}\gamma_{\nu}\gamma_{\rho}\gamma_{\sigma}\gamma_{\lambda}P_R\psi_l) 
\nonumber\\
& \hspace{4cm} - (16+128\epsilon)(\bar{\psi}_i\gamma^{\mu}P_L\psi_j)(\bar{\psi}_k\gamma_{\mu}P_R\psi_l) \;, 
\\[0.5mm] 
E^{S,RL}_{1} \;\sim\quad & (\bar{\psi}_i\sigma^{\mu\nu}P_R\psi_j)(\bar{\psi}_k\sigma_{\mu\nu}P_L\psi_l) + 6\epsilon(\bar{\psi}_iP_R\psi_j)(\bar{\psi}_kP_L\psi_l) \;, 
\\[0.5mm]  
E^{S,RL}_{2} \;\sim\quad & (\bar{\psi}_i\gamma^{\mu}\gamma^{\nu}\gamma^{\rho}\gamma^{\sigma}P_R\psi_j)(\bar{\psi}_k\gamma_{\mu}\gamma_{\nu}\gamma_{\rho}\gamma_{\sigma}P_L\psi_l) - (16 + 128\epsilon)(\bar{\psi}_i P_R\psi_j)(\bar{\psi}_k P_L\psi_l) \;, 
\\[0.5mm] 
E^{S,RL}_{3} \;\sim\quad & (\bar{\psi}_i\gamma^{\mu}\gamma^{\nu}\gamma^{\rho}\gamma^{\sigma}\gamma^{\lambda}\gamma^{\eta}P_R\psi_j)(\bar{\psi}_k\gamma_{\mu}\gamma_{\nu}\gamma_{\rho}\gamma_{\sigma}\gamma_{\lambda}\gamma_{\eta}P_L\psi_l) 
\nonumber\\
& \hspace{4cm} - (64 + 2880\epsilon)(\bar{\psi}_i P_R\psi_j)(\bar{\psi}_k P_L\psi_l) \;, 
\\[0.5mm] 
E^{S,RR}_{1} \;\sim\quad & (\bar{\psi}_i\sigma^{\mu\nu}P_R\psi_j)(\bar{\psi}_k\sigma_{\mu\nu}P_R\psi_l) + 4(\bar{\psi}_i P_R\psi_j)(\bar{\psi}_k P_R\psi_l) + 8(\bar{\psi}_i P_R\psi_l)(\bar{\psi}_k P_R\psi_j) \;, 
\\[0.5mm]  
E^{S,RR}_{2} \;\sim\quad & (\bar{\psi}_i\gamma^{\mu}\gamma^{\nu}\gamma^{\rho}\gamma^{\sigma}P_R\psi_j)(\bar{\psi}_k\gamma_{\mu}\gamma_{\nu}\gamma_{\rho}\gamma_{\sigma}P_R\psi_l) - (64-96\epsilon)(\bar{\psi}_i P_R\psi_j)(\bar{\psi}_k P_R\psi_l) 
\nonumber\\
& \hspace{4cm} + (16 - 8\epsilon)(\bar{\psi}_i\sigma^{\mu\nu}P_R\psi_j)(\bar{\psi}_k\sigma_{\mu\nu}P_R\psi_l) \;, 
\\[0.5mm] 
E^{S,RR}_{3} \;\sim\quad & (\bar{\psi}_i\gamma^{\mu}\gamma^{\nu}\gamma^{\rho}\gamma^{\sigma}\gamma^{\lambda}\gamma^{\eta}P_R\psi_j)(\bar{\psi}_k\gamma_{\mu}\gamma_{\nu}\gamma_{\rho}\gamma_{\sigma}\gamma_{\lambda}\gamma_{\eta}P_R\psi_l) 
\nonumber\\ 
& \hspace{4cm} - (1024-2944\epsilon)(\bar{\psi}_i P_R\psi_j)(\bar{\psi}_k P_R\psi_l) 
\nonumber\\
& \hspace{4cm} + (256 - 352\epsilon)(\bar{\psi}_i\sigma^{\mu\nu}P_R\psi_j)(\bar{\psi}_k\sigma_{\mu\nu}P_R\psi_l) \;.
\end{align}
The chirally-flipped sectors are obtained via $P_L \leftrightarrow P_R$.
The symbol $\sim$ is just an equal sign except in the case of four-quark operators, where 
each structure leads to two evanescent operators,
for example:
\begin{equation*}
E^{V,LL}_{1} \longrightarrow \left\{ \begin{array}{c}
E^{\text{V1,LL}}_{\substack{\vphantom{\ell} ud \\ ijkl}} = (\bar{d}_i^{\alpha}\gamma^{\mu}P_Lu_l^{\beta})(\bar{u}_k^{\beta}\gamma_{\mu}P_Ld_j^{\alpha}) - (\bar{d}_i^{\alpha}\gamma^{\mu}P_Ld_j^{\alpha})(\bar{u}_k^{\beta}\gamma_{\mu}P_Lu_l^{\beta}) 
\\[4mm]E^{\text{V2,LL}}_{\substack{\vphantom{\ell} ud \\ ijkl}} = (\bar{d}_i^{\alpha}\gamma^{\mu}P_Lu_l^{\alpha})(\bar{u}_k^{\beta}\gamma_{\mu}P_Ld_j^{\beta}) - (\bar{d}_i^{\alpha}\gamma^{\mu}P_Ld_j^{\beta})(\bar{u}_k^{\beta}\gamma_{\mu}P_Lu_l^{\alpha})
\end{array} \right. \quad.
\end{equation*}
An evanescent basis defined in this manner together with analogous structures for each flavor ensures that the basis is ``flavor-universal'', as argued in Section \ref{sec:Penguin Reconstruction}. Let us also note that any linear combination of the evanescent basis ($E_i' = M_{ij}E_j$) involving no physical operators, leaves the physical anomalous dimensions unchanged. Therefore, any choice of evanescent operators based on the above defined evanescent structures defines a completely equivalent renormalization scheme.

\section{Tables of Pole Coefficients}
\label{app:Pole Tables}

We collect all tables of pole coefficients present in the literature~\cite{Buras:1989xd,Buras:1992tc,Ciuchini:1993vr,Buras:2000if}, involved in the calculation of the two-loop four-fermion ADMs (in the NDR scheme), in Tables~\ref{tab:Tables CC}, \ref{tab:Tables CC2}, and \ref{tab:Tables P}. We have also recomputed some of them from scratch, as a cross-check. Just as in the literature, we refrain from including the color/charge factors. These poles depend only on the Dirac structure, regardless of the individual flavor indices (up to the selection rules for penguin diagrams), and thus they are written in terms of the bases in~\Eq{eq:Table Basis} for current-current and~\Eqs{eq:Lepton Basis}{eq:Quark Basis} for penguin diagrams. In this respect, the tables are presented here in a slightly different manner than in the literature. The chirally-flipped coefficients (VRR, VRL, SRL, SRR) are identical, and thus omitted in the tables. 

Diagrams are grouped and labeled according to the numbering in~\Reff{Buras:1992tc}. The multiplicity in each class is also included in the tables. Current-current diagrams are displayed in Tables~\ref{tab:Tables CC} and~\ref{tab:Tables CC2}, and penguin diagrams are shown in Table~\ref{tab:Tables P}. In case of classes 29, 30 and 31 of the current-current diagrams, we use the compact notation for the coefficient functions:
\eqa{
F_1(N_c,n_q) &= \frac{5}{2}N_c - n_q \ ; \qquad  
F_2(N_c, n_q) = \frac{2}{3}n_q - \frac{11}{3}N_c \ ; \qquad  
F_3(N_c, n_q) = \frac{26}{3}N_c - \frac{8}{3}n_q \ ; 
\nonumber\\ 
F_4(N_c,n_q) &= \frac{4}{9}n_q - \frac{16}{9}N_c \ ; \qquad  
F_5(N_c, n_q) = \frac{17}{72}N_c - \frac{1}{36}n_q \ ; 
\nonumber\\
F_6(N_c, n_q) &= \frac{134}{3}N_c - \frac{44}{3}n_q \ ; \qquad 
F_7(N_c, n_q) = \frac{62}{9}N_c - \frac{20}{9}n_q \ ;
\label{eq:QCD Tab-Functions}
}
which are relevant for the $O(\alpha_s^2)$ results. The respective functions for the $O(\alpha^2)$ coefficients can be obtained by substituting $N_c \to 0$ and $n_q \to 2(Q_d^2 N_c \, n_d + Q_u^2 N_c \, n_u + Q_e^2 \, n_e)$.
\begin{table}
\centering
\setlength{\tabcolsep}{14pt}
\renewcommand{\arraystretch}{1}
\begin{tabular}{@{}c c r r r r r r}
\toprule[0.5mm]
\multicolumn{2}{c}{Current-Current} & \multicolumn{2}{r}{VLL $\to$ VLL} & \multicolumn{2}{r}{VLR $\to$ VLR} & \multicolumn{2}{r}{SLR $\to$ SLR} \\
\midrule 
Diagram & Mult. &
$1/\epsilon^2$ & $1/\epsilon$ & $1/\epsilon^2$ & $1/\epsilon$ & $1/\epsilon^2$ & $1/\epsilon$ \\
\midrule[0.3mm] 
$4$ & $2$ & $-1$ & $5/2$ & $-1$ & $5/2$ & $-16$ & $16$ 
\\
\midrule 
$5$ & $2$ & $-16$ & $16$ & $-1$ & $5/2$ & $-1$ & $5/2$ 
\\
\midrule 
$6$ & $2$ & $-1$ & $5/2$ & $-16$ & $16$ & $-1$ & $5/2$ 
\\
\midrule 
$7$ & $2$ & $0$ & $-4$ & $0$ & $-4$ & $0$ & $-4$ 
\\
\midrule 
$8$ & $2$ & $0$ & $-4$ & $0$ & $-4$ & $0$ & $-4$ 
\\
\midrule 
$9$ & $2$ & $0$ & $-4$ & $0$ & $-4$ & $0$ & $-4$ 
\\
\midrule 
$10$ & $4$ & $-2$ & $3$ & $-2$ & $3$ & $-8$ & $-8$ 
\\
\midrule 
$11$ & $4$ & $8$ & $14$ & $2$ & $3$ & $2$ & $3$ 
\\
\midrule 
$12$ & $4$ & $-2$ & $3$ & $-8$ & $-8$ & $-2$ & $3$ 
\\
\midrule 
$13$ & $4$ & $2$ & $-3$ & $2$ & $-3$ & $8$ & $-4$ 
\\
\midrule 
$14$ & $4$ & $-8$ & $-2$ & $-2$ & $-3$ & $-2$ & $-3$ 
\\
\midrule 
$15$ & $4$ & $2$ & $-3$ & $8$ & $-4$ & $2$ & $-3$ 
\\
\midrule 
$16$ & $4$ & $8$ & $-14$ & $2$ & $-9$ & $8$ & $-16$ 
\\
\midrule 
$17$ & $4$ & $8$ & $14$ & $2$ & $3$ & $8$ & $16$ 
\\
\midrule 
$18$ & $4$ & $-2$ & $3$ & $-8$ & $8$ & $-8$ & $-8$ 
\\
\midrule 
$19$ & $4$ & $-2$ & $3$ & $-8$ & $-8$ & $-8$ & $8$ 
\\
\midrule 
$20$ & $4$ & $8$ & $14$ & $8$ & $16$ & $2$ & $3$ 
\\
\midrule 
$21$ & $4$ & $8$ & $-14$ & $8$ & $-16$ & $2$ & $-9$ 
\\
\midrule 
$22$ & $1$ & $-1$ & $0$ & $-1$ & $0$ & $-16$ & $0$ 
\\
\midrule 
$23$ & $1$ & $-16$ & $0$ & $-1$ & $0$ & $-1$ & $0$ 
\\
\midrule 
$24$ & $1$ & $-1$ & $0$ & $-16$ & $0$ & $-1$ & $0$ 
\\
\midrule 
$25$ & $4$ & $6$ & $-11$ & $6$ & $-11$ & $24$ & $-20$ 
\\
\midrule 
$26$ & $4$ & $-24$ & $2$ & $-6$ & $-7$ & $-6$ & $-7$ 
\\
\midrule 
$27$ & $4$ & $6$ & $-11$ & $24$ & $-20$ & $6$ & $-11$ 
\\
\midrule 
$28$ & $4$ & $0$ & $0$ & $0$ & $0$ & $0$ & $0$ 
\\
\midrule 
$29$ & $2$ & $0$ & $F_1$ & $0$ & $F_1$ & $-2F_1$ & $F_3$
\\
\midrule 
$30$ & $2$ & $2F_1$ & $F_2$ & $0$ & $F_1$ & $0$ & $F_1$
\\
\midrule 
$31$ & $2$ & $0$ & $F_1$ & $-2F_1$ & $F_3$ & $0$ & $F_1$ 
\\ 
\bottomrule[0.5mm]
\end{tabular}
\caption{\it Table of pole coefficients of the two-loop current-current diagrams with insertion of VLL, VLR and SLR operators. The multiplicity of each diagram is already taken into account in the coefficients: each coefficient combines already all diagrams in the same class.}
\label{tab:Tables CC}
\end{table}
\begin{table}
\centering
\setlength{\tabcolsep}{7pt}
\renewcommand{\arraystretch}{1}
\begin{tabular}{@{}c c r r r r r r r r}
\toprule[0.7mm]
\multicolumn{2}{c}{Current-Current} & \multicolumn{2}{r}{SRR $\to$ SRR} & \multicolumn{2}{r}{SRR $\to$ TRR} & \multicolumn{2}{r}{TRR $\to$ SRR} & \multicolumn{2}{r}{TRR $\to$ TRR} \\
\midrule 
Diagram & Mult. &
$1/\epsilon^2$ & $1/\epsilon$ & $1/\epsilon^2$ & $1/\epsilon$ & $1/\epsilon^2$ & $1/\epsilon$ & $1/\epsilon^2$ & $1/\epsilon$ \\
\midrule[0.7mm] 
$4$ & $2$ & $-16$ & $16$ & $0$ & $0$ & $0$ & $0$ & $0$ & $0$ 
\\
\midrule 
$5$ & $2$ & $-4$ & $9$ & $1$ & $-5/4$ & $48$ & $-76$ & $-12$ & $7$ 
\\
\midrule 
$6$ & $2$ & $-4$ & $9$ & $-1$ & $7/4$ & $-48$ & $52$ & $-12$ & $7$ 
\\
\midrule 
$7$ & $2$ & $0$ & $-4$ & $0$ & $0$ & $0$ & $0$ & $0$ & $4$ 
\\
\midrule 
$8$ & $2$ & $0$ & $2$ & $0$ & $1/2$ & $0$ & $24$ & $0$ & $-2$ 
\\
\midrule 
$9$ & $2$ & $0$ & $2$ & $0$ & $-1/2$ & $0$ & $-24$ & $0$ & $-2$ 
\\
\midrule 
$10$ & $4$ & $-8$ & $-8$ & $0$ & $0$ & $0$ & $0$ & $0$ & $4$ 
\\
\midrule 
$11$ & $4$ & $2$ & $0$ & $-1/2$ & $-5/4$ & $-24$ & $-20$ & $6$ & $8$ 
\\
\midrule 
$12$ & $4$ & $-2$ & $0$ & $-1/2$ & $-5/4$ & $-24$ & $-20$ & $-6$ & $-4$ 
\\
\midrule 
$13$ & $4$ & $8$ & $-4$ & $0$ & $0$ & $0$ & $0$ & $0$ & $0$ 
\\
\midrule 
$14$ & $4$ & $-2$ & $0$ & $1/2$ & $1/4$ & $24$ & $-28$ & $-6$ & $0$ 
\\
\midrule 
$15$ & $4$ & $2$ & $0$ & $1/2$ & $1/4$ & $24$ & $-28$ & $6$ & $-4$ 
\\
\midrule 
$16$ & $4$ & $8$ & $-4$ & $0$ & $0$ & $-96$ & $-64$ & $0$ & $0$ 
\\
\midrule 
$17$ & $4$ & $8$ & $4$ & $-2$ & $-2$ & $0$ & $0$ & $0$ & $0$ 
\\
\midrule 
$18$ & $4$ & $-8$ & $4$ & $0$ & $0$ & $-96$ & $-64$ & $0$ & $0$ 
\\
\midrule 
$19$ & $4$ & $-8$ & $-4$ & $-2$ & $-2$ & $0$ & $0$ & $0$ & $0$ 
\\
\midrule 
$20$ & $4$ & $-4$ & $10$ & $1$ & $-1$ & $-48$ & $64$ & $12$ & $2$ 
\\
\midrule 
$21$ & $4$ & $-4$ & $10$ & $-1$ & $2$ & $48$ & $-112$ & $12$ & $-22$ 
\\
\midrule 
$22$ & $1$ & $-16$ & $0$ & $0$ & $0$ & $0$ & $0$ & $0$ & $0$ 
\\
\midrule 
$23$ & $1$ & $-4$ & $5$ & $1$ & $-1/4$ & $48$ & $-28$ & $-12$ & $-5$ 
\\
\midrule 
$24$ & $1$ & $-4$ & $5$ & $-1$ & $3/4$ & $-48$ & $4$ & $-12$ & $-5$ 
\\
\midrule 
$25$ & $4$ & $24$ & $-20$ & $0$ & $0$ & $0$ & $0$ & $0$ & $0$ 
\\
\midrule 
$26$ & $4$ & $-6$ & $2$ & $3/2$ & $1/4$ & $72$ & $-108$ & $-18$ & $6$ 
\\
\midrule 
$27$ & $4$ & $6$ & $-2$ & $3/2$ & $1/4$ & $72$ & $-108$ & $18$ & $-18$ 
\\
\midrule 
$28$ & $4$ & $0$ & $0$ & $0$ & $-3$ & $0$ & $144$ & $0$ & $0$  
\\
\midrule 
$29$ & $4$ & $-2F_1$ & $F_3$ & $0$ & $0$ & $0$ & $0$ & $2F_1/3$ & $F_4$ 
\\
\midrule 
$30$ & $4$ & $0$ & $0$ & $-F_1/12$ & $F_5$ & $-8F_1$ & $F_6$ & $4F_1/3$ & $2F_4$ 
\\
\midrule 
$31$ & $4$ & $0$ & $0$ & $-F_1/12$ & $F_5$ & $-8F_1$ & $F_6$ & $-4F_1/3$ & $F_7$ 
\\
\bottomrule[0.7mm]
\end{tabular}
\caption{\it Table of pole coefficients of the two-loop current-current diagrams with insertion of SRR and TRR operators. The multiplicity of each diagram is already taken into account in the coefficients: each coefficient combines already all diagrams in the same class.}
\label{tab:Tables CC2}
\end{table}
\begin{table}
\centering
\setlength{\tabcolsep}{7pt}
\renewcommand{\arraystretch}{1}
\begin{tabular}{@{}c c r r r r r r r r}
\toprule[0.7mm]
\multicolumn{2}{c}{Penguin} & \multicolumn{2}{r}{VLL $\to$ VLL} & \multicolumn{2}{r}{VLL $\to$ VLR} & \multicolumn{2}{r}{SLR $\to$ VLL} & \multicolumn{2}{r}{SLR $\to$ VLR} \\
\midrule 
Diagram & Mult. &
$1/\epsilon^2$ & $1/\epsilon$ & $1/\epsilon^2$ & $1/\epsilon$ & $1/\epsilon^2$ & $1/\epsilon$ & $1/\epsilon^2$ & $1/\epsilon$ \\
\midrule[0.7mm] 
 $1$ & $1$ & $2/3$ & $0$ & $2/3$ & $0$ & $2/3$ & $0$ & $2/3$ & $0$ 
 \\
\midrule 
 $2$ & $1$ & $2/3$ & $-19/9$ & $2/3$ & $-19/9$ & $2/3$ & $-13/9$ & $2/3$ & $-13/9$ 
 \\
\midrule 
 $3$ & $2$ & $0$ & $-17/9$ & $0$ & $-17/9$ & $2$ & $10/9$ & $2$ & $10/9$ 
 \\
\midrule 
 $4$ & $2$ & $-2$ & $38/9$ & $-2$ & $38/9$ & $0$ & $23/9$ & $0$ & $23/9$ 
 \\
\midrule 
 $5$ & $2$ & $-2/3$ & $10/9$ & $-2/3$ & $10/9$ & $-2/3$ & $4/9$ & $-2/3$ & $4/9$ 
 \\
\midrule 
 $6$ & $1$ & $-4/3$ & $-29/9$ & $-4/3$ & $-29/9$ & $-4/3$ & $-11/9$ & $-4/3$ & $-11/9$ 
 \\
\midrule 
 $7$ & $2$ & $11/9$ & $35/54$ & $11/9$ & $35/54$ & $11/9$ & $-31/54$ & $11/9$ & $-31/54$ 
 \\
\midrule 
 $8$ & $2$ & $-4/9$ & $22/27$ & $-4/9$ & $22/27$ & $-4/9$ & $34/27$ & $-4/9$ & $34/27$ 
 \\
\midrule 
 $9$ & $2$ & $0$ & $0$ & $0$ & $0$ & $0$ & $0$ & $0$ & $0$ 
 \\
\midrule 
 $10$ & $2$ & $0$ & $0$ & $0$ & $0$ & $0$ & $0$ & $0$ & $0$ 
 \\
\midrule 
 $11$ & $1$ & $1$ & $-5/6$ & $1$ & $13/6$ & $1$ & $7/6$ & $1$ & $-11/6$ 
 \\
\midrule 
 $12$ & $1$ & $-1$ & $-13/6$ & $-1$ & $5/6$ & $-1$ & $11/6$ & $-1$ & $-7/6$ 
 \\
\midrule 
 $13$ & $2$ & $-2$ & $-4/3$ & $0$ & $-1$ & $-2$ & $2/3$ & $0$ & $-1$
 \\
\midrule 
 $14$ & $2$ & $0$ & $-1$ & $2$ & $-2/3$ & $0$ & $-1$ & $2$ & $-8/3$ 
 \\
\bottomrule[0.7mm]
\end{tabular}
\caption{\it Table of pole coefficients of the two-loop penguin diagrams (open-penguin type) with insertion of VLL, and SLR operators. The multiplicity of each diagram is already taken into account in the coefficients: each coefficient combines already all diagrams in the same class.}
\label{tab:Tables P}
\end{table}
%

\section{Penguin ADM Seeds}
\label{app:ADM Seeds}

In this appendix we provide the set of two-loop penguin ADM seeds required for the derivation of the complete four-fermion ADM at NLO in the JMS basis. They are given as matrices in the format of~\Eq{eq:Open Penguin ADM Seeds} and~\eqref{eq:Closed Penguin ADM Seeds}. Let us first generalize their definition in order to include both the lepton and down-quark bases in \Eqs{eq:Lepton Basis}{eq:Quark Basis}. The generalized $(n+m)$-loop ADMs for these bases read:
\begin{align}\hat\gamma^{(n,m)}_p\Bigg|_{\text{Lepton Basis}} = &
\arraycolsep=5pt
\def\arraystretch{1.25}
\begin{pmatrix}
0_{4\times4} & \hat O^{(n,m)}_{P_{e}} & \hat O^{(n,m)}_{P_{u}} & \hat O^{(n,m)}_{P_{d}} \\
0_{6\times4} & \hat{X}^{(n,m)}_e & \hat{X}^{(n,m)}_u & \hat{X}^{(n,m)}_d \\
\end{pmatrix} \;, \\
\hat\gamma^{(n,m)}_p\Bigg|_{\text{Down-Quark Basis}} = &
\arraycolsep=5pt 
\def\arraystretch{1.25}
\begin{pmatrix}
0_{6\times6} & \hat Q^{(n,m)}_{P_{e}} & \hat Q^{(n,m)}_{P_{u}} & \hat Q^{(n,m)}_{P_{d}} \\
0_{10\times6} & \hat{Y}^{(n,m)}_e & \hat{Y}^{(n,m)}_u & \hat{Y}^{(n,m)}_d \\
\end{pmatrix} \;,
\end{align}
while the corresponding results for the alternative bases in~Eqs.~\eqref{eq:Alternate Lepton Basis} and~\eqref{eq:Alternate Quark Basis} read:
\begin{align}
\hat\gamma^{(n,m)}_p\Bigg|_{\text{Alt. Lep. Basis}} = &
    \arraycolsep=5pt
    \def\arraystretch{1.25}
    \begin{pmatrix}
        0_{4\times4} & \hat C^{(n,m)}_{P_{e}} & \hat C^{(n,m)}_{P_{u}} & \hat C^{(n,m)}_{P_{d}} \\
        0_{6\times4} & \hat{X}^{(n,m)}_e & \hat{X}^{(n,m)}_u & \hat{X}^{(n,m)}_d \\
    \end{pmatrix} \;, \\
    \hat\gamma^{(n,m)}_p\Bigg|_{\text{Alt. Down-Quark Basis}} = &
    \arraycolsep=5pt 
    \def\arraystretch{1.25}
    \begin{pmatrix}
        0_{6\times6} & \hat K^{(n,m)}_{P_{e}} & \hat K^{(n,m)}_{P_{u}} & \hat K^{(n,m)}_{P_{d}} \\
        0_{10\times6} & \hat{Y}^{(n,m)}_e & \hat{Y}^{(n,m)}_u & \hat{Y}^{(n,m)}_d \\
    \end{pmatrix} \;.
\end{align}
Note that the structure of all penguin contributions mixing only into $\op_{5-10}^{(\ell)}$ or $\op_{7-16}^{(d)}$ is maintained. Due to the structure of the penguin operators the corresponding seeds depend the fermion charges $Q_f$ that characterize the $\Delta F = 1$ transition, $d_{i,j}$ or $\ell_{i,j}$ in \Eqs{eq:Lepton Basis}{eq:Quark Basis}, i.e. the ``outer flavors'', as well as on the charges of the ``inner'' flavors. Neutrinos are included in the leptonic case, but their charge should be put to zero ($Q_\ell \to Q_\nu = 0$).

The matrices $\hat X^{(n,m)}_{e,u,d}$ and $\hat Y^{(n,m)}_{e,u,d}$ are not relevant for our purposes, since they do not affect the change of basis. For completeness, we report them in the following as a function of the penguin seeds
\begin{align}
    \label{eq:X Matrix}
    \hat{X}^{(n,m)}_\psi = & 
    \arraycolsep=5pt 
    \def\arraystretch{1.5}
    \begin{pmatrix}
        \hat M_C \left(n_e \, \hat C^{(n,m)}_{P_\psi} + 2 \, \hat O^{(n,m)}_{P_\psi} \right) \Big|_{Q_f \to Q_e}\\
        n_u \hat N_C \, \hat C^{(n,m)}_{P_\psi} \Big|_{Q_f \to Q_u} \\
        n_d \hat N_C \, \hat C^{(n,m)}_{P_\psi} \Big|_{Q_f \to Q_d} \\
    \end{pmatrix}  \;, \\ 
    \label{eq:Y Matrix}
    \hat{Y}^{(n,m)}_\psi = &
    \arraycolsep=5pt 
    \def\arraystretch{1.5}
    \begin{pmatrix}
        \hat M_K \, \hat K^{(n,m)}_{P_\psi} \Big|_{Q_f \to Q_e} \\
        \hat N_K \left( n_u \, \hat K^{(n,m)}_{P_\psi} + 2 \, \delta_{q\hspace{0.5pt}u} \, \hat N_K \, \hat Q^{(n,m)}_{P_\psi} \right) \Big|_{Q_f \to Q_u} \\
        \hat N_K \left( n_d \, \hat K^{(n,m)}_{P_\psi} + 2 \, \delta_{q\hspace{0.5pt}d} \, \hat N_K \, \hat Q^{(n,m)}_{P_\psi} \right) \Big|_{Q_f \to Q_d} \\
    \end{pmatrix}\,,
\end{align}
where $\psi = e,u,d$ and where the matrices $\hat M_i$ and $\hat N_i$ are given by
\begin{equation}
\arraycolsep=5pt
\def\arraystretch{1.25}
\begin{array}{lcl}
     \hat M_C = \begin{pmatrix}
         \hat{\mathbb{1}}_{2\times 2} & 0_{2\times2}
     \end{pmatrix} \;, & \qquad &
     \hat M_K = \begin{pmatrix}
         \hat{\mathbb{1}}_{2\times2} & 0_{2\times 4} 
     \end{pmatrix} \;, \\
     \hat N_C =  
     \begin{pmatrix}
         0_{2\times2} & \hat{\mathbb{1}}_{2\times 2} 
     \end{pmatrix} \;, & \qquad &
     \hat N_K = 
     \begin{pmatrix}
         0_{4\times2} & \hat{\mathbb{1}}_{4\times 4} 
     \end{pmatrix} \;.
\end{array}
\end{equation}
The explicit lepton-basis and quark-basis penguin seeds,
$\hat O_{P_f}$, $\hat Q_{P_f}$, $\hat C_{P_f}$ and $\hat K_{P_f}$, 
are given in the supplementary material.

\newpage


\end{document}